\documentclass[11pt,fleqn]{article}

\usepackage[braket, qm]{qcircuit}
\usepackage{graphicx}

\usepackage{mathptmx}

\usepackage{amsmath}

\usepackage{pdfpages}

\usepackage{url}

\usepackage{enumitem}

\makeatletter
\renewcommand*\env@matrix[1][\arraystretch]{%
  \edef\arraystretch{#1}%
  \hskip -\arraycolsep
  \let\@ifnextchar\new@ifnextchar
  \array{*\c@MaxMatrixCols c}}
\makeatother

\usepackage{fancyhdr}
\date{}

\title{\bf Hands-on Quantum Programming Labs  \\ for EECS Students} 

\author{\normalsize Janche Sang  \\
        Chansu Yu \\
        Dept. of Electrical Engineering and Computer Science \\
        Cleveland State University \\
        Cleveland, OH 44115 \\
        USA
        }

\providecommand{\keywords}[1]{\textbf{\textit{Keywords ---}} #1}

\begin{document}

\maketitle

\vskip 0.25in

\begin{abstract}

This report presents a practical approach to teaching quantum computing to Electrical Engineering \& Computer Science (EECS) students
through dedicated hands-on programming labs.  The labs cover a diverse range of topics, encompassing fundamental elements, 
such as entanglement, quantum gates and circuits, as well as advanced algorithms including 
Quantum Key Distribution, Deutsch and Deutsch-Jozsa Algorithms, Simon's algorithm, and Grover's algorithm.
As educators, we aim to share our teaching insights and resources with fellow instructors in the field.
The full lab handouts and program templates are provided for interested instructors. 
Furthermore, the report elucidates the rationale behind the design of each experiment, enabling a deeper 
understanding of quantum computing. 

\end{abstract}

\vskip 0.25in


\newpage

\section{Introduction}

Quantum Computing is a rapidly growing field with many potential applications, 
including cryptography, optimization, pharmaceuticals, materials science, etc.
As more and more companies and organizations invest in quantum research and 
development, the demand for skilled professionals with expertise in quantum 
computing is likely to increase. 
That is, having an understanding of quantum computing and practical experiences
in quantum programming can be a valuable 
asset for Electrical Engineering \& Computer Science (EECS) 
students looking to pursue careers in industry or academia. 
Therefore, in Fall 2022,
we decided to offer an introductory-level Quantum Computing course
the first time for our EECS senior students at Cleveland State University(CSU).  

Recommended by Prof.\ Kiper at Miami University \cite{Miami_U},
we chose Bernhardt's book {\it "Quantum Computing for Everyone"}\cite{Bernhardt}
as the textbook. This book provides simple explanations of the challenging 
mathematics of quantum computing and gives elementary examples to illustrate 
what it means and how it is applied in problem solving.    
However, like many other quantum computing books, this textbook also lacks
the stuff of practical quantum programming, which would be helpful for
the EECS students because quantum programming expertise will likely be in high demand 
as quantum computing continues to advance.
  
Hands-on programming provides a means to bridge the gap between theoretical 
knowledge and practical implementation in quantum computing.
By engaging in hands-on programming, students can gain a deeper understanding of 
quantum concepts such as superposition, entanglement, quantum gates and 
quantum algorithms. For example, it might be easier for students to
understand how the Deutsch\cite{Deu85} and the Deutsch-Jozsa\cite{Deu92} algorithms work
via the quantum phase kickback circuit implementation than
the rigorous proof in math.  
Furthermore, actually writing code and running it on quantum computers or 
simulators, students can explore the real-world implications of quantum 
algorithms and discover novel ways to leverage quantum computing power.
As suggested in \cite{SIGCSE20}, a course with dedicated programming labs
would greatly improve students' learning experience in quantum programming.

Several open-source quantum software development kits, such as
Amazon's Braket, Google's Cirq, IBM's Qiskit, Microsoft's Q\# and QDK, etc. 
are available on-line. A good survey and comparison 
of these frameworks can be found in \cite{QC_frameworks}. 
For the dedicated hands-on labs, we focused on teaching students a single
quantum programming environment throughout the course.
We chose IBM's Qiskit because IBM is the only provider which allows 
unlimited free accesses to actual quantum computers. 
Furthermore, Cleveland Clinic \cite{Cleveland_Clinic} installed the 
IBM Quantum System One \cite{IBM_System_One},
which supports a 127-qubit Eagle processor,  
to accelerate biomedical discoveries.
Through the collaboration between CSU and the Cleveland Clinic\cite{CSU_QC},
our students will have opportunities to conduct experiments on the 
cutting-edge quantum computer.

Therefore, our goal of this work was to design hands-on experiments
for students to implement python-based 
quantum programs during seven dedicated lab sessions.  
Many of the experiments are tied to the textbook 
and students are asked to modify a program template
and then verify their experimental result with the corresponding
example described in the textbook. To give students more practice,
we selected and revised some experiments
from the Qiskit Textbook \cite{Qiskit_Textbook}, Qiskit Computing 
Labs \cite{Qiskit_Lab}, and from the Professors Yanofsky and 
Mannucci's book \cite{Yanofsky}.
Through this report, we would like to share our teaching experience and materials 
with other instructors teaching quantum computing. Interested readers can refer to 
the Appendices for all the hands-on lab handouts and program templates. Instructors 
may contact the authors for solutions to the experiments.


\section{Course Content and Schedule}
 
We outline the course content and schedule in Table~\ref{Content}.
The class is scheduled weekly for 3 consecutive hours
and a 75-minute hands-on lab section is carried out immediately after
the topic has been discussed in the class. 
Performing a lab immediately after the topic is taught ensures that the 
information is still fresh in students' minds.
During a hands-on lab, students can observe the results of their actions 
in real-time. This instant feedback allows them to identify and correct 
any misconceptions or errors, further reinforcing their learning and 
understanding of the material.

\begin{table}[htb]
\caption{Topics and Schedule}
\centerline{}
\begin{center}
\small
\begin{tabular}{|lll|}\hline
         &                        &                 \\
         & Topic 1                & Topic 2         \\
Week 1   & Introduction (Ch. 1)   & Overview: Qubit, Superposition,  \\ 
         &                        & Interference \& Entanglement \\ 
Week 2   & Linear Algebra (Ch. 2) & Linear Algebra (Ch. 2)   \\   
Week 3   & Spin \& Qubits (Ch. 3) & Hands-on Lab 1 (Python \& Jupyter) \\
Week 4   & Spin \& Qubits (Ch. 3) & Quantum Gates (Ch. 7) \\
Week 5   & Test 1                 & Entanglement (Ch. 4)  \\ 
Week 6   & Entanglement (Ch. 4)   & Hands-on Lab 2 (Entanglement) \\
Week 7   & Bell’s inequality (Ch. 5) & Hands-on Lab 3 (Quantum Gates and Circuits) \\
Week 8   & Ekert, QKD (Ch. 5),  Hands-on Lab 4 (QKD)  & Guest Speaker (Cleveland Clinic) \\
Week 9   & Test  2                   & Superdense code, Teleportation \\
         &                           & \& Error Correction (Ch. 7) \\
Week 10  & Deutsch \& Deutsch-Jozsa algorithms (Ch. 8) & Hands-on Lab 5 (Deutsch algorithm) \\
Week 12  & Simon’s algorithm (Ch. 8)  & Hands-on Lab 6 (Simon's algorithm)  \\
Week 14  & Shor’s algorithm \& Grover’s algorithm (Ch. 9) & Hands-on Lab 7 (Grover's algorithm and 3-SAT) \\
Week 15  & Complexity  (Ch. 8)               & Test 3 \\  
         &                        &                 \\ \hline
\end{tabular}
\label{Content}
\end{center}
\end{table}

\section{Hands-on Quantum Programming Labs}

The seven dedicated hands-on labs were conducted in our Linux cluster lab.
Each workstation has pre-installed the software packages and libraries, such as Python 3, Qiskit, and Jupyter, 
into a virtual environment. The installation commands can be found in the Appendices.
Note that the Jupyter Notebook \cite{Jupyter} is an interactive computing environment 
and its cell-based approach makes it easy to organize code and text, run code 
interactively, and create documents that combine executable code, visualizations, 
and explanations.

In the lab, students need to follow the procedure stated in the handout step by step
to implement the programs, record, and explain the results in their lab reports. 
This section gives a brief description of each lab and its objectives.
Some interesting experiments which are important for students to practice
are also presented below.  
Interested instructors can find the full handouts in the Appendices. 

\subsection{Hands-on Lab 1: Python and Jupyter}

\noindent {\bf Objectives: }
\begin{itemize}[noitemsep, topsep=-\parskip]
\item learn the features of Python
\item practice programming with Jupyter Notebook
\end{itemize} 

\vskip \parskip

In the first warm-up lab, each student will set up their working
environment (e.g. installing Qiskit tools, creating a dedicated sub-directory for this course, etc.),
and get familiar with some Linux commands. 
Students will also learn how to implement and run Python 3 programs in Jupyter Notebook's code cells, 
as well as how to load and execute program templates in cells.
Furthermore, they will review some features in the Python language,
such as using indentation instead of a pair of curly braces in C/C++/Java
to indicate a block of code, the dictionary data structure which uses
string as the array index, etc.  
Note that, in this lab, students are asked to copy and run
the quantum "Hello World" program which uses a Hadamard gate to set a qubit  
into a superposition state and then measures it. However, we will not
explain how the code works in detail until the next lab. 
This fosters a sense of curiosity that may lead to more profound 
and lasting learning outcomes.

\subsection{Hands-on Lab 2: Entanglement }

\noindent {\bf Objectives: }
\begin{itemize}[noitemsep, topsep=-\parskip]
\item Qiskit Terra and Qiskit Aer. 
\item Hadamard gate
\item two-qubit entanglement
\end{itemize}

\vskip \parskip

In this lab, students will learn how to write simple
quantum programs in Qiskit.  
A Qiskit program typically follows a structured pattern to firstly
construct the quantum circuits using the Qiskit Terra module and then 
execute the circuits either on a simulator (i.e. the Qiskit Aer module)
or on a real IBM quantum computer. 
Instructors should remind students that the Qiskit's qubits order 
is simply reversed as compared to the order used in
the textbook\cite{Bernhardt} (also in the reference\cite{Yanofsky}).
That is, Qiskit puts $q_0$ rightmost (e.g. $q_2 q_1 q_0$) while
the textbook puts $q_0$ leftmost (e.g. $q_0 q_1 q_2$).   
Students have to be careful while reading Qiskit documents,
Qiskit gate operation matrix, histogram output result, etc. 

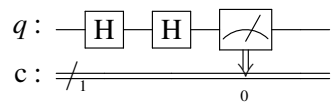
\begin{figure}[htb]
\centering
\scalebox{1.0}{
\Qcircuit @C=1.0em @R=0.2em @!R { \\
                \nghost{{q} :  } & \lstick{{q} :  } & \gate{\mathrm{H}} & \gate{\mathrm{H}} & \meter & \qw & \qw\\
                \nghost{\mathrm{{c} :  }} & \lstick{\mathrm{{c} :  }} & \lstick{/_{_{1}}} \cw & \cw & \dstick{_{_{\hspace{0.0em}0}}} \cw \ar @{<=} [-1,0] & \cw & \cw\\
\\ }}
\caption{Applying two H gates in succession}
\label{HH} 
\centerline{}
\end{figure}

Students will study in detail the quantum "Hello World" program template 
given in the previous lab. Next, they will learn that quantum randomness 
via the H gate is not simply like a classical random coin toss.
That is, as shown in Figure~\ref{HH}, students will modify the template
by applying two H gates in succession and observe that the result 
is not exactly the same as tossing the coin twice. They also need to
verify the result through the state vector calculation. 

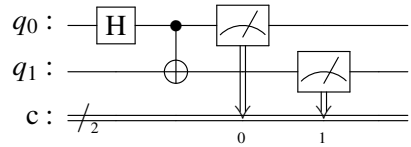
\begin{figure}
\centering
\scalebox{1.0}{
\Qcircuit @C=1.0em @R=0.2em @!R { \\
                \nghost{{q}_{0} :  } & \lstick{{q}_{0} :  } & \gate{\mathrm{H}} & \ctrl{1} & \meter & \qw & \qw & \qw\\
                \nghost{{q}_{1} :  } & \lstick{{q}_{1} :  } & \qw & \targ & \qw & \meter & \qw & \qw\\
                \nghost{\mathrm{{c} :  }} & \lstick{\mathrm{{c} :  }} & \lstick{/_{_{2}}} \cw & \cw & \dstick{_{_{\hspace{0.0em}0}}} \cw \ar @{<=} [-2,0] & \dstick{_{_{\hspace{0.0em}1}}} \cw \ar @{<=} [-1,0] & \cw & \cw\\
\\ }}

\caption{Two-qubit Entanglement: 00 or 11}
\label{Entangle}

\centerline{}
\end{figure}

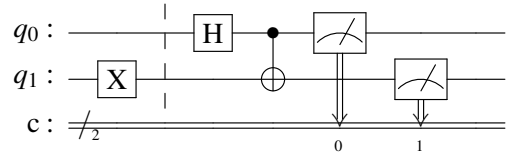
\begin{figure}
\centering
\scalebox{1.0}{
\Qcircuit @C=1.0em @R=0.2em @!R { \\
                \nghost{{q}_{0} :  } & \lstick{{q}_{0} :  } & \qw \barrier[0em]{1} & \qw & \gate{\mathrm{H}} & \ctrl{1} & \meter & \qw & \qw & \qw\\
                \nghost{{q}_{1} :  } & \lstick{{q}_{1} :  } & \gate{\mathrm{X}} & \qw & \qw & \targ & \qw & \meter & \qw & \qw\\
                \nghost{\mathrm{{c} :  }} & \lstick{\mathrm{{c} :  }} & \lstick{/_{_{2}}} \cw & \cw & \cw & \cw & \dstick{_{_{\hspace{0.0em}0}}} \cw \ar @{<=} [-2,0] & \dstick{_{_{\hspace{0.0em}1}}} \cw \ar @{<=} [-1,0] & \cw & \cw\\
\\ }}
\caption{Two-qubit Entanglement: 01 or 10}
\label{AreThey} 
\centerline{}
\end{figure}

The major theme of Lab 2 is entanglement. Students will construct the
two-qubit entanglement circuit shown in Figure~\ref{Entangle} and observe the entangled results
(i.e. either "00" or "11").
In the next experiment, they need to modify the circuit by initializing the qubit $q_1$ to be 1
like in the Figure~\ref{AreThey}. Students are asked to explain whether the result is entangled
or not in their reports. They can refer to page 59 in the textbook about the simple rule
to check the entanglement of two qubits. That is, assume that $r$, $s$, $t$, and $u$ represent the
probability amplitudes of two qubits $q_0 q_1$ states: 00, 01, 10, and 11, respectively.
These two qubits are entangled if $ru \neq st$. Otherwise, they are not entangled
and can be decomposed into the tensor product of two individual qubits.   

After finishing Lab 2, students will be given a homework assignment.
They are asked to run a two-qubit entanglement program on an actual quantum computer.
Each student needs to follow the steps to create a free account on the IBM website,
which allows the user to submit and execute quantum programs on IBM
quantum computers worldwide\cite{Qiskit_RealQC}. 
They need to explain in their report why the histogram result is a little different from 
the result obtained in the Lab2 Experiment using the simulator.

\subsection{Hands-on Lab 3: Quantum Gates and Circuits}

\noindent {\bf Objectives: }
\begin{itemize}[noitemsep, topsep=-\parskip]
\item entanglement with more than 2 qubits in Qiskit
\item constructing circuits with quantum gates such as X, H, CNOT, and Toffoli
\item use of quantum gates to emulate classical gates (AND, NAND, and OR)
\end{itemize}

\vskip \parskip

This lab starts with doing two experiments: one is to entangle three qubits from the same control 
qubit and the other is to entangle four qubits in a chain manner. 
Next, students are asked to use quantum gates such as X, H, CNOT, and Toffoli, 
to construct circuits for emulating the classical AND, NAND, and OR gates. 
Some students may have learned NP-completeness already. Instructors could 
tell them that, for NP-complete problems, such as SAT, quantum computers can give 
quadratic speedup over the classical computers. Learning how to emulate the
classical gates will help them build the oracle circuits 
used in the Grover's search algorithm\cite{Grover} for solving the SAT problem. 
These SAT-related experiments will be conducted in Lab 7 in the future.

To emulate the classical AND gate function, we can simply use
a Toffoli gate {\tt ccx(c0,c1,t)}, where {\tt c0} and {\tt c1} are the control 
qubits and {\tt t} is the target qubit. When both {\tt c0} and {\tt c1} are
set to 1, the target qubit {\tt t} will be flipped to 1.
To implement the NAND gate function, we just need to add
a NOT gate (i.e. the quantum X gate)
to flip the target qubit {\tt t} of the Toffoli gate at the end.

\begin{figure}[htb]
\centering
\scalebox{1.0}{
\Qcircuit @C=1.0em @R=0.8em @!R { \\
	 	\nghost{{q}_{0} :  } & \lstick{{q}_{0} :  } & \ctrl{2} & \qw & \ctrl{1} & \qw & \qw\\
	 	\nghost{{q}_{1} :  } & \lstick{{q}_{1} :  } & \qw & \ctrl{1} & \ctrl{1} & \qw & \qw\\
	 	\nghost{out :  } & \lstick{out :  } & \targ & \targ & \targ & \qw & \qw\\
\\ }}
\caption{Implementing the two-qubit OR function with CNOT and Toffoli gates}
\label{OR1}
\centerline{}
\end{figure}
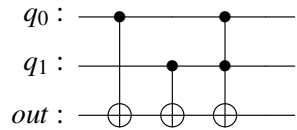

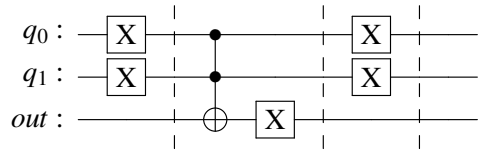
\begin{figure}[htb]
\centering
\scalebox{1.0}{
\Qcircuit @C=1.0em @R=0.2em @!R { \\
	 	\nghost{{q}_{0} :  } & \lstick{{q}_{0} :  } & \gate{\mathrm{X}} \barrier[0em]{2} & \qw & \ctrl{1} & \qw \barrier[0em]{2} & \qw & \gate{\mathrm{X}} \barrier[0em]{2} & \qw & \qw & \qw\\
	 	\nghost{{q}_{1} :  } & \lstick{{q}_{1} :  } & \gate{\mathrm{X}} & \qw & \ctrl{1} & \qw & \qw & \gate{\mathrm{X}} & \qw & \qw & \qw\\
	 	\nghost{out :  } & \lstick{out :  } & \qw & \qw & \targ & \gate{\mathrm{X}} & \qw & \qw & \qw & \qw & \qw\\
\\ }}
\caption{An alternative implementation of the two-qubit OR function}
\label{OR2}
\centerline{}
\end{figure}

Note that there are two different methods to implement the classical OR gate.   
One is to use two CNOT gates and one Toffoli gate, as shown in Figure~\ref{OR1}. 
If either one of the two inputs is 1, it uses the CNOT to toggle the output 
result to be 1. If both inputs are 1's, output result will be toggled twice 
back to 0 and hence we need the Toffoli gate to toggle the output qubit again. 
The other method is to adopt X and AND gates based on the following equation: \\
\mbox{\hskip 0.5in } $q_0 \; \vee \; q_1 \; = \; \neg ( \neg q_0 \; \wedge \neg q_1 )  $ \\
That is, as shown in Figure~\ref{OR2}, we invert both inputs and apply the NAND function.
Finally, we have to invert the inputs back to their original values.  
Note that it has been shown that at least five two-qubit gates are needed to implement
the Toffolli gate\cite{Tofolli}. 
Consequently, both of their circuit depths, which calculates the longest path between the data input and 
the output, appear to be 7.
In other words, there is no significant difference between these two methods for ORing two qubits.

However, if we extend the Boolean expression to OR three qubits: $q_0 \vee q_1 \vee q_2$, 
the first method will have a longer circuit depth than the second approach.
Figure\ref{OR1_3qbits} shows a naive implementation based on the first method.
It is worth noting that we could use an auxiliary qubit $anc$ to obtain $q_0 \vee q_1$ first
and then perform $anc \vee q_2$ to improve the first method, as depicted in Figure\ref{OR1_anc_3qbits}. 
Unfortunately, it still requires additional operations to uncompute and, consequently, free the ancilla qubit.
The implementation using on the second NAND-based approach can be found in Figure\ref{OR2_3qbits}.
Note that the {\tt cccx()} operation in Figure\ref{OR2_3qbits} may also need three Tofolli gates,
but it does not require several CNOT gates as in Figure\ref{OR1_anc_3qbits}.
Therefore, we recommend using the NAND-based method to implement the three-qubit OR function, especially
in the 3-CNF-SAT problem.

\begin{figure}[htb]
\centering
\scalebox{1.0}{
\Qcircuit @C=1.0em @R=0.2em @!R { \\
	 	\nghost{{q}_{0} :  } & \lstick{{q}_{0} :  } & \ctrl{3} & \qw & \qw & \ctrl{1} & \ctrl{2} & \qw & \ctrl{1} & \qw & \qw\\
	 	\nghost{{q}_{1} :  } & \lstick{{q}_{1} :  } & \qw & \ctrl{2} & \qw & \ctrl{2} & \qw & \ctrl{1} & \ctrl{1} & \qw & \qw\\
	 	\nghost{{q}_{2} :  } & \lstick{{q}_{2} :  } & \qw & \qw & \ctrl{1} & \qw & \ctrl{1} & \ctrl{1} & \ctrl{1} & \qw & \qw\\
	 	\nghost{out :  } & \lstick{out :  } & \targ & \targ & \targ & \targ & \targ & \targ & \targ & \qw & \qw\\
\\ }}
\caption{An implementation of the three-qubit OR function }
\label{OR1_3qbits}
\centerline{}
\scalebox{1.0}{
\Qcircuit @C=1.0em @R=0.8em @!R { \\
	 	\nghost{{q}_{0} :  } & \lstick{{q}_{0} :  } & \ctrl{3} & \qw & \ctrl{1} \barrier[0em]{4} & \qw & \qw & \qw & \qw \barrier[0em]{4} & \qw & \ctrl{1} & \qw & \ctrl{3} & \qw & \qw\\
	 	\nghost{{q}_{1} :  } & \lstick{{q}_{1} :  } & \qw & \ctrl{2} & \ctrl{2} & \qw & \qw & \qw & \qw & \qw & \ctrl{2} & \ctrl{2} & \qw & \qw & \qw\\
	 	\nghost{{q}_{2} :  } & \lstick{{q}_{2} :  } & \qw & \qw & \qw & \qw & \ctrl{2} & \qw & \ctrl{1} & \qw & \qw & \qw & \qw & \qw & \qw\\
	 	\nghost{{anc} :  } & \lstick{{anc} :  } & \targ & \targ & \targ & \qw & \qw & \ctrl{1} & \ctrl{1} & \qw & \targ & \targ & \targ & \qw & \qw\\
	 	\nghost{out :  } & \lstick{out :  } & \qw & \qw & \qw & \qw & \targ & \targ & \targ & \qw & \qw & \qw & \qw & \qw & \qw\\
\\ }}
\caption{An implementation of the three-qubit OR function with an ancilla qubit}
\label{OR1_anc_3qbits}
\centerline{}
\scalebox{1.0}{
\Qcircuit @C=1.0em @R=0.2em @!R { \\
	 	\nghost{{q}_{0} :  } & \lstick{{q}_{0} :  } & \gate{\mathrm{X}} \barrier[0em]{3} & \qw & \ctrl{1} & \qw \barrier[0em]{3} & \qw & \gate{\mathrm{X}} & \qw & \qw\\
	 	\nghost{{q}_{1} :  } & \lstick{{q}_{1} :  } & \gate{\mathrm{X}} & \qw & \ctrl{1} & \qw & \qw & \gate{\mathrm{X}} & \qw & \qw\\
	 	\nghost{{q}_{2} :  } & \lstick{{q}_{2} :  } & \gate{\mathrm{X}} & \qw & \ctrl{1} & \qw & \qw & \gate{\mathrm{X}} & \qw & \qw\\
	 	\nghost{out :  } & \lstick{out :  } & \qw & \qw & \targ & \gate{\mathrm{X}} & \qw & \qw & \qw & \qw\\
\\ }}
\caption{An alternative implementation of the three-qubit OR function}
\label{OR2_3qbits}
\centerline{}
\end{figure}

\begin{figure}[htb]
\centering
\scalebox{1.0}{
\Qcircuit @C=1.0em @R=0.8em @!R { \\
	 	\nghost{{q}_{0} :  } & \lstick{{q}_{0} :  } & \qw & \ctrl{1} & \targ & \ctrl{1} &  \qw & \qw\\
	 	\nghost{{q}_{1} :  } & \lstick{{q}_{1} :  } & \qw & \targ & \ctrl{-1} & \targ   &  \qw & \qw\\
\\ }}
\caption{Swap the states in $q_0$ and $q_1$} 
\label{SWAP}
\centerline{}
\end{figure}
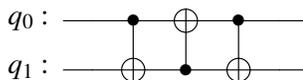

There are two slightly more complicated quantum circuits that are useful for students to study.
The first circuit is shown in Figure~\ref{SWAP}. After implementing and executing the program,
students will find out, for the four possible inputs $q_0 q_1$: 00, 01, 10, and 11, the
corresponding outputs will be: 00, 10, 01, and 11, respectively. They need to deduce that the circuit
swaps the states between $q_0$ and $q_1$.
 
\begin{figure}[htb]
\centering
\scalebox{1.0}{
\Qcircuit @C=1.0em @R=0.2em @!R { \\
	 	\nghost{{q}_{0} :  } &  \lstick{{q}_{0} :  } & \qw & \gate{\mathrm{H}} & \ctrl{1} & \gate{\mathrm{H}}  & \qw & \qw & \;\;\;\;\;\;\;\;\; \equiv \;\;\;\;\;\; &
                         \nghost{{q}_{0} :  } &  \lstick{{q}_{0} :  } & \qw & \targ  & \qw & \qw \\
	 	\nghost{{q}_{1} :  } &  \lstick{{q}_{1} :  } & \qw & \gate{\mathrm{H}} & \targ & \gate{\mathrm{H}}     & \qw & \qw &                                   &
                         \nghost{{q}_{1} :  } &  \lstick{{q}_{1} :  } & \qw & \ctrl{-1}  & \qw & \qw \\
\\ }}
\vskip 0.1in
$ \frac{1}{2}
  \begin{bmatrix}
\phantom{-}1 & \phantom{-}1 & \phantom{-}1 & \phantom{-}1 \\
\phantom{-}1 & -1           & \phantom{-}1 & -1 \\
\phantom{-}1 & \phantom{-}1 & -1 & -1 \\
\phantom{-}1 & -1 & -1 &  \phantom{-}1
  \end{bmatrix}
\times
  \begin{bmatrix}
\phantom{-}1 & \phantom{-}0 & \phantom{-}0 & \phantom{-}0 \\
\phantom{-}0 & \phantom{-}1 & \phantom{-}0 & \phantom{-}0 \\
\phantom{-}0 & \phantom{-}0 & \phantom{-}0 & \phantom{-}1 \\
\phantom{-}0 & \phantom{-}0 & \phantom{-}1 & \phantom{-}0 \\
  \end{bmatrix}
\times
  \frac{1}{2}
  \begin{bmatrix}
\phantom{-}1 & \phantom{-}1 & \phantom{-}1 & \phantom{-}1 \\
\phantom{-}1 & -1           & \phantom{-}1 & -1 \\
\phantom{-}1 & \phantom{-}1 & -1 & -1 \\
\phantom{-}1 & -1 & -1 &  \phantom{-}1
  \end{bmatrix}
\;
  \equiv
\;
  \begin{bmatrix}
\phantom{-}1 & \phantom{-}0 & \phantom{-}0 & \phantom{-}0 \\
\phantom{-}0 & \phantom{-}0 & \phantom{-}0 & \phantom{-}1 \\
\phantom{-}0 & \phantom{-}0 & \phantom{-}1 & \phantom{-}0 \\
\phantom{-}0 & \phantom{-}1 & \phantom{-}0 & \phantom{-}0 \\
  \end{bmatrix}
$

\caption{Phase Kickback and Corresponding Matrix Operations}
\label{PhaseKB}
\centerline{}
\end{figure}
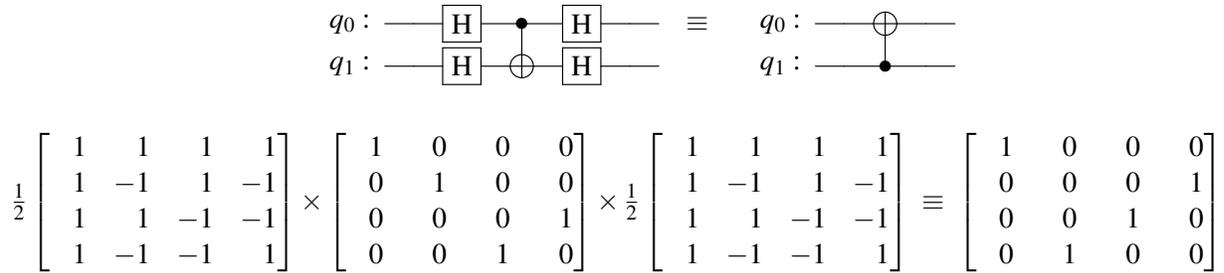

The second circuit, known as Phase Kickback and depicted on the upper left of Figure~\ref{PhaseKB}, will generate the outputs   
$q_0 q_1$ as 00, 11, 10, and 01, corresponding to the inputs: 00, 01, 10, and 11, respectively.
Namely, the circuit acts like the $q_1$ is the control while the $q_0$ becomes the target
of a CNOT gate, as shown in the upper right of Figure~\ref{PhaseKB}.  
Note that the two H gates in $q_0$'s wire are not successive due to the presence of a 
CNOT connection in between. 
Students are recommended to use matrix operations, as shown in the bottom part of Figure~\ref{PhaseKB},
to verify the circuit equivalence.
The Phase Kickback circuit has been used in many quantum algorithms.
In our experience, comprehending the Phase Kickback circuit would be helpful for beginners 
in elucidating the workings of the Deutsch and Deutsch-Jozsa algorithms.

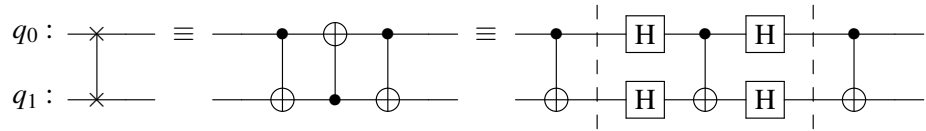
\begin{figure}[htb] 
\centering
\scalebox{1.0}{
\Qcircuit @C=1.0em @R=1.0em @!R { \\
	 	\nghost{{q}_{0} :  } & \lstick{{q}_{0} :  } & \qswap & \qw & \qw     & \;\;\;\;\;\;\;\;\; \equiv \;\;\;\;\;\;\;\;\; & 
                         & \qw & \ctrl{1} & \targ & \ctrl{1} &  \qw & \qw & \;\;\;\;\;\;\;\;\; \equiv \;\;\;\;\;\;\;\;\ &
                         & \ctrl{1} \barrier[0em]{1} & \qw & \gate{\mathrm{H}} & \ctrl{1} & \gate{\mathrm{H}} \barrier[0em]{1} & \qw & \ctrl{1} & \qw & \qw\\
	 	\nghost{{q}_{1} :  } & \lstick{{q}_{1} :  } & \qswap \qwx[-1] & \qw & \qw                                     &                                        & 
                         & \qw & \targ & \ctrl{-1} & \targ   &  \qw & \qw &                                        &    
                         & \targ & \qw & \gate{\mathrm{H}} & \targ & \gate{\mathrm{H}} & \qw & \targ & \qw & \qw\\
\\ }}
\caption{Swap the states in $q_0$ and $q_1$ if $q_1$ cannot be the control qubit} 
\label{SWAP2}
\centerline{}
\end{figure}

It is also worth mentioning that, though not included in the Lab3, 
the circuits in Figure~\ref{SWAP} and Figure~\ref{PhaseKB} can be used together.
The current IBM quantum computer adopts the heavy-hexagon lattice topology to connect qubits\cite{HeavyHexgon}.
Hence, the swap gate is needed to facilitate interactions between non-adjacent 
(i.e. not directly connected) qubits. 
Moreover, the control qubits are not adjacent in the heavy-hexagon lattice and this limitation makes the circuit in Figure~\ref{SWAP} 
impossible if $q_0$ and $q_1$ are directly connected together. If the qubit $q_1$ in Figure~\ref{SWAP} cannot be the
control qubit, we can simply use the Phase Kickback to solve the problem\cite{MappingQC}.
The solution is depicted in the rightmost part in Figure~\ref{SWAP2}. 
This can be left as a homework exercise for students.

\begin{figure}
\centering
\scalebox{1.0}{
\Qcircuit @C=1.0em @R=0.8em @!R { \\
	 	\nghost{{q}_{0} :  } & \lstick{{q}_{0} :  } & \ctrl{2} & \qw & \;\;\;\;\;\;\;\;\; \equiv \;\;\;\;\;\;\;\;\ & & \qw & \ctrl{1} & \qw & \ctrl{1} & \qw & \qw\\
	 	\nghost{{q}_{1} :  } & \lstick{{q}_{1} :  } & \qw      & \qw & & & \ctrl{1} & \targ & \ctrl{1} & \targ & \qw & \qw\\
	 	\nghost{{q}_{2} :  } & \lstick{{q}_{2} :  } & \targ    & \qw & & & \targ & \qw & \targ & \qw & \qw & \qw\\
\\ }}

\caption{The Bridge gate: The CNOT gate (control: $q_0$, target: $q_2$) by way of $q_1$}
\label{BRIDGE}
\centerline{}
\end{figure}
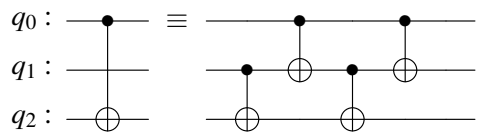

In addition to the swap gate, there is another interesting quantum circuit, called the bridge gate\cite{BridgeGate},
which lets the CNOT gate operate on non-adjacent qubits. For example, if the control qubit $q_0$
and the target qubit $q_2$ are not adjacent, they can use their common neighbor $q_1$ as the bridge.
This involves four direct-connect CNOT gates, as shown in Figure~\ref{BRIDGE}.
The circuit can be comprehended using the following instructions in sequence:
\begin{align*}
      q_2 &= q_{1.org} \oplus  q_{2.org}  \\
      q_1 &= q_{0.org} \oplus  q_{1.org}  \\
      q_2 &= q_1 \oplus  q_2 = (q_{0.org} \oplus q_{1.org}) \oplus (q_{1.org} \oplus  q_{2.org}) = q_{0.org} \oplus q_{2.org}   \\
      q_1 &= q_0 \oplus  q_1 = q_{0.org} \oplus (q_{0.org} \oplus  q_{1.org}) = q_{1.org}  \\
\end{align*}
Note that the notation $q_{org}$ denotes the original value of the qubit $q$.
It can be seen that the value in $q_2$, the target qubit of the CNOT gate, achieves its desired result,
while the intermediate qubit $q_1$ remains unchanged at the end.
We plan to include this example as a new experiment in the future.

\subsection{Hands-on Lab 4: Quantum Key Distribution}

\noindent {\bf Objectives: }
\begin{itemize}[noitemsep, topsep=-\parskip]
\item the BB84 protocol (see textbook page 53)
\item the Ekert protocol (see textbook page 87)
\end{itemize}

\vskip \parskip

Quantum Key Distribution (QKD) is a cryptographic technique that utilizes quantum mechanics 
to enable two communication parties to produce a shared random secret key known only to them.
The secret key can be used for encrypting and decrypting messages.
There are two QKD protocols discussed in the textbook: BB84 and Ekert91.
In this lab, students are asked to implement Qiskit programs to emulate
the behaviors of these two protocols and verify the output results of the probabilities
from their programs with the corresponding descriptions in the textbook.  

The BB84 protocol exploits the principles of quantum mechanics, utilizing the properties of quantum states 
to ensure the security of the key exchange process. As described in the textbook, Alice prepares  
a sequence of $4n$ qubits, encodes each qubit randomly in one of two bases (i.e. either vertical or horizontal)
and then sends them to Bob. 
Bob measures each of these qubits using randomly chosen bases. Then, they publicly communicate which bases they used 
for each qubit and only keep the qubits corresponding to the times when they both use the same basis.
The length of these kept qubits is roughly about $2n$. These qubits will be identical if Eve is not intercepting them.
Therefore, Alice and Bob can compare half of these qubits to detect the presence of any Eve's eavesdropping attempts. 
If they agree on all of them, they know Eve is not listening in and they can use the other $n$ qubits as the shared key 
to encrypt/decrypt messages in the future communications.

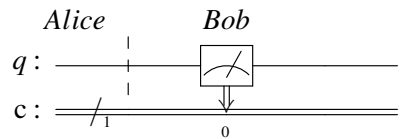
\begin{figure}
\centering
\scalebox{1.0}{
\Qcircuit @C=2.5em @R=0.2em @!R { \\
                                 &  \;\;\;\;\;\; Alice                       &     &  Bob\\
	 	\nghost{{q} :  } &   \lstick{{q} :  } \barrier[0em]{0} & \qw & \meter & \qw & \qw\\
	 	\nghost{\mathrm{{c} :  }} & \lstick{\mathrm{{c} :  }} & \lstick{/_{_{1}}} \cw & \dstick{_{_{\hspace{0.0em}0}}} \cw \ar @{<=} [-1,0] & \cw & \cw\\
\\ }}

\caption{BB84 emulation: Both Alice and Bob choose the Vertical direction}
\label{BB84_V}
\centerline{}
\end{figure}

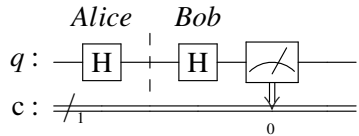
\begin{figure}
\centering
\scalebox{1.0}{

\Qcircuit @C=1.0em @R=0.2em @!R { \\
                                 &                  & Alice                              &     & Bob               &        &
     &    \\
	 	\nghost{{q} :  } & \lstick{{q} :  } & \gate{\mathrm{H}} \barrier[0em]{0} & \qw & \gate{\mathrm{H}} & \meter & \qw & \qw\\
	 	\nghost{\mathrm{{c} :  }} & \lstick{\mathrm{{c} :  }} & \lstick{/_{_{1}}} \cw & \cw & \cw & \dstick{_{_{\hspace{0.0em}0}}} \cw \ar @{<=} [-1,0] & \cw & \cw\\
\\ }}

\caption{BB84 emulation: Both Alice and Bob choose the Horizontal direction}
\label{BB84_H}
\centerline{}
\end{figure}

The first simple experiment is to let students observe that Alice and Bob will obtain the same qubit value 
if they choose the same basis -- either the vertical or horizontal direction, as shown in Figure~\ref{BB84_V} and Figure~\ref{BB84_H},
respectively. In the next experiment, students are asked to run a template program which emulates the BB84 protocol except 
detecting any interception. The template initially sets the $n$ to be 4, a small value so students can trace the executions
easily. Based on the program output, they need to mark, in their report, the bit positions where Alice and Bob use the same bases and
then extract the corresponding bits from ”Alice bits” and ”Bob results” into two bit sequences.
Finally, they need to add the code to select half of the same-basis bits and compare them to determine whether they are the same or not. 
If the bits comparison is True, students can increase the variable $n$ to a larger value, say 1000, and run the program
again to make sure that their program works for a larger size of bit sequence.

\begin{figure}
\centering
\scalebox{1.0}{
\Qcircuit @C=1.0em @R=0.2em @!R { \\
                                 &                  &  Alice                             &     & Eve                     &
  & Bob               &        &     & \\ 
	 	\nghost{{q} :  } & \lstick{{q} :  } & \gate{\mathrm{H}} \barrier[0em]{0} & \qw & \meter \barrier[0em]{0} & \qw & \gate{\mathrm{H}} & \meter & \qw & \qw\\
	 	\nghost{\mathrm{{c} :  }} & \lstick{\mathrm{{c} :  }} & \lstick{/_{_{1}}} \cw & \cw & \dstick{_{_{\hspace{0.0em}0}}} \cw \ar @{<=} [-1,0] & \cw & \cw & \dstick{_{_{\hspace{0.0em}0}}} \cw \ar @{<=} [-1,0] & \cw & \cw\\
\\ }}

\caption{BB84 emulation with Eve’s eavesdropping and guessing wrong}
\label{BB84_E_wrong}
\centerline{}
\end{figure}
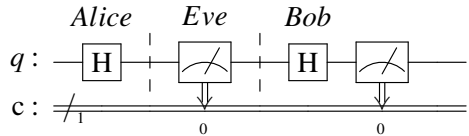

In the third experiment, students are asked to add lines of code to their program implemented in the first experiment
to emulate Eve’s interception. It can be observed that if Alice, Eve, and Bob choose the same basis,
they will all get the same bit and hence Eve’s eavesdropping cannot be detected. Figure~\ref{BB84_E_right} is an example 
in which Alice, Eve, and Bob all choose the horizontal basis. Note that after Eve guesses the measurement basis,
she will send the qubit to Bob using the same basis.  If Eve chooses the wrong basis, 
as shown in Figure~\ref{BB84_E_wrong}, the emulated program will output that Bob gets the 
right bit with only a 50\% chance.   

\begin{figure}
\centering
\scalebox{1.0}{
\Qcircuit @C=1.0em @R=0.2em @!R { \\
                                &                  &      Alice                    &     &              &  Eve      & 
                            &     & Bob               &        &     &   \\
	 	\nghost{{q} :  } & \lstick{{q} :  } & \gate{\mathrm{H}} \barrier[0em]{0} & \qw & \gate{\mathrm{H}} & \meter & \gate{\mathrm{H}} \barrier[0em]{0} & \qw & \gate{\mathrm{H}} & 
\meter & \qw & \qw\\
	 	\nghost{\mathrm{{c} :  }} & \lstick{\mathrm{{c} :  }} & \lstick{/_{_{1}}} \cw & \cw & \cw & \dstick{_{_{\hspace{0.0em}0}}} \cw \ar @{<=} [-1,0] & \cw & \cw & \cw & \dstick{_
{_{\hspace{0.0em}0}}} \cw \ar @{<=} [-1,0] & \cw & \cw\\
\\ }}

\caption{BB84 emulation with Eve’s eavesdropping and guessing right}
\label{BB84_E_right}
\centerline{}
\end{figure}
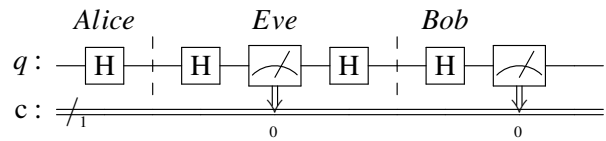

In Experiment 4, students need to modify the code implemented in Experiment 2 to emulate Eve's actions. That is, Eve firstly generates a random choice 
of basis for each bit. Next, she measures each qubit based on the generated random choice of basis. Finally, she encodes the message again based
on the generated random choice of basis and then sends the message to Bob. Students are also asked to add some code to find out the
percentage of the $n$ sample bits in which Alice and Bob disagree. As described in the textbook, if Alice and Bob disagree about a quarter
of the sample bits (i.e. $n/4$), they know that Eve's intercepting. Students can check their program's output whether it is $1/4$ or not. 

Experiments 5, 6, and 7 are about the Ekert91 protocol which utilizes entangled qubits as in the Bell's test.
There are a few variations of the protocol.  Based on the version presented in the textbook, there are $3n$ pairs of
entangled qubits. For each pair, Alice receives one and randomly chooses a direction to measure her qubit,
while Bob gets the other and also randomly selects a direction to measure it. 
There are three directions Alice and Bob can use:  $0^{\circ}$, $120^{\circ}$, or $240^{\circ}$.
Therefore, they will agree around $n$ bits when both choose the same direction.
Because the bases they chose will be revealed over a public insecure line,    
they can use these same-basis $n$ bits as the shared secure key if Eve is not intercepting.   
Note that to detect interception,
they use the sequence of qubits that come from the times when they chose different bases. 

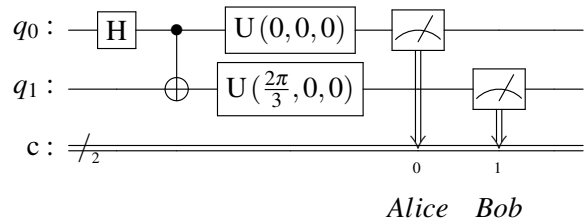
\begin{figure}
\centering
\scalebox{1.0}{

\Qcircuit @C=1.0em @R=0.2em @!R { \\
	 	\nghost{{q}_{0} :  } & \lstick{{q}_{0} :  } & \gate{\mathrm{H}} & \ctrl{1} & \gate{\mathrm{U}\,(\mathrm{0,0,0})} & \meter & \qw & \qw & \qw\\
	 	\nghost{{q}_{1} :  } & \lstick{{q}_{1} :  } & \qw & \targ & \gate{\mathrm{U}\,(\mathrm{\frac{2\pi}{3},0,0})} & \qw & \meter & \qw & \qw\\
	 	\nghost{\mathrm{{c} :  }} & \lstick{\mathrm{{c} :  }} & \lstick{/_{_{2}}} \cw & \cw & \cw & \dstick{_{_{\hspace{0.0em}0}}} \cw \ar @{<=} [-2,0] & \dstick{_{_{\hspace{0.0em}1
}}} \cw \ar @{<=} [-1,0] & \cw & \cw\\
                                     &                      &                   &          &                                    & Alice       & Bob    &     &        \\
\\ }}

\caption{Ekerti91 emulation: Measuring two entangled qubits with different bases }
\label{Ekert91}
\centerline{}
\end{figure}

The template program given in Experiment 5 simply tests for one pair of entangled qubits
when Alice and Bob measure the qubits using different directions via the Qiskit U-gate (see Figure~\ref{Ekert91}).
From the program output, students need to calculate the probability both Alice and Bob's qubits agree (i.e.
combining the percentages of ’00’ and ’11’) and the probability they disagree (i.e. adding the
percentages of ’10’ and ’01’). In addition to $0^{\circ}$ and $120^{\circ}$ different directions,
students also need to test the other five cases such as $0^{\circ}$ and $240^{\circ}$,
$120^{\circ}$ and $240^{\circ}$, etc. Students will find out that for any case,
the probability Alice and Bob agree is approximate 0.25 and the probability they disagree is around 0.75.   
The theoretic proof can be found in the section about Bell's inequality in the textbook. 
 
In Experiment 6, students need to study, modify and run another template program which emulates Alice's and Bob's behaviors
based on the Ekert protocol. This program outputs the probability that Alice and Bob agree when they
use different bases.  In the next experiment, they are asked to modify and extend this template program to detect whether 
Eve is intercepting or not. If Eve is eavesdropping, the probability that Alice and Bob agree when they
use different bases will raise from 1/4 (as in the Experiment 6) to 3/8. 
Students can check whether their results are consistent with the description in the textbook. 
Note that the textbook does not derive how to get the probability 3/8 in detail.
This can be left as an homework assignment for students.

\subsection{Hands-on Lab 5: Deutsch and Deutsch-Jozsa Algorithms }

\noindent {\bf Objectives: }
\begin{itemize}[noitemsep, topsep=-\parskip]
\item the Deutsch algorithm
\item the Deutsch-Jozsa algorithm
\end{itemize}
 
\vskip \parskip

The Deutsch algorithm is the first algorithm to show that a quantum algorithm could be faster than a classic one to solve a specific problem. 
It specifically addresses the task of determining whether a black-box function $ f: \{0,1\} \longrightarrow \{0,1\}$ is constant or balanced. 
Note that a function is called balanced if it sends half of its inputs to 0 and the other half to 1. 
There are four such functions. Two are constant: the constant 0 function ($f(0)\!=\!f(1)\!=\!0$), or the constant 1 function ($f(0)\!=\!f(1)\!=\!1$),
while the other two are balanced: the identity function  ($f(0)\!=\!0, \: f(1)\!=\!1$), or the inversion function  ($f(0)\!=\!1, \: f(1)\!=\!0$).
In classical computing, it would typically require two function evaluations to tell whether a black-box function is constant or balanced.
By employing quantum principles such as superposition and interference, the quantum algorithm requires only a single query to the function, 
providing a speedup over the classical method.

\begin{figure}
\centering
\scalebox{1.0}{
\Qcircuit @C=1.0em @R=0.2em @!R { 
                                 &                  &                                                             &  & F & & & & & & & \;\;\;\;\;\;\;\;\;\; 
                                 &                  &                                                             &  & F \\
	 	\nghost{{x} :  } & \lstick{{x} : \ket{0} } &  \gate{\mathrm{H}} \barrier[0em]{1} & \qw  & \qw \barrier[0em]{1} & \qw & \gate{\mathrm{H}} &  
 \meter & \qw & \qw  &        & 
	 	\nghost{{x} :  } & \lstick{{x} : \ket{0} } &  \gate{\mathrm{H}} \barrier[0em]{1} & \qw & \qw \barrier[0em]{1} & \qw & \gate{\mathrm{H}} & \meter & \qw & \qw \\ 
	 	\nghost{{y} :  } & \lstick{{y} : \ket{1} } &  \gate{\mathrm{H}} & \qw & \qw & \qw & \qw & \qw & \qw & \qw    & &   
	 	\nghost{{y} :  } & \lstick{{y} : \ket{1} } &  \gate{\mathrm{H}} &  \qw & \gate{\mathrm{X}} & \qw & \qw & \qw & \qw & \qw\\
	 	\nghost{\mathrm{{c} :  }} & \lstick{\mathrm{{c} :  }} & \lstick{/_{_{1}}} \cw &  \cw & \cw & \cw & \cw & \dstick{_{_{\hspace{0.0em}0}}} \cw \ar @{<=} [-2,0] & \cw & \cw  & & 
	 	\nghost{\mathrm{{c} :  }} & \lstick{\mathrm{{c} :  }} & \lstick{/_{_{1}}} \cw &  \cw & \cw & \cw & \cw & \dstick{_{_{\hspace{0.0em}0}}} \cw \ar @{<=} [-2,0] & \cw & \cw\\
\\ }}

\caption{Deutsch -– the Constant 0 function (left) and the Constant 1 function (right) }
\label{DeutschConst}
\centerline{}
\end{figure}
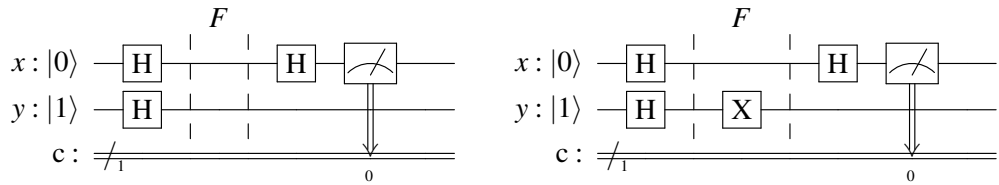

In the first four experiments of this lab, students learned how to construct the circuits for each of the four black-box functions.
For a quantum system, every operation must be unitary and thus reversible. For achieving this, each function, say $F$, needs another input $y$ to
map the state $\ket{x,y}$ to the state $\ket{x, y \oplus f(x)}$. Because $\ket{x, (y \oplus f(x)) \oplus f(x)} = \ket{x, y \oplus 0} = \ket{x, y}$, 
the function $F$ is its own inverse.
To construct the circuit for the constant 0 function, where $f(x) = 0$, the output of the bottom qubit becomes $y \oplus f(x) = y \oplus 0 = y$.
Namely, there are only two straight wires in the black box $F$ and there is no connection between the qubit $x$ and the qubit $y$, as shown
in the left part of Figure~\ref{DeutschConst}.     
Hence, the top qubit $x$ goes through two H gates in succession. Based on the experience we learned from Experiment 2 in Lab 2,
the qubit $x$ will be restored back to its initial value 0 before measurement. 
Similarly, the circuit for the constant 1 function, where $f(x) = 1$, can be constructed as in the right part of Figure~\ref{DeutschConst}. 
This is because $y \oplus f(x) = y \oplus 1 = \overline{y}$. Same as the constant 0 function, there is no connection between the qubit $x$ and the qubit $y$.
The qubit $x$ also goes through two successive H gates and its value will be restored back to its initial value 0.   
Therefore, both constant functions yield the value 0 in the qubit $x$.

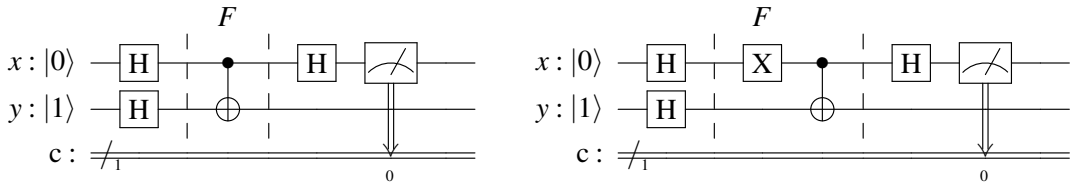
\begin{figure}
\centering
\scalebox{1.0}{
\Qcircuit @C=1.0em @R=0.2em @!R { 
                                 &                  &                                                              &  & F & & & & & & & \;\;\;\;\;\;\;
                                 &                  &                                                              &  & F \\
	 	\nghost{{x} :  } & \lstick{{x} : \ket{0} } &  \gate{\mathrm{H}} \barrier[0em]{1} & \qw & \ctrl{1} \barrier[0em]{1} & \qw & \gate{\mathrm{H}} & \meter & \qw & \qw & &
	 	\nghost{{x} :  } & \lstick{{x} : \ket{0} } &  \gate{\mathrm{H}} \barrier[0em]{1} & \qw & \gate{\mathrm{X}} & \ctrl{1} \barrier[0em]{1} & \qw & \gate{\mathrm{H}} & \meter & \qw & \qw\\
	 	\nghost{{y} :  } & \lstick{{y} : \ket{1} } &  \gate{\mathrm{H}} & \qw & \targ & \qw & \qw & \qw & \qw & \qw & &
	 	\nghost{{y} :  } & \lstick{{y} : \ket{1} } &  \gate{\mathrm{H}} & \qw & \qw & \targ & \qw & \qw & \qw & \qw & \qw\\
	 	\nghost{\mathrm{{c} :  }} & \lstick{\mathrm{{c} :  }} & \lstick{/_{_{1}}} \cw & \cw & \cw & \cw & \cw & \dstick{_{_{\hspace{0.0em}0}}} \cw \ar @{<=} [-2,0] & \cw & \cw & &
	 	\nghost{\mathrm{{c} :  }} & \lstick{\mathrm{{c} :  }} & \lstick{/_{_{1}}} \cw & \cw & \cw & \cw & \cw & \cw & \dstick{_{_{\hspace{0.0em}0}}} \cw \ar @{<=} [-2,0] & \cw & \cw
\\ }}

\caption{Deutsch –- the balanced functions; either Identity (left) or Inversion (right) }
\label{DeutschBal}
\centerline{}
\end{figure}

The left portion of Figure~\ref{DeutschBal} shows how to build the circuit of the balanced identity function, where $f(x) = x$. 
That is, the function $F$ adopts a CNOT gate to map the input state $\ket{x,y}$ to the output state $\ket{x, y \oplus f(x)} = \ket{x, y \oplus x}$.
From the Experiment 7 in the Lab 3 (i.e. the phase kickback circuit), we know that bottom qubit $y$ becomes the control bit of the CNOT gate.
Furthermore, another tricky setting is the initial value of the qubit $y$. The value 1 in $y$ will let the qubit $y$ toggle the value in the qubit $x$
from 0 to 1, which is different than the result of the constant functions.  
Likewise, the circuit of the balanced inversion function, where $f(x) = \overline{x}$, can be constructed as in the right portion of Figure~\ref{DeutschBal}.
The function $F$ maps the state from $\ket{x,y}$ to $\ket{x, y \oplus f(x)} = \ket{x, y \oplus \overline{x}}$. 
To implement it, we need to apply an X gate for the qubit $x$ to invert its value.
However, this X gate will not change the superposition state $\frac{1}{\sqrt{2}} \, (\ket{0} + \ket{1})$ in $x$
because \ket{0} becomes \ket{1} and \ket{1} becomes \ket{0}.   
Same as the balanced identity function, the value 1 in $y$ will flip the value in the qubit $x$ from 0 to 1.
Hence, if we measure the qubit $x$ and its value is 1, we know that the function is balanced. 

\begin{figure}
\centering
\scalebox{1.0}{
\Qcircuit @C=1.0em @R=0.2em @!R { 
                                     &                  &                                                                &  & F \\
	 	\nghost{{x}_{0} :  } & \lstick{{x}_{0} : \ket{0} } &  \gate{\mathrm{H}} \barrier[0em]{3} & \qw & \qw \barrier[0em]{3} & \qw & \gate{\mathrm{H}} & \meter & \qw & \qw & \qw & \qw\\
	 	\nghost{{x}_{1} :  } & \lstick{{x}_{1} : \ket{0} } &  \gate{\mathrm{H}} & \qw & \qw & \qw & \gate{\mathrm{H}} & \qw & \meter & \qw & \qw & \qw\\
	 	\nghost{{x}_{2} :  } & \lstick{{x}_{2} :\ket{0}  } &  \gate{\mathrm{H}} & \qw & \qw & \qw & \gate{\mathrm{H}} & \qw & \qw & \meter & \qw & \qw\\
	 	\nghost{{y} :  } & \lstick{{y} : \ket{1} }         &  \gate{\mathrm{H}} & \qw & \qw & \qw & \qw & \qw & \qw & \qw & \qw & \qw\\
	 	\nghost{\mathrm{{c} :  }} & \lstick{\mathrm{{c} :  }} & \lstick{/_{_{3}}} \cw & \cw & \cw & \cw & \cw & \dstick{_{_{\hspace{0.0em}0}}} \cw \ar @{<=} [-4,0] & \dstick{_{_{\hspace{0.0em}1}}} \cw \ar @{<=} [-3,0] & \dstick{_{_{\hspace{0.0em}2}}} \cw \ar @{<=} [-2,0] & \cw & \cw\\
\\ }}

\caption{Deutsch-Jozsa -- the constant 0 function $f(x_0,x_1,x_2) = 0 $ }
\label{DJC}
\centerline{}
\end{figure}
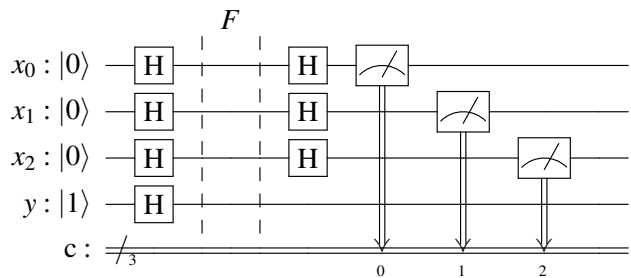

In the class lecture, students learned that the Deutsch algorithm doesn't truly demonstrate the power of quantum parallelism because it employs a single query to 
distinguish between two cases: constant or balanced. The Deutsch-Jozsa algorithm generalizes the Deutsch problem with a larger domain.
It determines whether a black-box function $ f: \{0,1\}^n \longrightarrow \{0,1\}$ is constant (i.e. it always outputs the same value for all $2^n$ inputs) or balanced
(i.e. half of the inputs go to 0 and the other half go to 1). 
The Deutsch-Jozsa algorithm shows the power of quantum parallelism by using only one query to the oracle, 
as compared to the worst case scenario $2^{n-1}+1$ function evaluations needed in the classical method.  
In Experiment 5, students are asked to load and run a template program involving a constant 0 function with $n=3$,
as depicted in Figure~\ref{DJC}. Next, they are instructed to add an X gate on the qubit $y$ to make the black box $F$ a constant 1 function.
The instructor would remind students that there is no connection between the $x$ qubits and the qubit $y$ for both constant functions. 
Hence, just like the constant functions in the Deutsch's algorithm, these three qubits $x_0 x_1 x_2$ go through two successive H gates 
and the value in $x_0 x_1 x_2$ will be restored back to "000".   

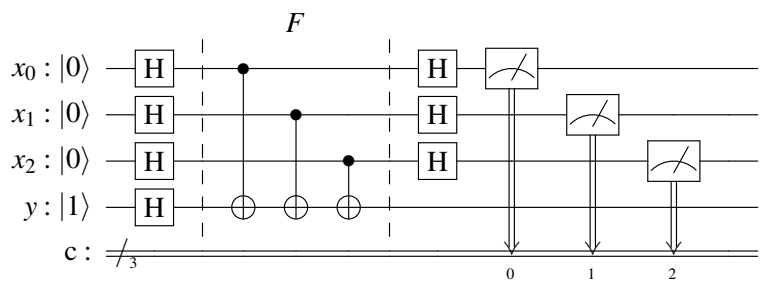
\begin{figure}
\centering
\scalebox{1.0}{

\Qcircuit @C=1.0em @R=0.2em @!R { 
                                     &                  &                           &                                    &  &    F \\
	 	\nghost{{x}_{0} :  } & \lstick{{x}_{0} :  \ket{0} } & \gate{\mathrm{H}} \barrier[0em]{3} & \qw & \ctrl{3} & \qw & \qw \barrier[0
em]{3} & \qw & \gate{\mathrm{H}} & \meter & \qw & \qw & \qw & \qw\\
	 	\nghost{{x}_{1} :  } & \lstick{{x}_{1} :  \ket{0} } & \gate{\mathrm{H}} &  \qw & \qw & \ctrl{2} & \qw & \qw & \gate{\mathrm{H}} & \qw & \meter & \qw & \qw & \qw\\
	 	\nghost{{x}_{2} :  } & \lstick{{x}_{2} :  \ket{0} } & \gate{\mathrm{H}} &  \qw & \qw & \qw & \ctrl{1} & \qw & \gate{\mathrm{H}} & \qw & \qw & \meter & \qw & \qw\\
	 	\nghost{{y} :  } & \lstick{{y} :  \ket{1} }         & \gate{\mathrm{H}} &  \qw & \targ & \targ & \targ & \qw & \qw & \qw & \qw & \qw & \qw &
 \qw\\
	 	\nghost{\mathrm{{c} :  }} & \lstick{\mathrm{{c} :  }} & \lstick{/_{_{3}}} \cw & \cw & \cw & \cw & \cw & \cw & \cw & \dstick{_{_{\hspace{0.0em}0}}} \cw \ar @{<=} [-4,0] & \dstick{_{_{\hspace{0.0em}1}}} \cw \ar @{<=} [-3,0] & \dstick{_{_{\hspace{0.0em}2}}} \cw \ar @{<=} [-2,0] & \cw & \cw\\
\\ }}

\caption{ Deutsch-Jozsa -- the balanced function $f(x_0,x_1,x_2) = x_0 \oplus x_1 \oplus x_2$}
\label{DJB1}
\end{figure}

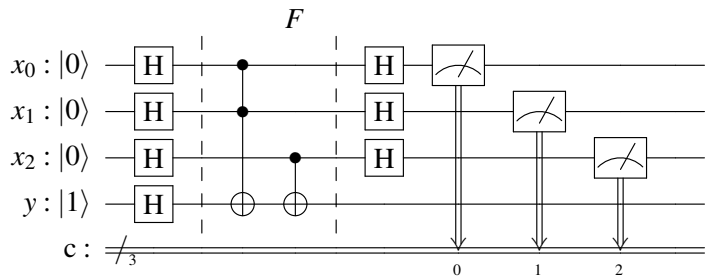
\begin{figure}
\centering
\scalebox{1.0}{
\Qcircuit @C=1.0em @R=0.2em @!R { \\
                                     &                  &                           &                                    &  &    F \\
	 	\nghost{{x}_{0} :  } & \lstick{{x}_{0} : \ket{0}  } &  \gate{\mathrm{H}} \barrier[0em]{3} & \qw & \ctrl{1} & \qw \barrier[0em]{3} & \qw & \gate{\mathrm{H}} & \meter & \qw & \qw & \qw & \qw\\ 
                \nghost{{x}_{1} :  } & \lstick{{x}_{1} : \ket{0}  } &  \gate{\mathrm{H}} &  \qw & \ctrl{2} & \qw & \qw & \gate{\mathrm{H}} & \qw & \meter &  \qw & \qw & \qw\\
	 	\nghost{{x}_{2} :  } & \lstick{{x}_{2} : \ket{0}  } &  \gate{\mathrm{H}} &  \qw & \qw & \ctrl{1} & \qw & \gate{\mathrm{H}} & \qw & \qw    &  \meter & \qw & \qw\\
	 	\nghost{{y} :      } & \lstick{{y} :     \ket{1}  } &  \gate{\mathrm{H}} &  \qw & \targ & \targ & \qw & \qw & \qw & \qw & \qw & \qw & \qw\\
	 	\nghost{\mathrm{{c} :  }} & \lstick{\mathrm{{c} :  }} & \lstick{/_{_{3}}} \cw & \cw & \cw & \cw & \cw & \cw & \dstick{_{_{\hspace{0.0em}0}}} \cw \ar @{<=} [-4,0] & \dstick{_{_{\hspace{0.0em}1}}} \cw \ar @{<=} [-3,0] & \dstick{_{_{\hspace{0.0em}2}}} \cw \ar @{<=} [-2,0] & \cw & \cw\\
\\ }}

\caption{ Deutsch-Jozsa -- the balanced function $f(x_0,x_1,x_2) = x_0 x_1 \oplus x_2$}
\label{DJB2}
\centerline{}
\end{figure}

In Experiment 6 and Experiment 7, students are asked to construct the circuits of the balanced functions 
$ f(x_0,x_1,x_2) = x_0 \oplus x_1 \oplus x_2$ and $f(x_0,x_1,x_2) = x_0  x_1 \oplus x_2$, respectively.  
As shown in Figure~\ref{DJB1} and Figure~\ref{DJB2}, there is at least one CNOT connection between the $x$ qubits and the qubit $y$. 
Since the phase kickback causes the input $x$ qubits to be flipped through the connection(s),
the value in $x_0 x_1 x_2$ will not be "000" anymore.
That's why we can determine the function is constant or balanced by examining the $x$ qubits. 
 
We believe that these hands-on programming experiments will help students easily understand the Deutsch and the Deutsch-Jozsa algorithms 
from a different perspective, rather than relying solely on the rigorous mathematical proofs presented in the textbook. 
The phase kickback circuit, along with the $y$ qubit's initial value set to 1, will flip certain values of the $x$-qubit for the balanced functions. 
This flipping will not occur for the constant functions due to the lack of any connection.
However, while the mathematical derivation might be lengthy, it clearly demonstrates the relationship between the
top $x$-qubit(s) and the bottom $y$ qubit. As emphasized twice in the textbook, once for the Deutsch algorithm and once for the Deutsch-Jozsa algorithms,
the $x$-qubit(s) and the $y$ qubit are not entangled because their 
joint state can be expressed as a tensor product decomposition.

\subsection{Hands-on Lab 6: Simon's Algorithm}
\noindent {\bf Objectives: }
\begin{itemize}[noitemsep, topsep=-\parskip]
\item the Simon's algorithm\cite{Simon}
\item the Oracle construction method for the Simon's algorithm
\end{itemize}

\vskip \parskip

Simon's algorithm solves the problem of finding the hidden string in a black-box function $ f: \{0,1\}^n \longrightarrow \{0,1\} ^n$ .
The black-box 2-to-1 function has the property that there is a secret non-zero binary string $s$ of length $n$, (i.e.$  s \neq 00 \dots 0$), s.t.
$f(x) = f(y)$ if and only if $y=x$ or $y = x \oplus s $. 
Simon's algorithm contains quantum procedures as well as classical post-processing procedures.
As described in the textbook, we can run Simon's circuit $n + N$ times, where $N$ does not depend on $n$, to
get $n+N$ equations. These equations contain the $n-1$ independent vectors, so we can use a classical algorithm
(e.g. Gaussian elimination) to recover the secret string $s$. 

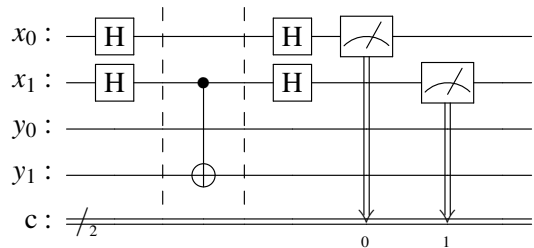
\begin{figure}
\centering
\scalebox{1.0}{
\Qcircuit @C=1.0em @R=0.2em @!R { \\
	 	\nghost{{x}_{0} :  } & \lstick{{x}_{0} :  } & \gate{\mathrm{H}} \barrier[0em]{3} & \qw & \qw \barrier[0em]{3} & \qw & \gate{\mathrm{H}} & \meter & \qw & \qw & \qw\\
	 	\nghost{{x}_{1} :  } & \lstick{{x}_{1} :  } & \gate{\mathrm{H}} & \qw & \ctrl{2} & \qw & \gate{\mathrm{H}} & \qw & \meter & \qw & \qw\\
	 	\nghost{{y}_{0} :  } & \lstick{{y}_{0} :  } & \qw & \qw & \qw & \qw & \qw & \qw & \qw & \qw & \qw\\
	 	\nghost{{y}_{1} :  } & \lstick{{y}_{1} :  } & \qw & \qw & \targ & \qw & \qw & \qw & \qw & \qw & \qw\\
	 	\nghost{\mathrm{{c} :  }} & \lstick{\mathrm{{c} :  }} & \lstick{/_{_{2}}} \cw & \cw & \cw & \cw & \cw & \dstick{_{_{\hspace{0.0em}0}}} \cw \ar @{<=} [-4,0] & \dstick{_{_{\hspace{0.0em}1}}} \cw \ar @{<=} [-3,0] & \cw & \cw\\
\\ }}
\caption{A Simple Example of Simon's Algorithm Circuit with $s=10$ }
\label{SimonExp1}
\centerline{}
\end{figure}

Students begin this lab by running a 2-qubit template program with the hidden string $s = 10$.
Figure~\ref{SimonExp1} depicts the circuit of this program.
Students should be reminded that, similar to the Deutsch-Jozsa algorithm, the top qubits $x$
will be measured instead of the bottom qubits. 
This template outputs the top $x$ qubits either 00 or 01.
Note that the template does not call Gaussian elimination to find $s$ and hence 
students need to manually solve the $n-1$ linearly independent equations. 
This will help them know how the whole algorithm works.
Because the output string '00' which wouldn't give us any information, students
need to use the other output string 01 to calculate its dot product with the secret 
string $s$ (i.e. $s_0 s_1$) and the result should be 0. That is, they will get\\
\mbox{\hskip 0.5in } $ 0 \times s_0 + 1 \times s_1 = 0 $. \\
They have to figure out that $s_1 = 0$ and hence $s=10$ because not all of the digits in
$s$ can be 0. In fact, this example is from the textbook and has been discussed in the class lecture.
Students now have the opportunity to practice it once again.
In Experiment 2 and Experiment 3, students will modify the oracle circuit in the program based on the given diagrams
in the handout. They may need to run their programs a few times to get $n-1$ different non-zero strings and then
use these strings to derive the bit pattern of the secret string $s$. 

In Experiment 4 and Experiment 5, students will learn a simple method, as presented in \cite{Simon_Oracle},
for constructing an oracle circuit $F$ using a given string $s$.
Figure~\ref{SimonOracle} shows the steps to construct such a 2-to-1 mapping function.   
Note that in the first step, because each qubit in the register $y$ is initialized to be 0, 
the content of each qubit in the register $x$, which is the classical information encoded as
either a $\ket{0}$ or a $\ket{1}$, will be copied through the CNOT gate
to the corresponding qubit in the register $y$ . 
In the second step, if $x_k == 0$, the register $y$ remains unchanged.
Otherwise, the register $y$ will be XORed with $s$, i.e. $ y \longleftarrow y \oplus s$. 

\begin{figure}
\centering
\rule[1ex]{\textwidth-1.2in}{0.1pt}
\begin{minipage}{5in} 
\begin{enumerate}
\item Apply the CNOT gates from qubits of the first register (i.e. $x$)
           to qubits of the second register (i.e. $y$). 
          
\item Because the bits in the string $s$ cannot be all 0, find the least
            index $k$ such that $s_k = 1$. Next, apply the CNOT gates from the qubit $x_k$ 
            to any qubit $y_i$ if $s_i = 1$. \\
\end{enumerate}
\end{minipage}
\rule[1ex]{\textwidth-1.2in}{0.1pt}
\centering
\caption{A Simon's Oracle Construction Algorithm}
\label{SimonOracle}
\centerline{}
\end{figure}

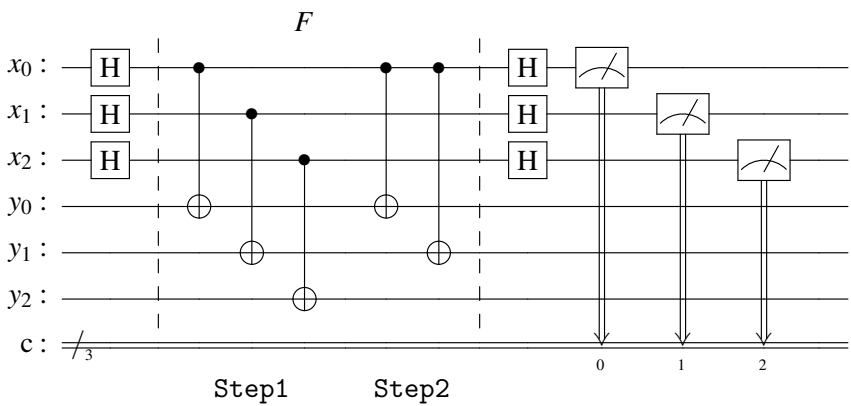
\begin{figure}
\centering
\scalebox{1.0}{
\Qcircuit @C=1.0em @R=0.2em @!R { \\
                                     &                      &                                    &     &          &     &   F \\
	 	\nghost{{x}_{0} :  } & \lstick{{x}_{0} :  } & \gate{\mathrm{H}} \barrier[0em]{5} & \qw & \ctrl{3} & \qw & \qw & \qw & \ctrl{3} & \ctrl{4} \barrier[0em]{5} & \qw & \gate{\mathrm{H}} & \meter & \qw & \qw & \qw & \qw\\
	 	\nghost{{x}_{1} :  } & \lstick{{x}_{1} :  } & \gate{\mathrm{H}} & \qw & \qw & \ctrl{3} & \qw & \qw & \qw & \qw & \qw & \gate{\mathrm{H}} & \qw & \meter & \qw & \qw & \qw\\
	 	\nghost{{x}_{2} :  } & \lstick{{x}_{2} :  } & \gate{\mathrm{H}} & \qw & \qw & \qw & \ctrl{3} & \qw & \qw & \qw & \qw & \gate{\mathrm{H}} & \qw & \qw & \meter & \qw & \qw\\
	 	\nghost{{y}_{0} :  } & \lstick{{y}_{0} :  } & \qw & \qw & \targ & \qw & \qw & \qw & \targ & \qw & \qw & \qw & \qw & \qw & \qw & \qw & \qw\\
	 	\nghost{{y}_{1} :  } & \lstick{{y}_{1} :  } & \qw & \qw & \qw & \targ & \qw & \qw & \qw & \targ & \qw & \qw & \qw & \qw & \qw & \qw & \qw\\
	 	\nghost{{y}_{2} :  } & \lstick{{y}_{2} :  } & \qw & \qw & \qw & \qw & \targ & \qw & \qw & \qw & \qw & \qw & \qw & \qw & \qw & \qw & \qw\\
	 	\nghost{\mathrm{{c} :  }} & \lstick{\mathrm{{c} :  }} & \lstick{/_{_{3}}} \cw & \cw & \cw & \cw & \cw & \cw & \cw & \cw & \cw & \cw & \dstick{_{_{\hspace{0.0em}0}}} \cw \ar @{<=} [-6,0] & \dstick{_{_{\hspace{0.0em}1}}} \cw \ar @{<=} [-5,0] & \dstick{_{_{\hspace{0.0em}2}}} \cw \ar @{<=} [-4,0] & \cw & \cw\\
                                     &                      &                                    &     &          &   {\tt Step 1}   &  &    & \;\;\;\;\;\;\;\; {\tt Step 2} \\
\\ }}

\caption{An Example of Constructing Simon's Algorithm Oracle Circuit with $s=110$ }
\label{SimonExp4}
\centerline{}
\end{figure}

Figure~\ref{SimonExp4} illustrates an example of using these two steps to
construct the Simon's Algorithm Oracle Circuit with $s=110$.
Note that in this example, $k$ is 0 because $s_0 = 1$.
If the secret string $s$ is 011, $k$ is 1 since $s_1 = 1$
and hence the CNOT control bit is $x_1$ in the Step 2. 

Note that the oracle $F$ constructed using the method in Figure~\ref{SimonOracle} implements
the function:\\
\mbox{\hskip 1in}  $f(x) = \;$ if $\; x_k == 1 \;$ return $\; (x \oplus s) \;$ else $\;$  return $\; x $ \\
We need to show that this function has the property $f(x) = f(x \oplus s)$. Assume that the value in the $x$ register is $b$
when entering $F$. If $b_k$ is 0, the register $y$ has the value $b$ when leaving $F$.
Otherwise, it has the value $b \oplus s$. Consider another input value $d$, where $d = b \oplus s$,
in the $x$ register when entering $F$. Note that $s_k$ is 1, so $d_k = b_k \oplus s_k = \overline{b_k}$.
There are two cases we need to verify. When $b_k = 0$, $d_k = \overline{b_k} = 1$ and hence the register 
$y$ will have the value $d \oplus s$ = $(b \oplus s) \oplus s = b$. 
If $b_k = 1$, then $d_k$ is 0 and the register $y$ will the value $d$ which is $b \oplus s$.  
Therefore, for both cases, $f(b) = f(d) = f(b \oplus s)$.

\begin{figure}
\centering
\scalebox{1.0}{
\Qcircuit @C=1.0em @R=0.2em @!R { \\
                                     &                      &                                    &     &          &     F \\
	 	\nghost{{x}_{0} :  } & \lstick{{x}_{0} :  } & \qw \barrier[0em]{5} & \qw & \qw & \qw & \targ \barrier[0em]{5} & \qw & \qw & \meter & \qw & \qw & \qw & \qw\\
	 	\nghost{{x}_{1} :  } & \lstick{{x}_{1} :  } & \qw & \qw & \targ & \qw & \qw & \qw & \qw & \qw & \meter & \qw & \qw & \qw\\
	 	\nghost{{x}_{2} :  } & \lstick{{x}_{2} :  } & \qw & \qw & \qw & \targ & \qw & \qw & \qw & \qw & \qw & \meter & \qw & \qw\\
	 	\nghost{{y}_{0} :  } & \lstick{{y}_{0} :  } & \qw  & \qw & \qw & \qw & \qw & \qw & \qw & \qw & \qw & \qw & \qw & \qw\\
	 	\nghost{{y}_{1} :  } & \lstick{{y}_{1} :  } & \gate{\mathrm{H}} & \qw & \ctrl{-3} & \qw & \ctrl{-4} & \qw & \gate{\mathrm{H}} & \qw & \qw & \qw & \qw & \qw\\
	 	\nghost{{y}_{2} :  } & \lstick{{y}_{2} :  } & \gate{\mathrm{H}} & \qw & \qw & \ctrl{-3} & \qw & \qw & \gate{\mathrm{H}} & \qw & \qw & \qw & \qw & \qw\\
	 	\nghost{\mathrm{{c} :  }} & \lstick{\mathrm{{c} :  }} & \lstick{/_{_{3}}} \cw & \cw & \cw & \cw & \cw & \cw & \cw & \dstick{_{_{\hspace{0.0em}0}}} \cw \ar @{<=} [-6,0] & \dstick{_{_{\hspace{0.0em}1}}} \cw \ar @{<=} [-5,0] & \dstick{_{_{\hspace{0.0em}2}}} \cw \ar @{<=} [-4,0] & \cw & \cw\\
\\ }}

\caption{ Circuit Equivalence Translation via Phase Kickback in Simon's Algorithm Circuit }
\label{SimonExp4_KB2}
\centerline{}
\end{figure}
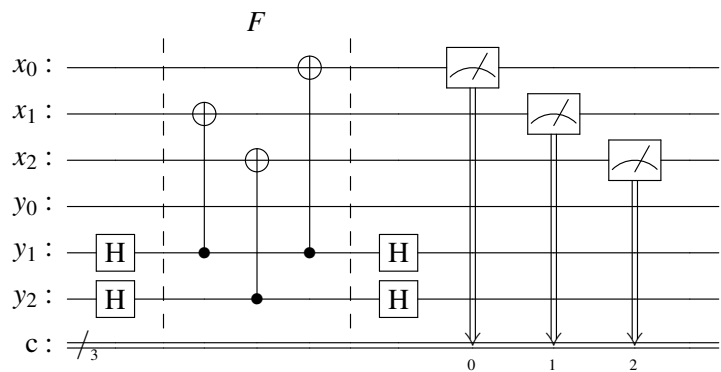

There is another way to explain why the measured outputs of the top qubits are limited to certain patterns.
Here, we employ the circuit equivalence translation along with the phase kickback property.
We use the circuit in Figure~\ref{SimonExp4} as an example. 
Firstly, we can simplify the circuit by removing the two successive CNOT gates from $x_0$ to $y_0$
because CNOT is its own inverse.
Next, we add two successive H gates in several places 
to yield the circuit as shown in Figure~\ref{SimonExp4_KB1}.
Adding two successive H gates does not change the circuit's property.
Now, we can use the the phase kickback circuit pattern, as shown in  
in Figure~\ref{PhaseKB}, to reverse the control-target positions in each CNOT gate. 
Note that there are two successive H gates added in the $y_1$ wire between the CNOT gate controlled from
$x_1$ and the CNOT gate controlled from $x_0$.  
The first H gate added will be used as the ending H gate in the phase kickback pattern, reversing the control position 
of the CNOT gate from original $x_1$ to $y_1$.
The second added H gate will serve as the initial H gate in the phase kickback pattern 
to reverse the control position of the next CNOT gate from $x_0$ to $y_1$. 
The translated circuit can be found in Figure~\ref{SimonExp4_KB2}. 
It shows that $x_0$ and $x_1$ are entangled via $y_1$.
Hence, the measured output $c_0 c_1 c_2$ must satisfy the property $c_0 \oplus c_1 = 0$, while
$c_2$ is independent of $c_0$ and $c_1$.

Note that we also need to add additional H gates in the $x_k$ wire if there are 3 or more 1's in $s$.
For example, if $s=111$, two successive H gates should be inserted in the $x_0$ wire 
between the second CNOT gate and the third CNOT gate in Step 2.
In general, if the secret string $s$ has the non-zero bits indexed at locations $k$, $l$, $m$, $\dotsc$,
the measured output should satisfy the property $c_k \oplus c_l \oplus c_m \oplus \dotsc = 0$. 
Namely, even number of these bits, $c_k$,  $c_l$, $c_m$, $\dots$, will have the value 1 in the measured output.
  
\begin{figure}
\centering
\scalebox{1.0}{
\Qcircuit @C=1.0em @R=0.2em @!R { \\
                                     &                      &                                    &     &          &     & &  \;\;\;  F \\
	 	\nghost{{x}_{0} :  } & \lstick{{x}_{0} :  } & \qw & \gate{\mathrm{H}} \barrier[0em]{5} & \qw & \ctrl{3} & \qw & \qw & \ctrl{3} & \qw & \ctrl{4} \barrier[0em]{5} & \qw & \gate{\mathrm{H}} & \qw & \meter & \qw & \qw & \qw & \qw\\
	 	\nghost{{x}_{1} :  } & \lstick{{x}_{1} :  } & \qw & \gate{\mathrm{H}}  & \qw & \qw & \ctrl{3} & \qw & \qw & \qw & \qw & \qw & \gate{\mathrm{H}} & \qw & \qw & \meter & \qw & \qw & \qw\\
	 	\nghost{{x}_{2} :  } & \lstick{{x}_{2} :  } & \qw & \gate{\mathrm{H}}  & \qw & \qw & \qw & \ctrl{3} & \qw & \qw & \qw & \qw & \gate{\mathrm{H}} & \qw & \qw & \qw & \meter & \qw & \qw\\
	 	\nghost{{y}_{0} :  } & \lstick{{y}_{0} :  } & \gate{\mathrm{H}} & \gate{\mathrm{H}} & \qw & \targ & \qw & \qw & \targ & \qw & \qw & \qw & \gate{\mathrm{H}} & \gate{\mathrm{H}} & \qw & \qw & \qw & \qw & \qw & \\
	 	\nghost{{y}_{1} :  } & \lstick{{y}_{1} :  } & \gate{\mathrm{H}} & \gate{\mathrm{H}} & \qw & \qw & \targ & \qw &  \gate{\mathrm{H}} & \gate{\mathrm{H}} & \targ & \qw & \gate{\mathrm{H}} & \gate{\mathrm{H}} & \qw & \qw & \qw & \qw & \qw\\
	 	\nghost{{y}_{2} :  } & \lstick{{y}_{2} :  } & \gate{\mathrm{H}} & \gate{\mathrm{H}} & \qw & \qw & \qw & \targ & \qw & \qw & \qw & \qw & \gate{\mathrm{H}} & \gate{\mathrm{H}} & \qw & \qw & \qw & \qw & \qw\\
	 	\nghost{\mathrm{{c} :  }} & \lstick{\mathrm{{c} :  }} & \lstick{/_{_{3}}} \cw & \cw & \cw & \cw & \cw & \cw & \cw & \cw & \cw & \cw & \cw & \cw & \dstick{_{_{\hspace{0.0em}0}}} \cw \ar @{<=} [-6,0] & \dstick{_{_{\hspace{0.0em}1}}} \cw \ar @{<=} [-5,0] & \dstick{_{_{\hspace{0.0em}2}}} \cw \ar @{<=} [-4,0] & \cw & \cw\\
\\ }}
\caption{Adding Successive H gates before Circuit Equivalence Translation }
\label{SimonExp4_KB1}
\centerline{}
\end{figure}
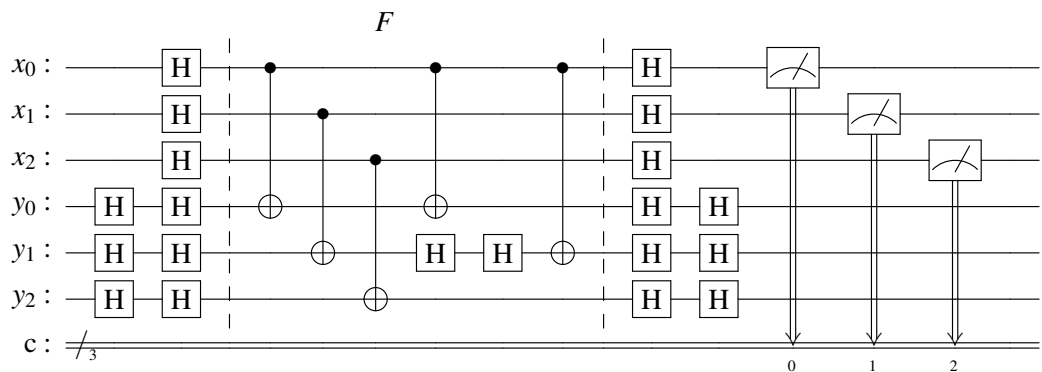

It's worth noting that the topic of constructing Simon's Algorithm oracle, while intriguing, 
falls beyond the scope of the textbook's contents.
Its intricate details are reserved for advanced students who have already grasped the fundamental 
principles discussed in the labs.
Whether this topic will be addressed depends on the instructor's discretion, allowing flexibility to 
adapt to the pace and progress of the lab.

\subsection{Hands-on Lab 7: Grover's Search Algorithm}

\noindent {\bf Objectives: }
\begin{itemize}[noitemsep, topsep=-\parskip]
\item the Grover's algorithm, including the Oracle construction and the Amplitude magnification method
\item the Qiskit program for solving the 3-CNF-SAT problem via Grover's Search
\end{itemize}

\vskip \parskip

As described in the textbook, Grover's search algorithm is a quantum computing technique designed to 
accelerate the search for the specific item(s) in an unstructured database.   
In the algorithm, an initial quantum state is prepared by applying a Hadamard transform to all possible input states. 
Subsequently, a specialized quantum oracle is employed to mark the target state by inverting its phase (i.e. flipping
the sign of its probability amplitude). The next step involves a process of applying a diffusion operator, 
which amplifies the probability amplitude of the target state while reducing the amplitudes of other states. 
The phase inversion oracle and the amplitude amplification (also called amplitude magnification) procedure may need to be applied repeatedly 
approximately $\sqrt{m}$ times, where $m$ is the number of items in the database.
For example, in the case of an NP-Complete problem 3-CNF-SAT with $n$ bits, 
there are $m=2^n$ possible assignments in the database.
Grover's search algorithm can reduce the run-time complexity
from $O(2^n)$ to $O(\sqrt{m})$, which is equivalent to $O(2^{n/2})$.
While quadratic speedup may not be as impressive as exponential speedup, it remains valuable for handling massive 
data sets.

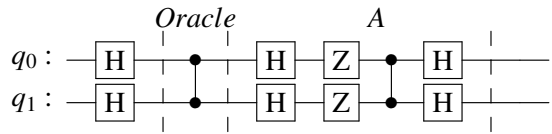
\begin{figure}
\centering
\scalebox{1.0}{
\Qcircuit @C=1.0em @R=0.2em @!R { \\
                                       &                      &                                    &        & Oracle &   &      &  \;\;\;\;\;\;\;\;\;\;\; A \\
	 	\nghost{{q}_{0} :  } & \lstick{{q}_{0} :  } & \gate{\mathrm{H}} \barrier[0em]{1} & \qw & \ctrl{1} \barrier[0em]{1} & \qw & \gate{\mathrm{H}} & \gate{\mathrm{Z}} & \ctrl{1} & \gate{\mathrm{H}} \barrier[0em]{1} & \qw & \qw & \qw\\
	 	\nghost{{q}_{1} :  } & \lstick{{q}_{1} :  } & \gate{\mathrm{H}} & \qw & \control\qw & \qw & \gate{\mathrm{H}} & \gate{\mathrm{Z}} & \control\qw & \gate{\mathrm{H}} & \qw & \qw & \qw\\
\\ }} 
 
\caption{Grover's Algorithm with the Boolean expression $q_0 \wedge q_1$ }
\label{Grover_Exp1}
\centerline{}
\end{figure}

The first experiment in this lab asks students to load and run a template program
which finds the desired element location 11 among 2-bit strings: 00, 01, 10, and 11.
In fact, this is a very simple SAT problem that determines the satisfiability of the expression $q_0 \wedge q_1$. 
Figure~\ref{Grover_Exp1} shows the circuit which includes the initial Hadamard transforms, the
phase inversion oracle, and the amplitude amplification. The phase inversion oracle can be realized
using a Controlled-Z gate {\tt cz(c,t)}, which flips the phase of the target qubit 
{\tt t} if the control qubit {\tt c} is in the $\ket{1}$ state.    
That is, the operation matrix of Controlled-Z is
\[
\begin{bmatrix}
1 & 0 & 0 & 0 \\
0 & 1 & 0 & 0 \\
0 & 0 & 1 & 0 \\
0 & 0 & 0 & -1 
\end{bmatrix}
\] 
It's symmetric and it doesn't matter which qubit is the controlled or target. 
As described in \cite{ProgQC}, the Controlled-Z gate is a phase-logic operation which performs phase AND (denoted as pAND).
In this experiment, students are asked to use the Operator class defined in the Qiskit {\tt quantum\_info} library 
in their program to display the oracle operation matrix and the amplitude amplification matrix.
They will find out that the operation matrix of amplitude amplification is 
\[
\begin{bmatrix}
-0.5 & 0.5 & 0.5 & 0.5 \\
0.5 & -0.5 & 0.5 & 0.5 \\
0.5 & 0.5 & -0.5 & 0.5 \\
0.5 & 0.5 & 0.5 & -0.5 
\end{bmatrix}
\]
and check whether it is the same as in the textbook.
Students need to look at the output state vector (i.e. probability amplitude) to find the
location with the highest probability.

\begin{figure*}
\centering
\includegraphics[width=2.5in]{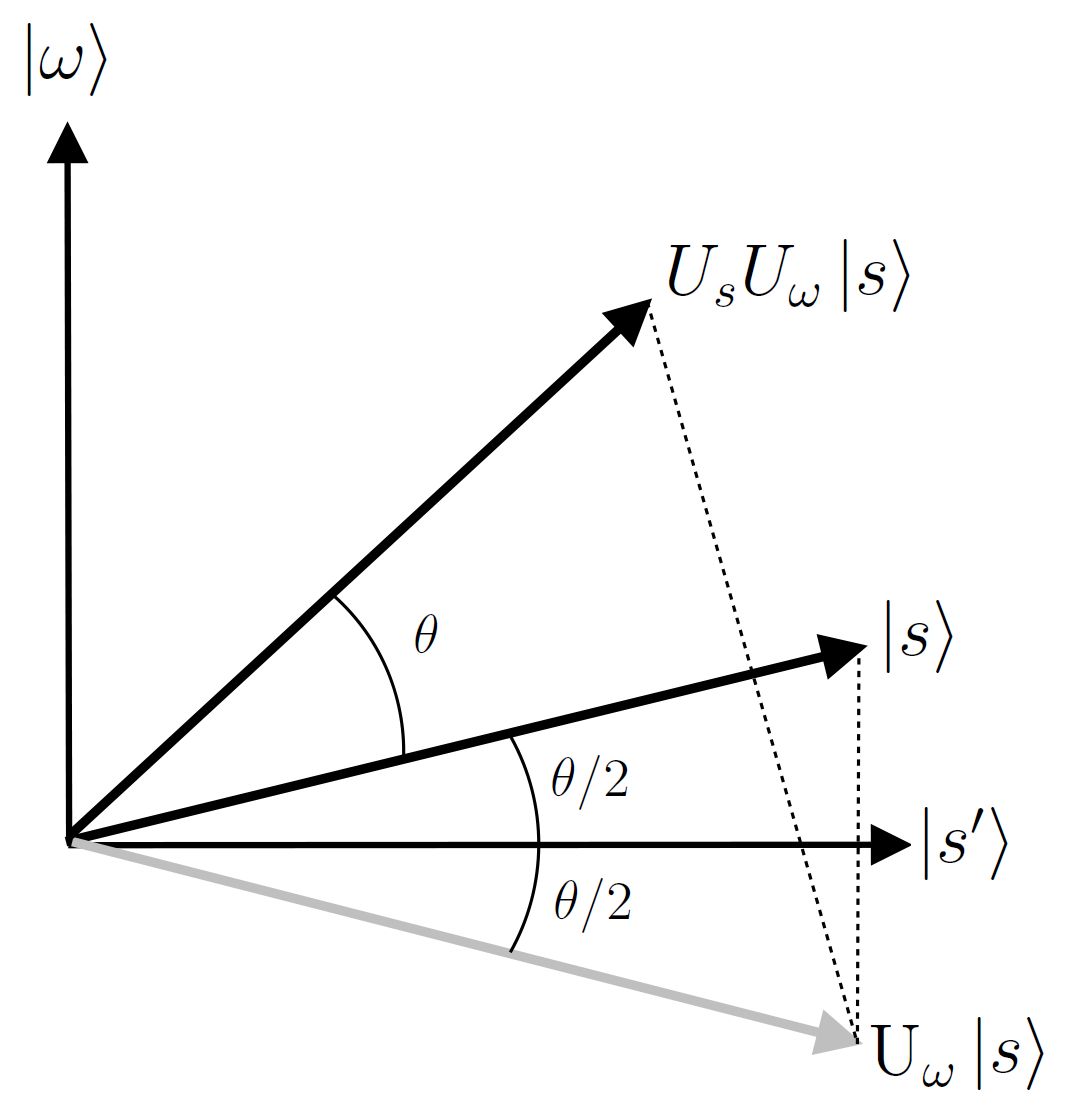}
\caption{Geometric interpretation of the first iteration of Grover's algorithm (Courtesy of \cite{Geo})}
\label{Visual}
\end{figure*}

Note that neither the textbook nor the reference book presents how to construct the amplitude amplification 
circuit.  
The geometric visualization of the Grover's algorithm, as shown in Figure~\ref{Visual}, can assist students in 
understanding how the algorithm works, particularly the construction of amplitude amplification.
Below is a brief description.
Remember that the system is initialized to the uniform superposition over all states\\
\mbox{\hskip 0.5in}  $\ket{s} = H^{\otimes n} \ket{0^n} = \frac{1}{\sqrt{m}} \sum_{x=0}^{m-1} \ket{x}$, where $m=2^n$ \\
prior to entering the phase inversion oracle. The operator of the oracle $U_\omega$, which flips the sign of the
probability amplitude associated with the target location $\ket{\omega}$, will transform the state to $U_\omega \, \ket{s}$
before amplitude magnification.   
The amplitude amplification operator, $U_s = 2 \ket{s}\!\bra{s} - I$, reflects the state about $\ket{s}$
from $U_\omega \, \ket{s}$ to $U_s \, U_\omega \, \ket{s}$, which is closer to the target state $\ket{\omega}$.  
The trick is how to solve the reflection $2 \ket{s}\!\bra{s} \; - \; I$.
Because $\ket{s} = H^{\otimes n} \ket{0^n}$, we can firstly apply the Hadamard gates to transform $\ket{s}$ to $\ket{0^n}$, i.e.,\\
\mbox{\hskip 0.5in} $H^{\otimes n} \ket{s} = H^{\otimes n} \, H^{\otimes n} \ket{0^n} = \ket{0^n} $. \\
Now we can do a reflection about the zero state via $ 2 \ket{0^n}\!\bra{0^n} \; - \;  I$, which can be implemented
by using the Z gates and the Controlled-Z gate. 
Finally, we apply the Hadamard gates to transform it back. In summary, the amplitude amplification operation is represented as\\ 
\mbox{\hskip 0.5in} $U_s = 2 \ket{s}\!\bra{s} - I = H^{\otimes n} \; ( 2 \ket{0^n}\!\bra{0^n} \; - \;  I ) \;H^{\otimes n} $, \\
which explains the amplitude amplification circuit shown in Figure~\ref{Grover_Exp1}. 
For advanced students, the instructor may guide them to consult the Qiskit textbook or Nielsen's and Chuang's book\cite{Nielsen}
for more detailed explanations. 

In Experiment 2, students will work on another example searching for the location indexed at 01. 
Once more, this can be seen as a SAT problem that determines the satisfiability of the expression $\neg q_0 \wedge q_1$. 
Students can modify the previous template program by adding two X-gates, one is before the Controlled-Z and the other 
is after the Controlled-Z, on the qubit $q_0$ in the oracle. The X-gate before the Controlled-Z performs the negation of $q_0$,
while the second X-gate uncomputes it back. Theoretically, this pair of X-gates, one swapping columns and
the other swapping rows in the Controlled-Z operation matrix, moves the value -1 in the matrix to the desired element location.     

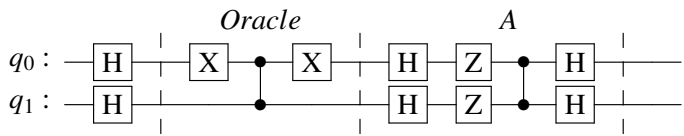
\begin{figure}
\centering
\scalebox{1.0}{
\Qcircuit @C=1.0em @R=0.2em @!R { \\
                                     &                      &                                    &     &                   & Oracle   &                                    &     &     & \;\;\;\;\;\;\;\;\;\;\; A  \\
	 	\nghost{{q}_{0} :  } & \lstick{{q}_{0} :  } & \gate{\mathrm{H}} \barrier[0em]{1} & \qw & \gate{\mathrm{X}} & \ctrl{1} & \gate{\mathrm{X}} \barrier[0em]{1} & \qw & \gate{\mathrm{H}} & \gate{\mathrm{Z}} & \ctrl{1} & \gate{\mathrm{H}} \barrier[0em]{1} & \qw & \qw & \qw\\
	 	\nghost{{q}_{1} :  } & \lstick{{q}_{1} :  } & \gate{\mathrm{H}} & \qw & \qw & \control\qw & \qw & \qw & \gate{\mathrm{H}} & \gate{\mathrm{Z}} & \control\qw & \gate{\mathrm{H}} & \qw & \qw & \qw\\
\\ }}
\caption{Grover's Search Algorithm with the Boolean expression $\neg q_0 \wedge q_1$ }
\label{Grover_Exp2}
\centerline{}
\end{figure}
 
In the next experiment, students will load and run a template program that includes a 3-bit oracle with the desired element location at 111.
Because Qiskit does not support the ccz gate, the template program uses a ccx gate instead, but it requires placing a pair of H-gates 
before and after the X gate in the ccx operation. It's important to note that operation of HXH is equivalent to a Z gate.
The program outputs two distinct probability amplitudes. The higher one is located at 111, while the lower one is situated
at other locations. Students are asked to compare these two probability amplitudes with the two probability amplitudes mentioned in the textbook. 
As mentioned earlier, in order to achieve a higher probability of obtaining the correct answer, it may be necessary to apply the oracle 
and amplitude amplification repeatedly. 
To repeat the process, we do not recommend that students manually compose the oracle and amplitude amplification circuits again or 
use a loop construct in Python. This is because it may result in the number of gate operations 
growing exponentially to $O(2^{n/2})$. Therefore, students are instructed to adapt the {\tt repeat()}
method in Qiskit, which generates some kind of branch instruction like in \cite{eQASM} to jump back to the oracle for repetition.

In this experiment, it is also interesting for students to observe that if the oracle and the amplitude amplification tasks are performed three times,
the probability amplitude of the desired location will decrease. In other words, it becomes overcooked. 
 
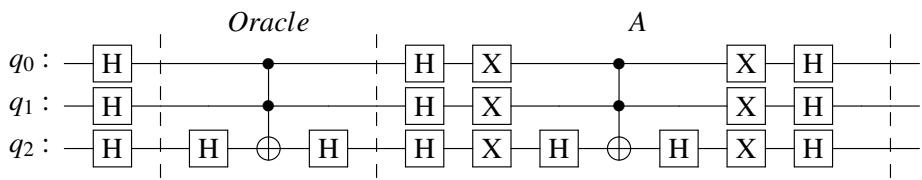
\begin{figure}
\centering
\scalebox{1.0}{
\Qcircuit @C=1.0em @R=0.2em @!R { \\
                                       &                      &                                  &  &        & Oracle &   &  &  & & &    \;\;\;\;\;\; A \\
	 	\nghost{{q}_{0} :  } & \lstick{{q}_{0} :  } & \gate{\mathrm{H}} \barrier[0em]{2} & \qw & \qw & \ctrl{1} & \qw \barrier[0em]{2} & \qw & \gate{\mathrm{H}} & \gate{\mathrm{X}} & \qw & \ctrl{1} & \qw & \gate{\mathrm{X}} & \gate{\mathrm{H}} & \qw \barrier[0em]{2} & \qw & \qw\\
	 	\nghost{{q}_{1} :  } & \lstick{{q}_{1} :  } & \gate{\mathrm{H}} & \qw & \qw & \ctrl{1} & \qw & \qw & \gate{\mathrm{H}} & \gate{\mathrm{X}} & \qw & \ctrl{1} & \qw & \gate{\mathrm{X}} & \gate{\mathrm{H}} &  \qw & \qw & \qw\\
	 	\nghost{{q}_{2} :  } & \lstick{{q}_{2} :  } & \gate{\mathrm{H}} & \qw & \gate{\mathrm{H}} & \targ & \gate{\mathrm{H}} & \qw & \gate{\mathrm{H}} & \gate{\mathrm{X}} & \gate{\mathrm{H}} & \targ & \gate{\mathrm{H}} & \gate{\mathrm{X}} & \gate{\mathrm{H}} & \qw & \qw & \qw\\
\\ }}

\caption{Grover's Algorithm with the Boolean expression $q_0 \wedge q_1 \wedge q_2 $ }
\label{Grover_Exp3}
\centerline{}
\end{figure}

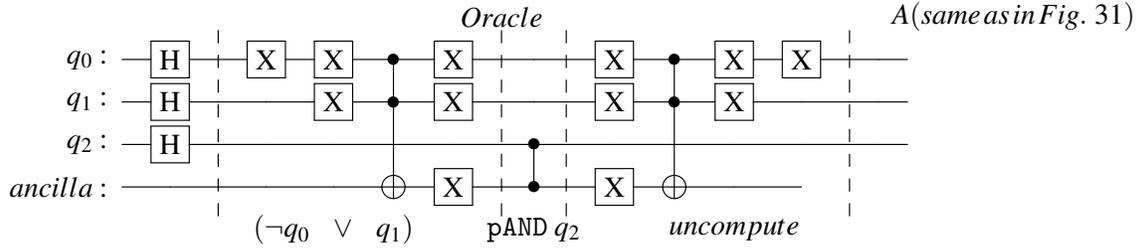
\begin{figure}
\scalebox{1.0}{
\Qcircuit @C=1.0em @R=0.2em @!R { \\
                                       &   & & & & & &                                  &           Oracle &   &  &  & & & & & & & & &  \;\;\;\;\;\; A (same\,as\,in\,Fig.~\ref{Grover_Exp3}) \\
	 	\nghost{{q}_{0} :  } & \lstick{{q}_{0} :  } & \gate{\mathrm{H}} \barrier[0em]{3} & \qw & \gate{\mathrm{X}} & \gate{\mathrm{X}} & \ctrl{1} & \gate{\mathrm{X}} \barrier[0em]{3} & \qw & \qw \barrier[0em]{3} & \qw & \gate{\mathrm{X}} & \ctrl{1} & \gate{\mathrm{X}} & \gate{\mathrm{X}} \barrier[0em]{3} & \qw & \qw & \qw & &\\
	 	\nghost{{q}_{1} :  } & \lstick{{q}_{1} :  } & \gate{\mathrm{H}} & \qw & \qw & \gate{\mathrm{X}} & \ctrl{2} & \gate{\mathrm{X}} & \qw & \qw & \qw & \gate{\mathrm{X}} & \ctrl{2} & \gate{\mathrm{X}} & \qw & \qw & \qw & \qw & & \\
	 	\nghost{{q}_{2} :  } & \lstick{{q}_{2} :  } & \gate{\mathrm{H}} & \qw & \qw & \qw & \qw & \qw & \qw & \ctrl{1} & \qw & \qw & \qw & \qw & \qw & \qw & \qw & \qw & & \\
	 	\nghost{{ancilla} :  } & \lstick{{ancilla} :  } & \qw & \qw & \qw & \qw & \targ & \gate{\mathrm{X}} & \qw & \control\qw & \qw & \gate{\mathrm{X}} & \targ & \qw & \qw & & & \\
                                     &                     &                    &   & &   (\neg q_0 \;\;\; \vee \;\;\;  q_1) & & & &  {\tt pAND} \; q_2  & & & & uncompute  \\ 
\\ }}

\caption{Grover's Algorithm with the Boolean expression $(\neg q_0 \vee q_1 ) \wedge q_2 $ }
\label{Grover_New}
\centerline{}
\end{figure}

We plan to add another experiment with a slightly more complicated Boolean expression, e.g. $(\neg q_0 \vee q_1 ) \wedge q_2 $, in the future.
As shown in Figure~\ref{Grover_New}, after negating $q_0$, we use the implementation in Figure~\ref{OR2} to perform
the OR operation, which places the result of the clause $(\neg q_0 \wedge q_1)$ into the ancilla qubit. 
Next, we use the Controlled-Z gate to 
phase AND the ancilla qubit with $q_2$. Finally, we have to perform uncomputation to clean up temporary effects on the
ancilla bit so that it can be re-used. 
An even more complicated example, which contains several clauses in a Boolean formula, can be found in \cite{ProgQC}.

\begin{figure}
\centering
\includegraphics[width=6.25in]{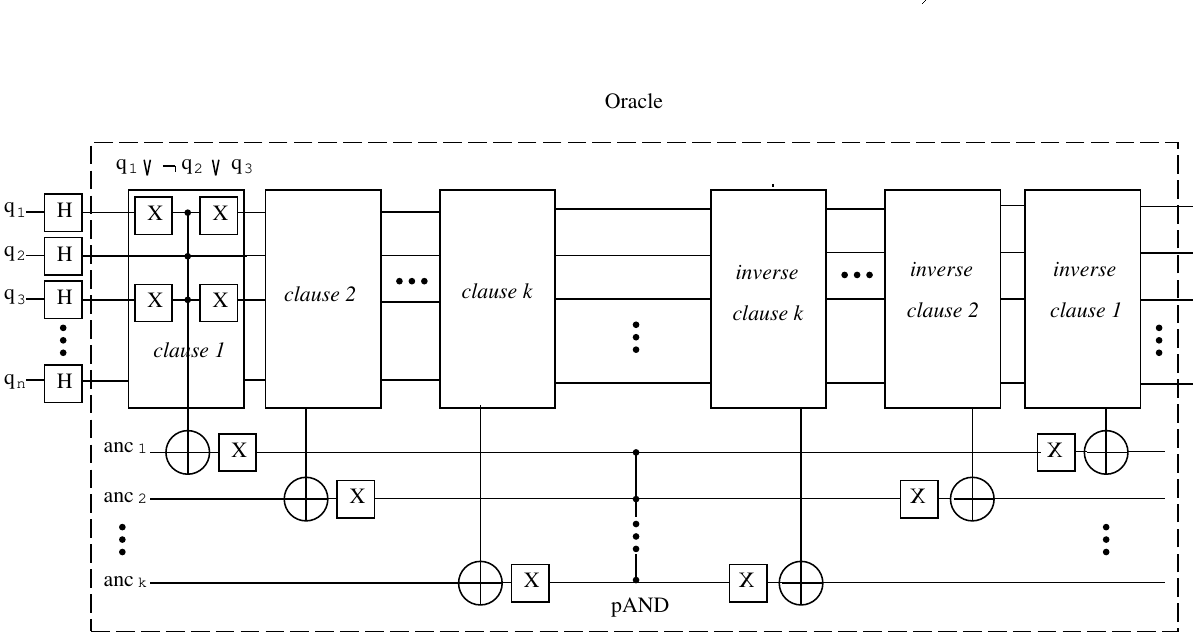}
\caption{A General Structure of the 3-CNF-SAT Oracle}
\label{3SAT}
\end{figure}

Figure~\ref{3SAT} depicts a general structure of the 3-CNF-SAT oracle with $k$ clauses.
For each clause $i$, where $1 \leq i \leq k$, we can utilize the NAND-based approach to implement the OR operation
of the literals, and the result is stored in the corresponding ancilla qubit $anc_i$. An example can be found
in the clause 1 of Figure~\ref{3SAT} which implements the Boolean expression $q_1 \vee \neg q_2 \vee q_3$. 
After obtaining the results of all clauses, Next, we proceed to use 
phase AND on all the ancilla qubits. Similar to the previous example, the inverse computation is
performed to free the ancilla qubits for future use. 
These experiments justify the necessity for students to learn
how to use quantum gates to emulate classical AND and OR gates in Lab 3.

\section{Conclusion}

A collection of engaging laboratory experiments has been developed to serve as 
supplementary resources for the textbook.
Many of the experiments are closely linked with the content of the textbook.
These experiments capture students' interest while also offering a concrete platform 
for validating their comprehension through the comparison of experimental outcomes with 
textbook examples. The feedback collected from the students has, overall, been very positive. 
The students indicated that these
practical labs help them not only improve their hands-on skills in quantum programming, 
but also understand the rationale behind them.
We believe that this report is valuable for utilization as an instructor's manual for 
Quantum programming education.

\section*{Acknowledgment}

This work was supported in part by the Cleveland Innovation District grant 
funded by JobsOhio, a private non-profit corporation.

\bibliographystyle{plain}
\bibliography{qc}

\newpage

\setlength{\voffset}{0cm}
\setlength{\hoffset}{0cm}

\fancyhead[L]{\textbf{\textit{Appendix} Qiskit Virtual Environment Installation}}
\includepdf[pages=-,pagecommand={}]{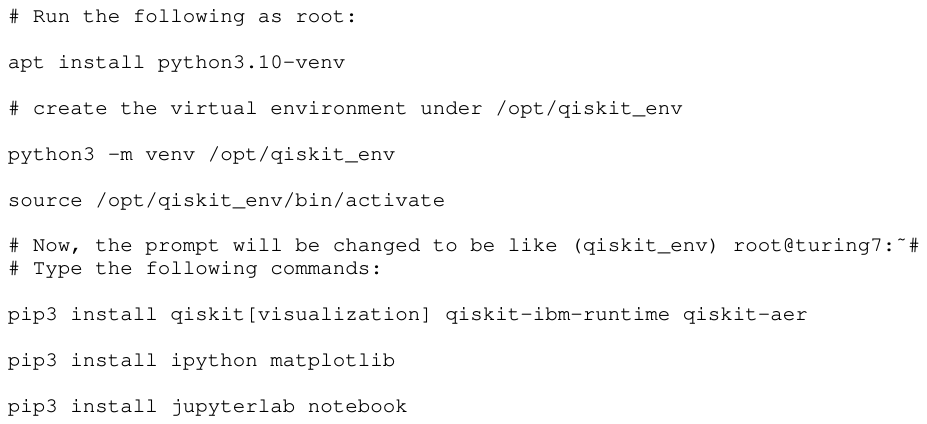}

\fancyhead[L]{\textbf{\textit{Appendix} Lab1: Python and Jupyter}}
\includepdf[pages=-,pagecommand={}]{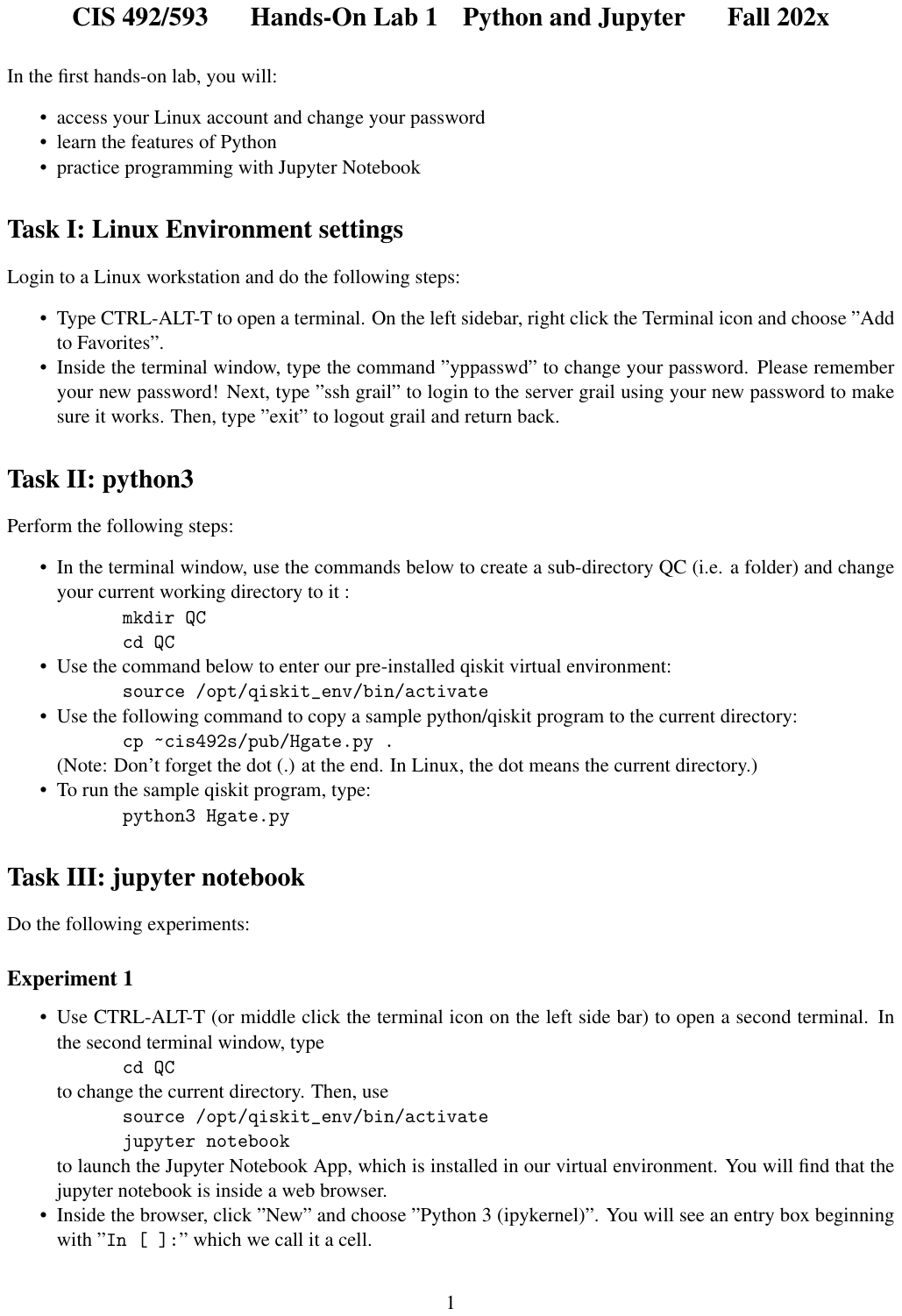}



\fancyhead[L]{\textbf{\textit{Appendix} Template: Hgate.py}}
\includepdf[pages=-,pagecommand={},width=\textwidth]{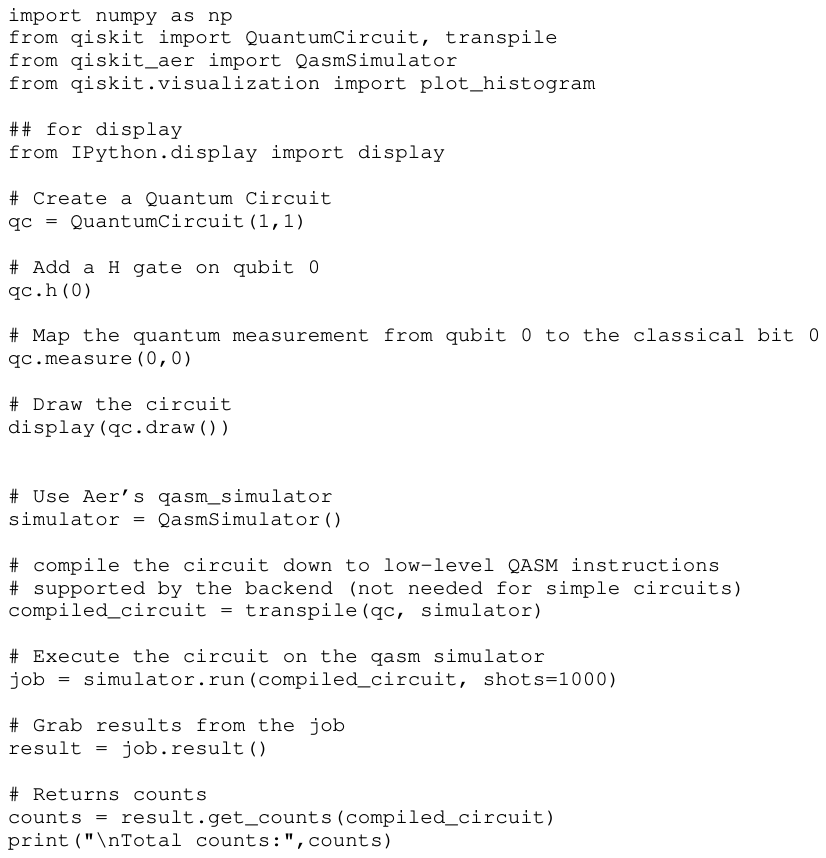}

\fancyhead[L]{\textbf{\textit{Appendix} Lab2: Entanglement }}
\includepdf[pages=-,pagecommand={}]{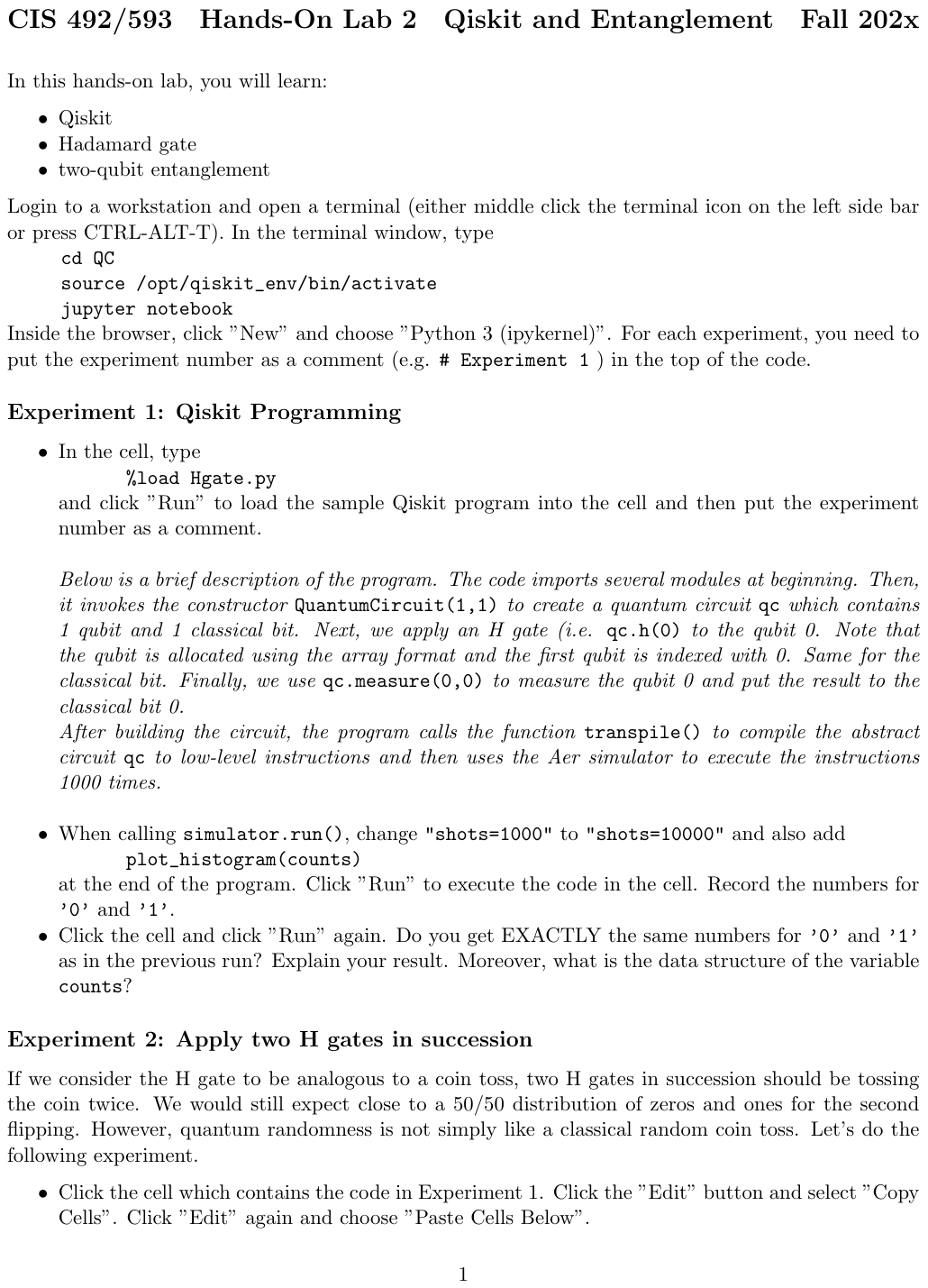}

\fancyhead[L]{\textbf{\textit{Appendix} Homework: Running Qiskit Programs on an IBM Quantum Computer }}
\includepdf[pages=-,pagecommand={}]{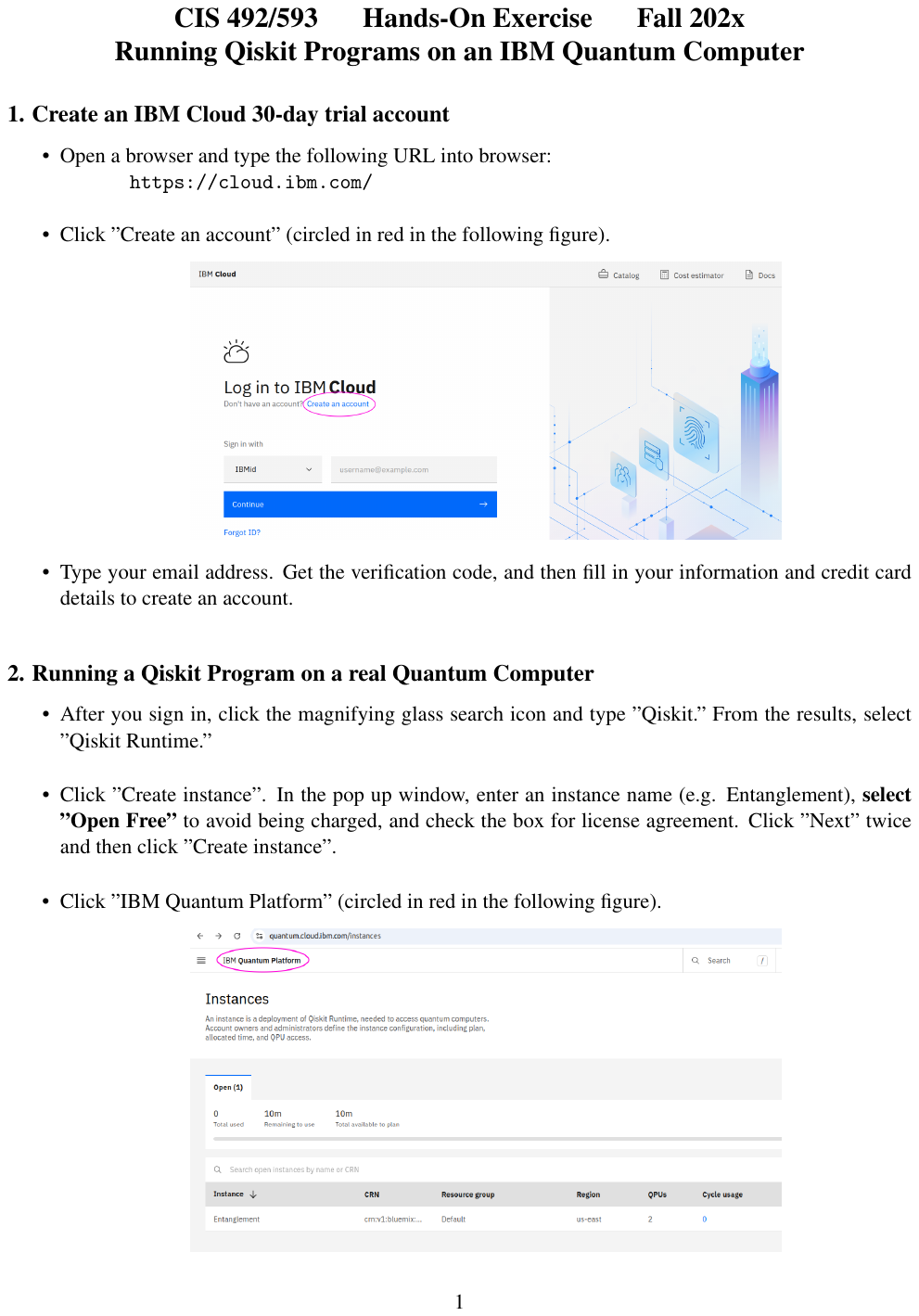}

\fancyhead[L]{\textbf{\textit{Appendix} Template: IBMQ\_v2\_1.py}}
\includepdf[pages=-,pagecommand={},width=\textwidth]{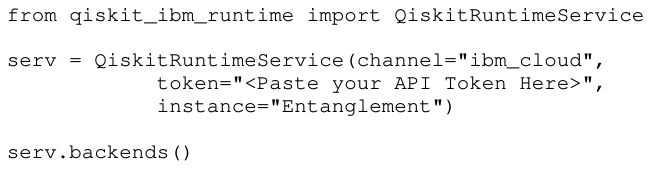}

\fancyhead[L]{\textbf{\textit{Appendix} Template: IBMQ\_v2\_2.py}}
\includepdf[pages=-,pagecommand={},width=\textwidth]{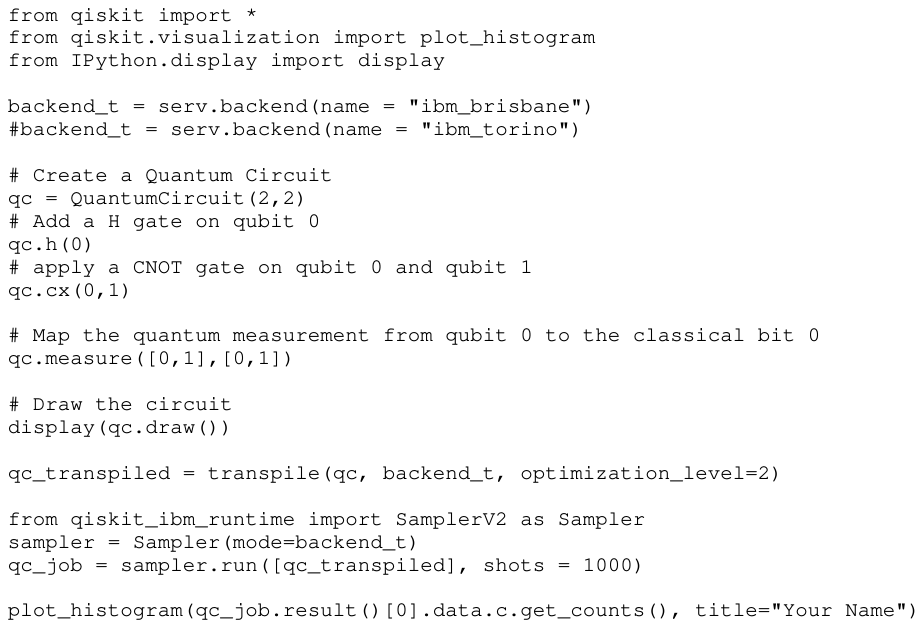}

\fancyhead[L]{\textbf{\textit{Appendix} Lab3: Quantum Gates and Circuits}}
\includepdf[pages=-,pagecommand={}]{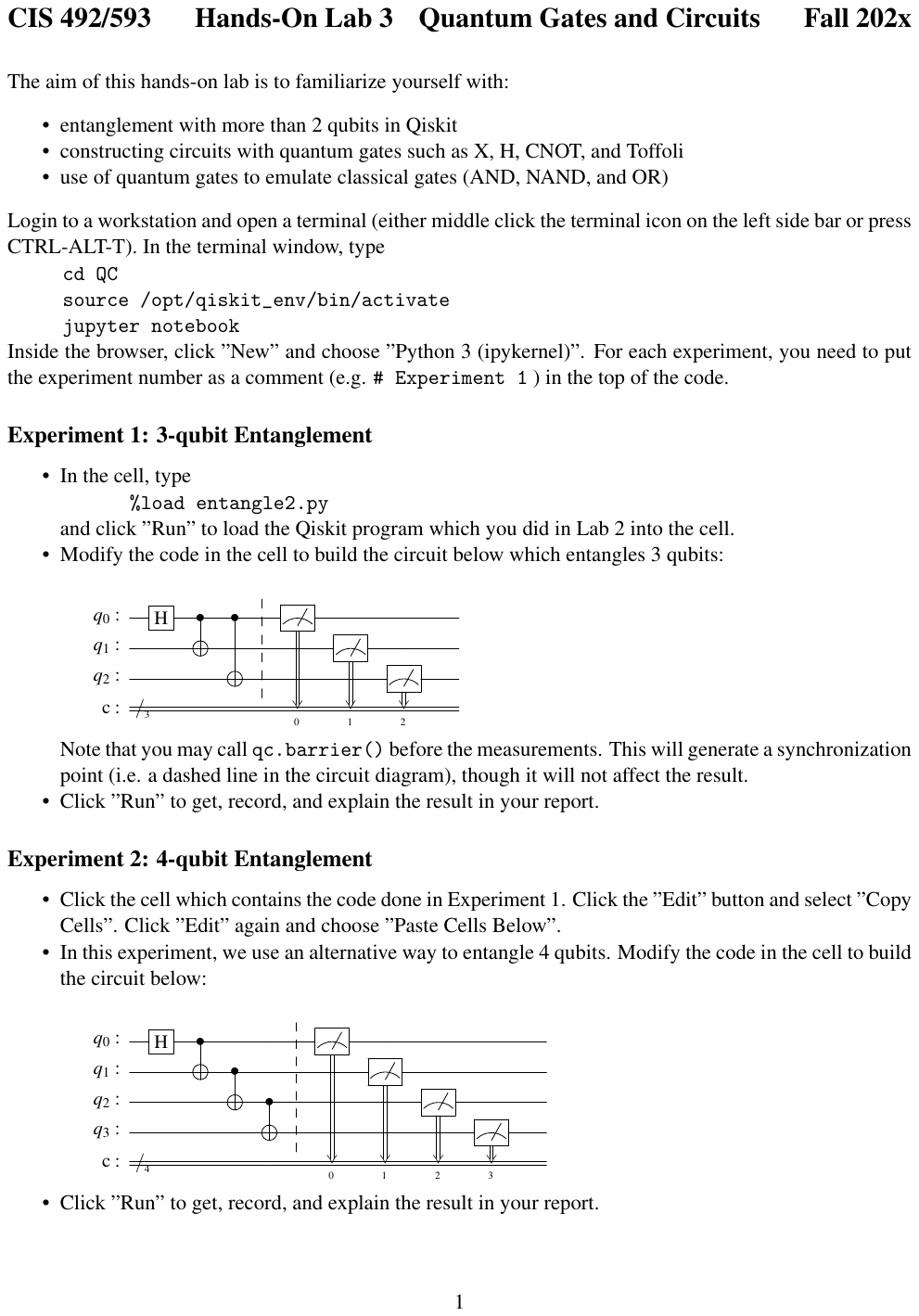}

\fancyhead[L]{\textbf{\textit{Appendix} Template: ANDgate.py}}
\includepdf[pages=-,pagecommand={},width=\textwidth]{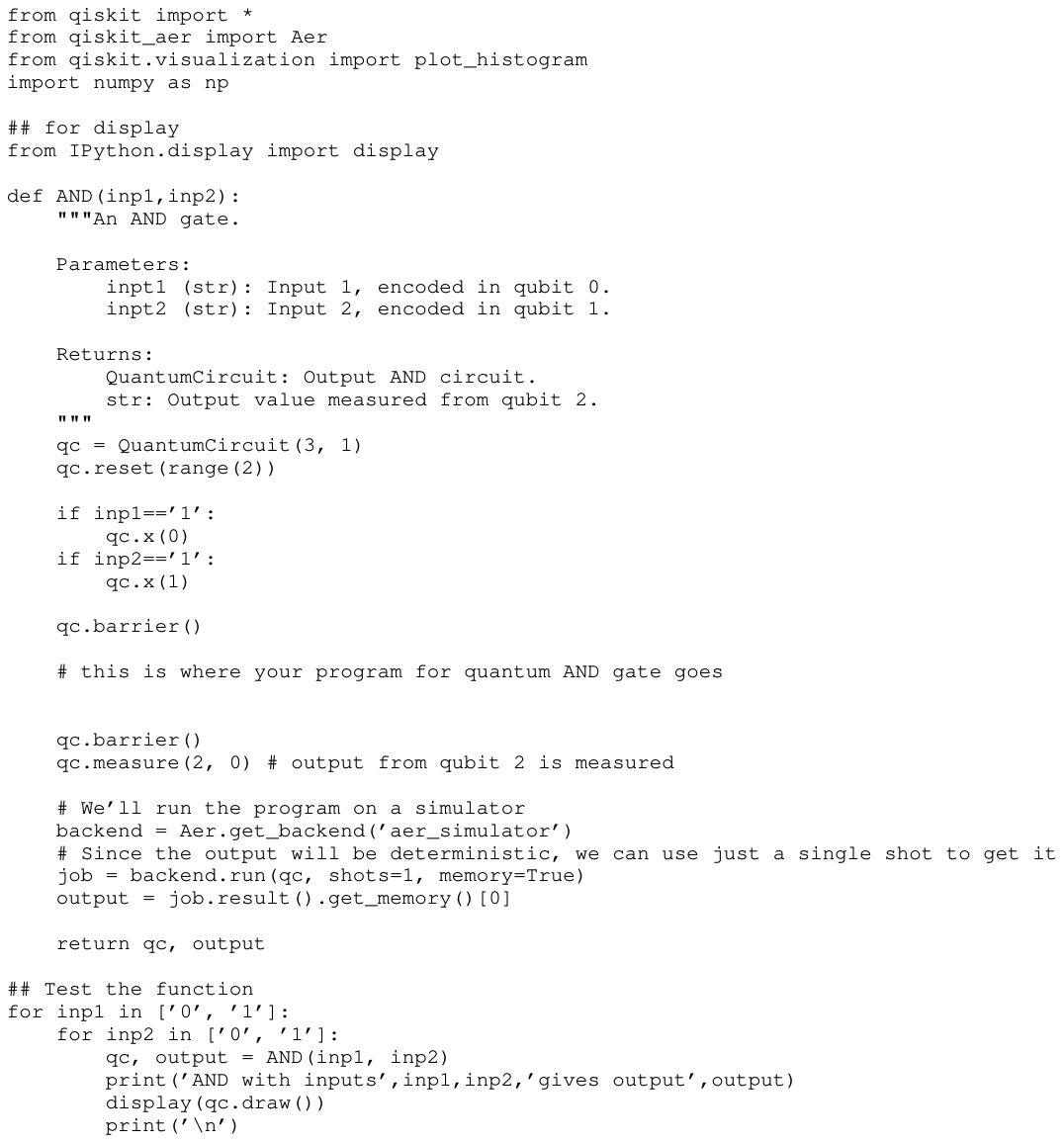}

\fancyhead[L]{\textbf{\textit{Appendix} Template: ORgate.py}}
\includepdf[pages=-,pagecommand={},width=\textwidth]{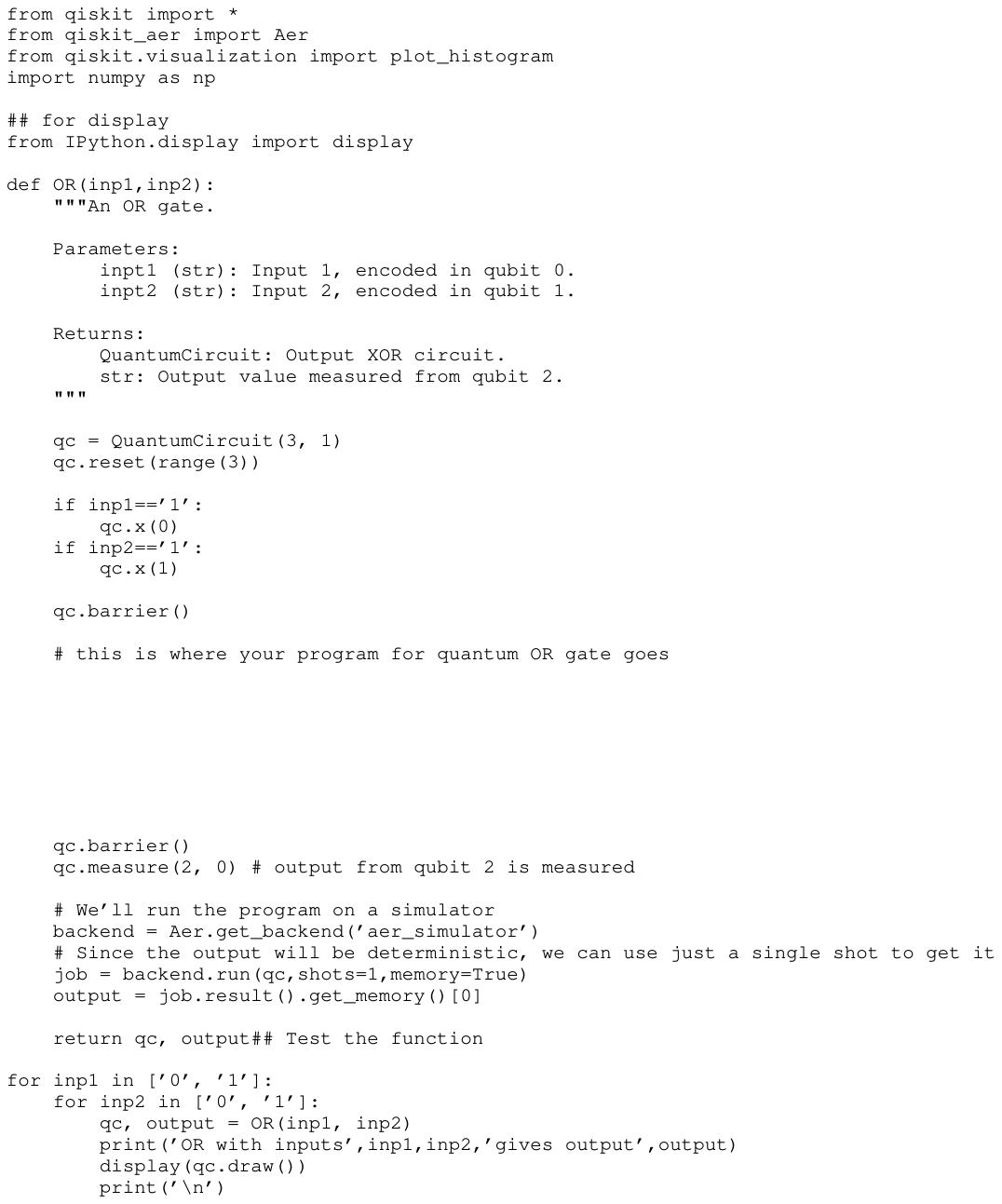}

\fancyhead[L]{\textbf{\textit{Appendix} Template: UNKNOWNfunc.py}}
\includepdf[pages=-,pagecommand={},width=\textwidth]{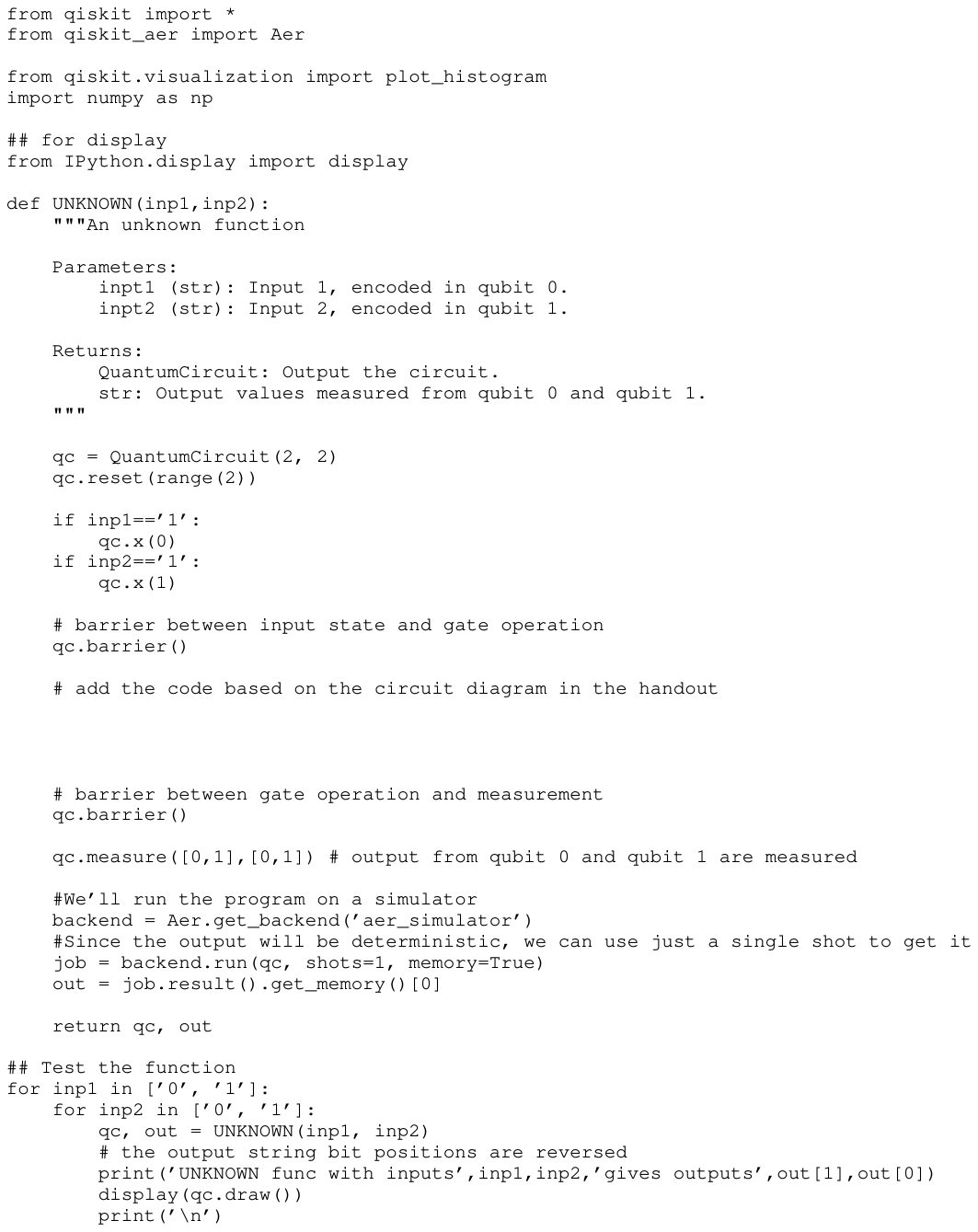}

\fancyhead[L]{\textbf{\textit{Appendix} Lab4: Quantum Key Distribution }}
\includepdf[pages=-,pagecommand={}]{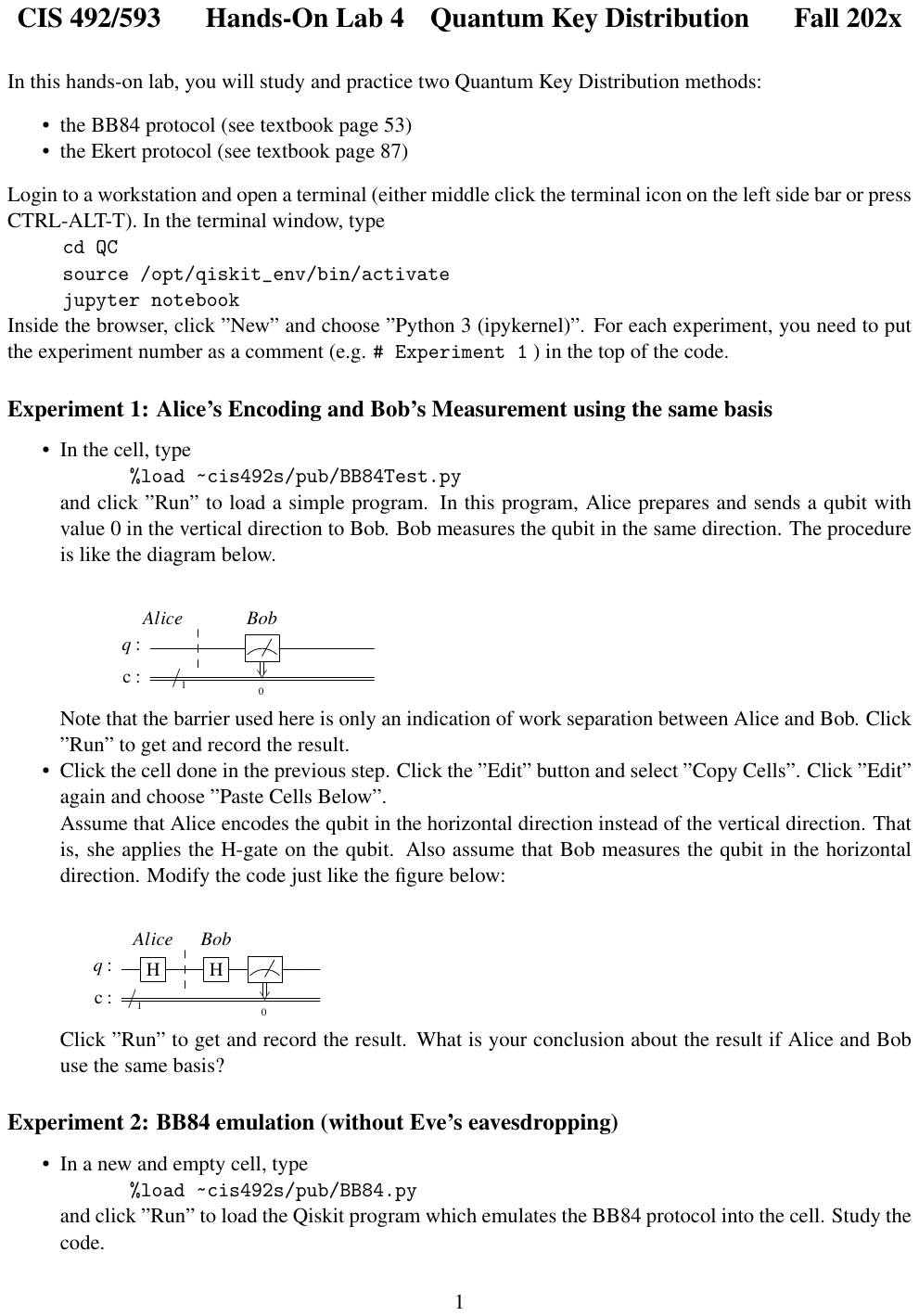}

\fancyhead[L]{\textbf{\textit{Appendix} Template: BB84Test.py}}
\includepdf[pages=-,pagecommand={},width=\textwidth]{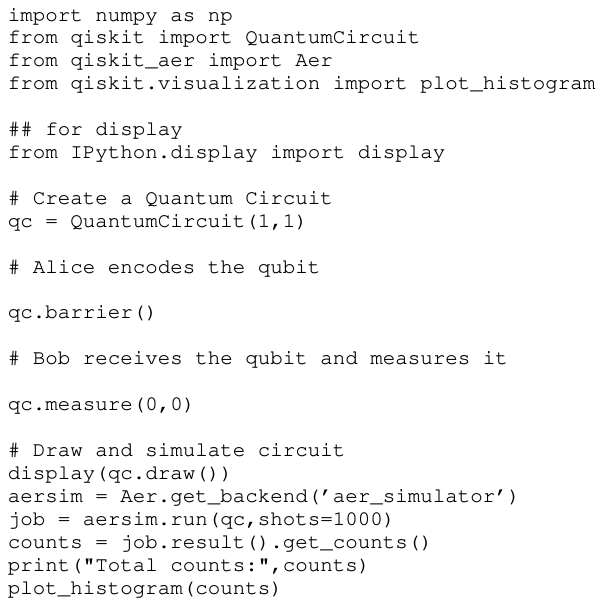}

\fancyhead[L]{\textbf{\textit{Appendix} Template: BB84.py}}
\includepdf[pages=-,pagecommand={},width=\textwidth]{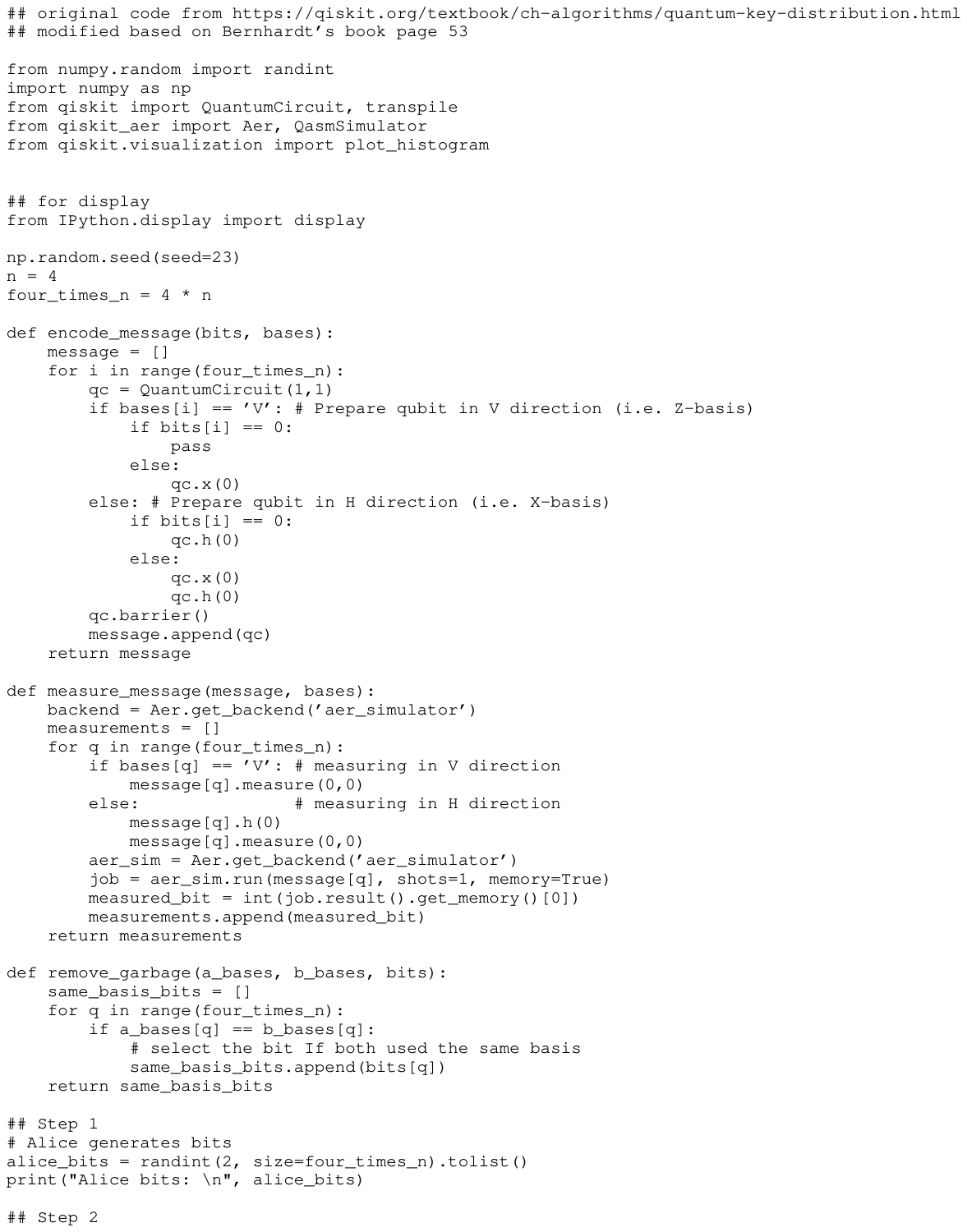}

\fancyhead[L]{\textbf{\textit{Appendix} Template: EkertTest.py}}
\includepdf[pages=-,pagecommand={},width=\textwidth]{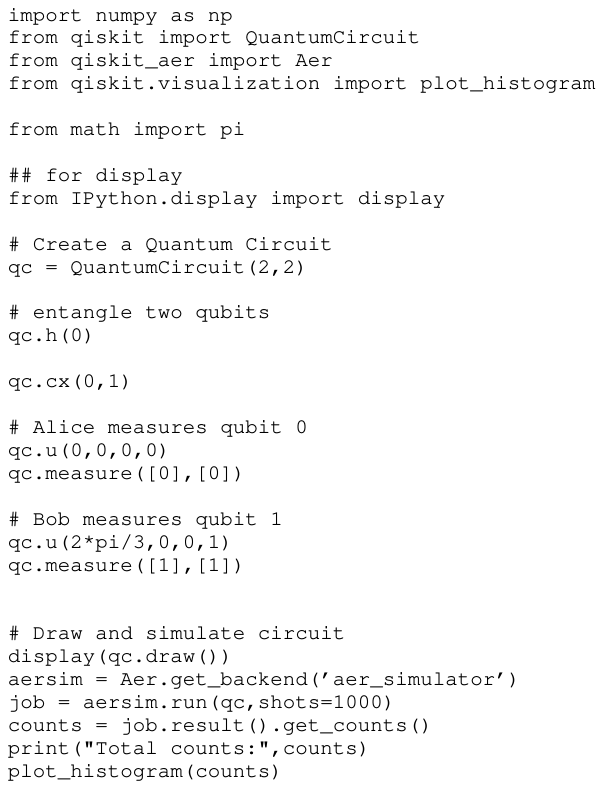}

\fancyhead[L]{\textbf{\textit{Appendix} Template: Ekert.py}}
\includepdf[pages=-,pagecommand={},width=\textwidth]{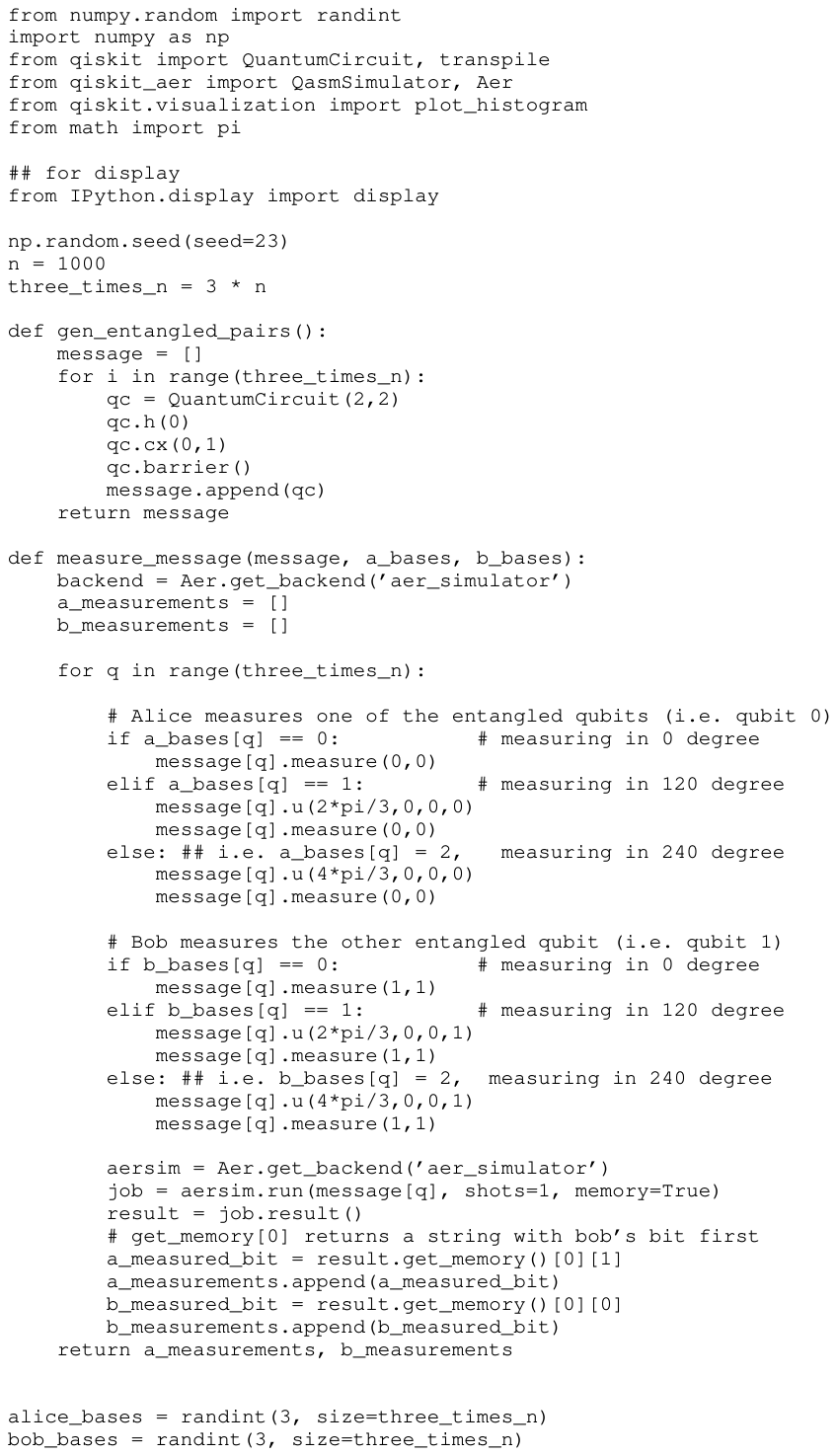}

\fancyhead[L]{\textbf{\textit{Appendix} Lab5: Deutsch and Deutsch-Jozsa Algorithms }}
\includepdf[pages=-,pagecommand={}]{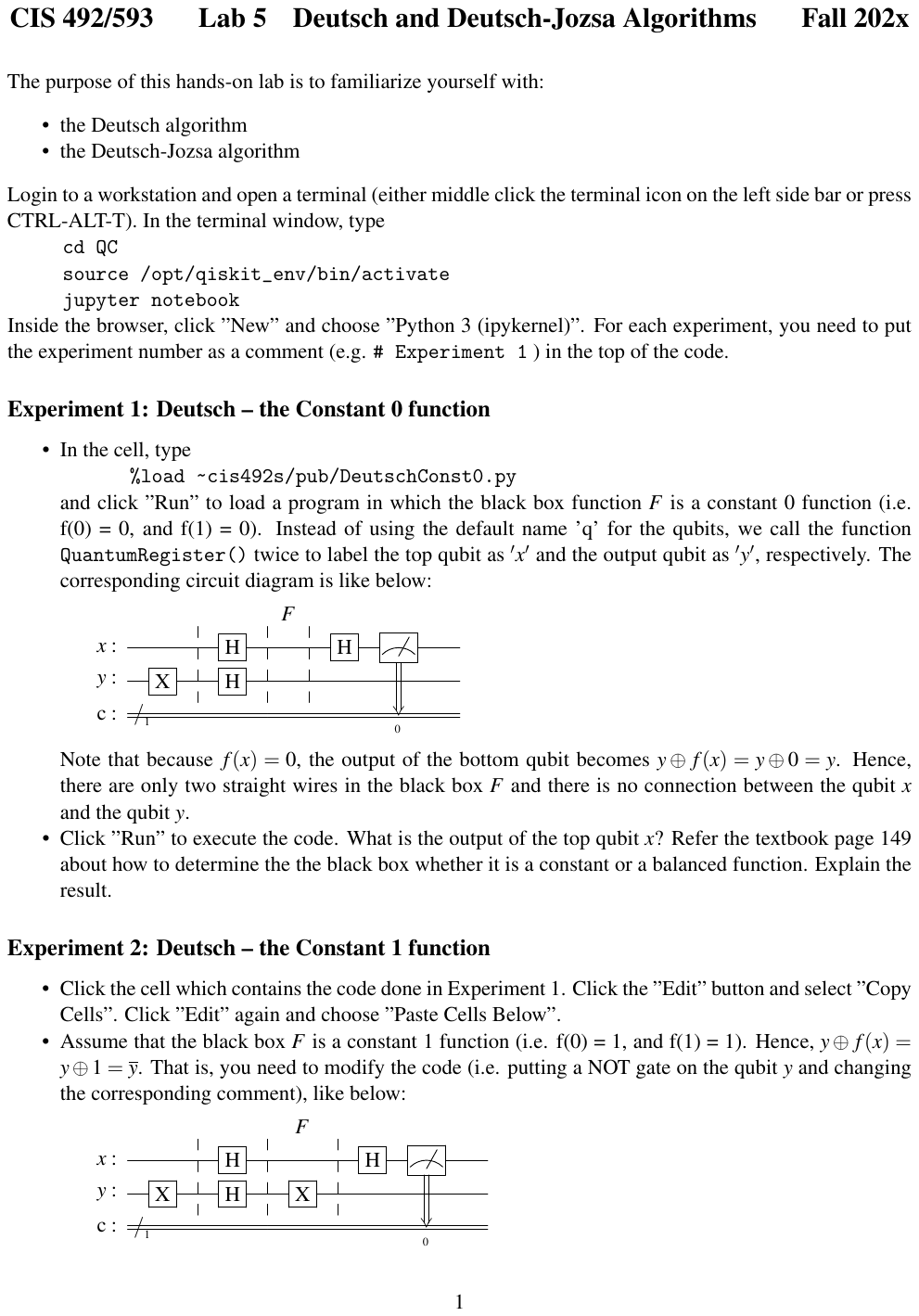}

\fancyhead[L]{\textbf{\textit{Appendix} Template: DeutschConst0.py}}
\includepdf[pages=-,pagecommand={},width=\textwidth]{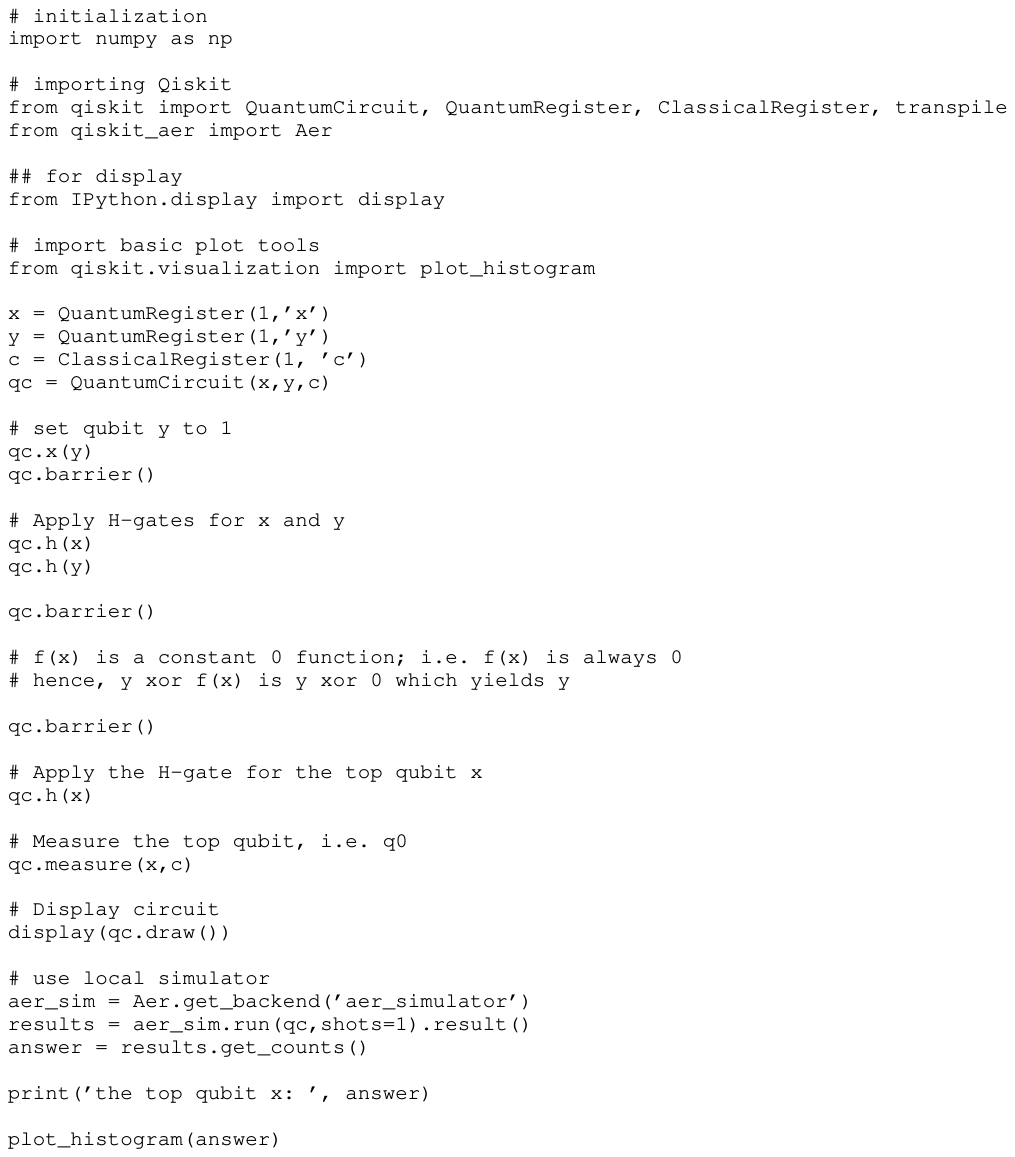}

\fancyhead[L]{\textbf{\textit{Appendix} Template: DJ-Const0.py}}
\includepdf[pages=-,pagecommand={},width=\textwidth]{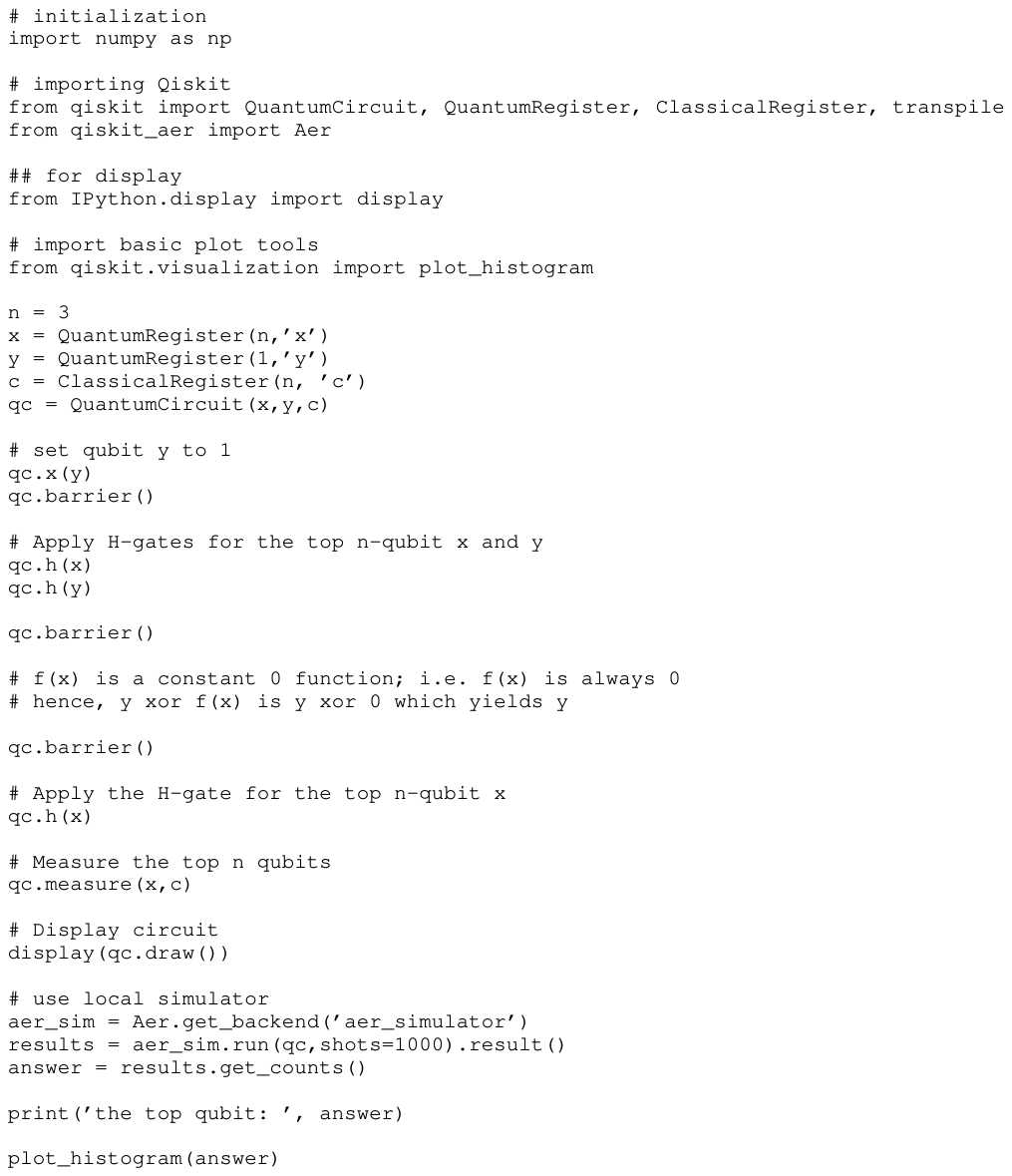}

\fancyhead[L]{\textbf{\textit{Appendix} Lab6: Simon’s Algorithm}}
\includepdf[pages=-,pagecommand={}]{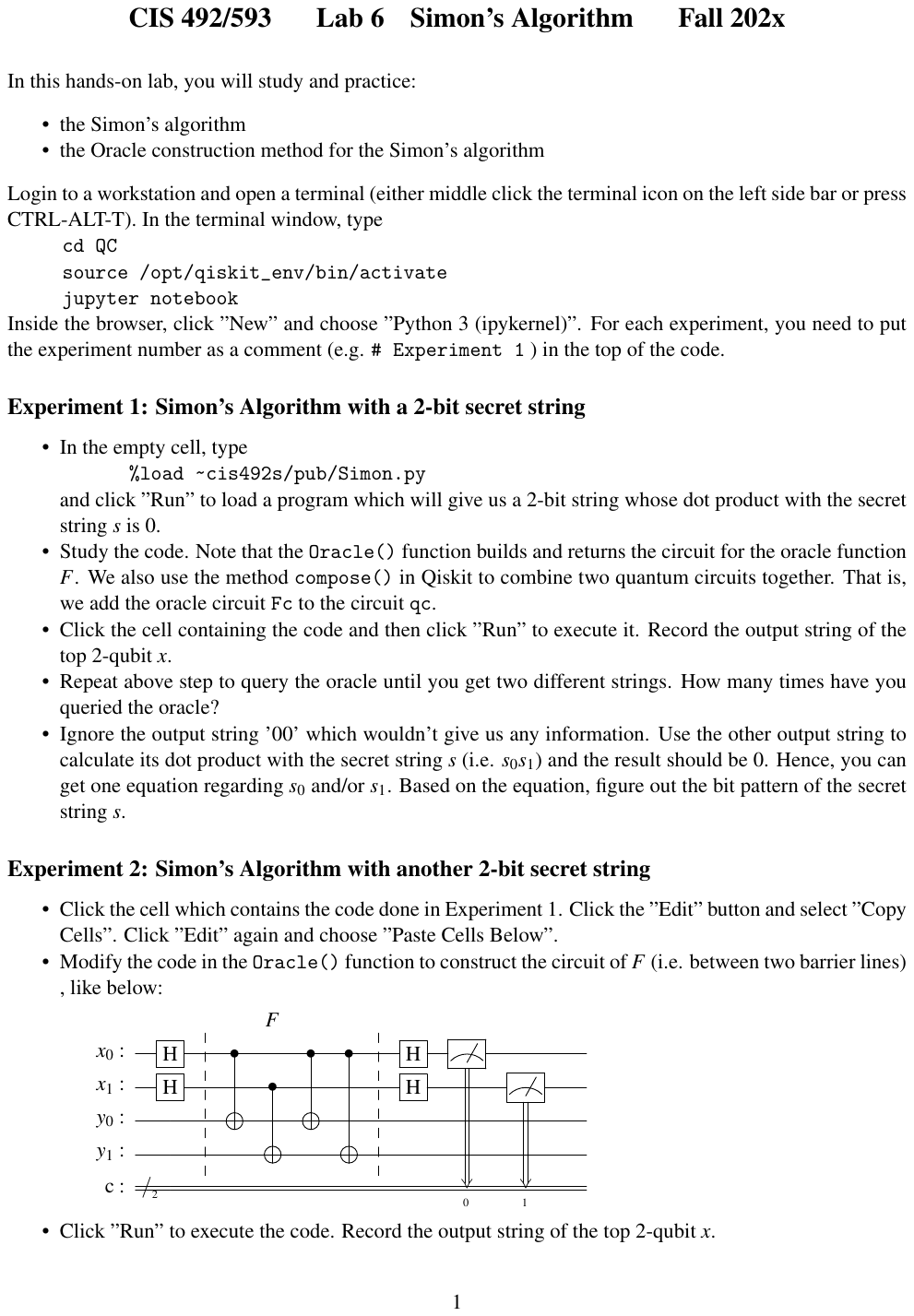}

\fancyhead[L]{\textbf{\textit{Appendix} Template: Simon.py}}
\includepdf[pages=-,pagecommand={},width=\textwidth]{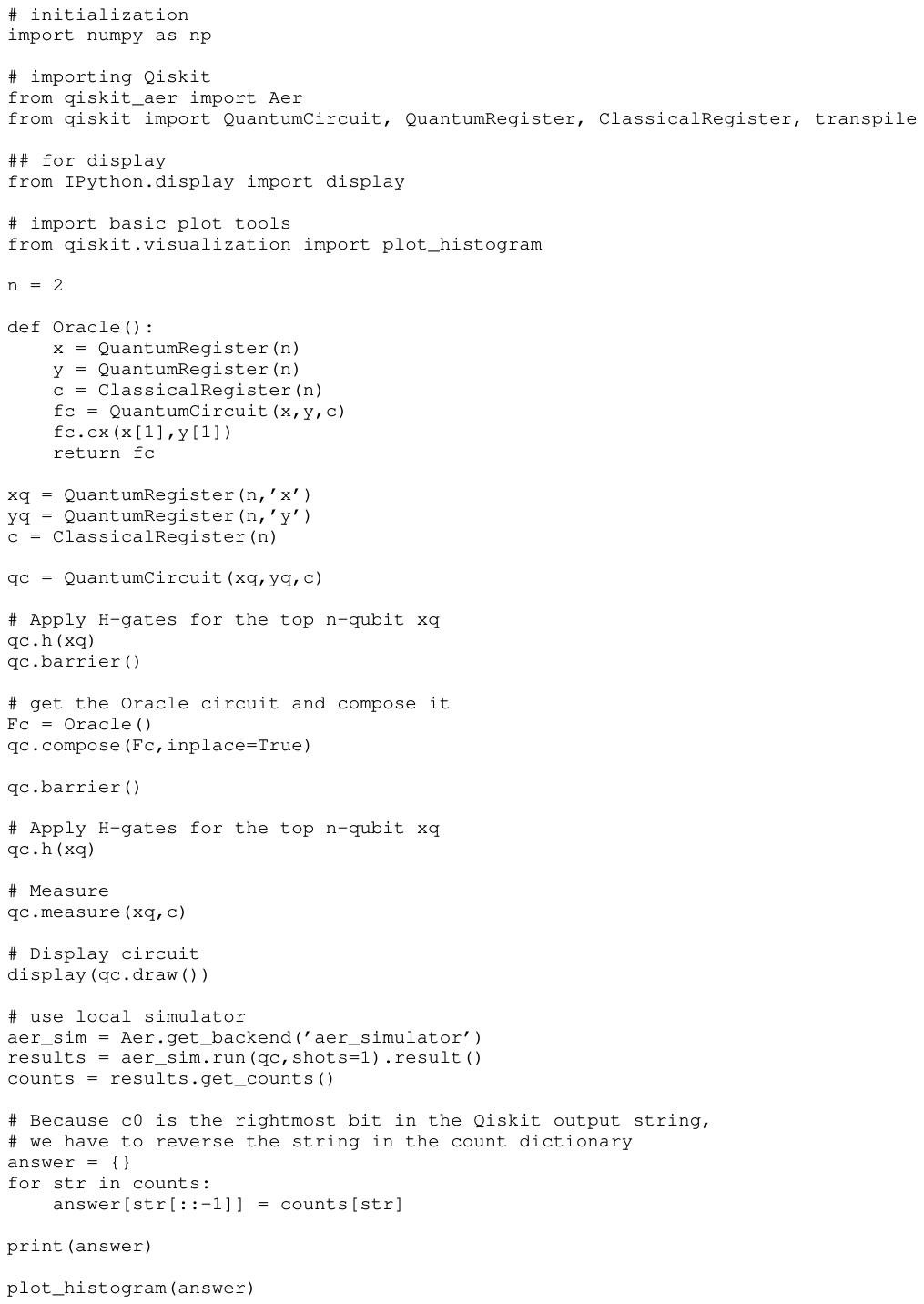}

\fancyhead[L]{\textbf{\textit{Appendix} Lab7: Grover’s Search Algorithm }}
\includepdf[pages=-,pagecommand={}]{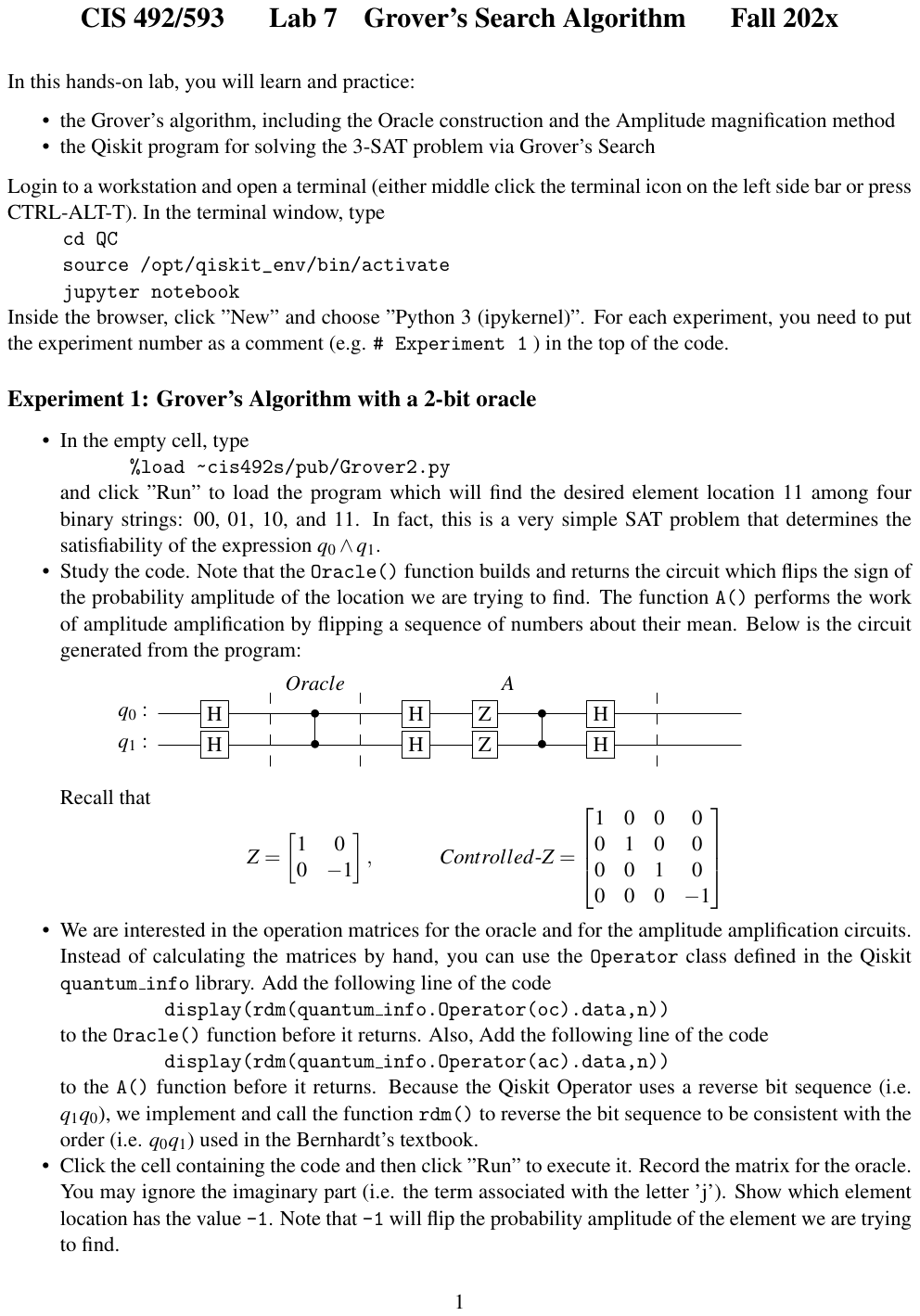}

\fancyhead[L]{\textbf{\textit{Appendix} Template: Grover2.py}}
\includepdf[pages=-,pagecommand={},width=\textwidth]{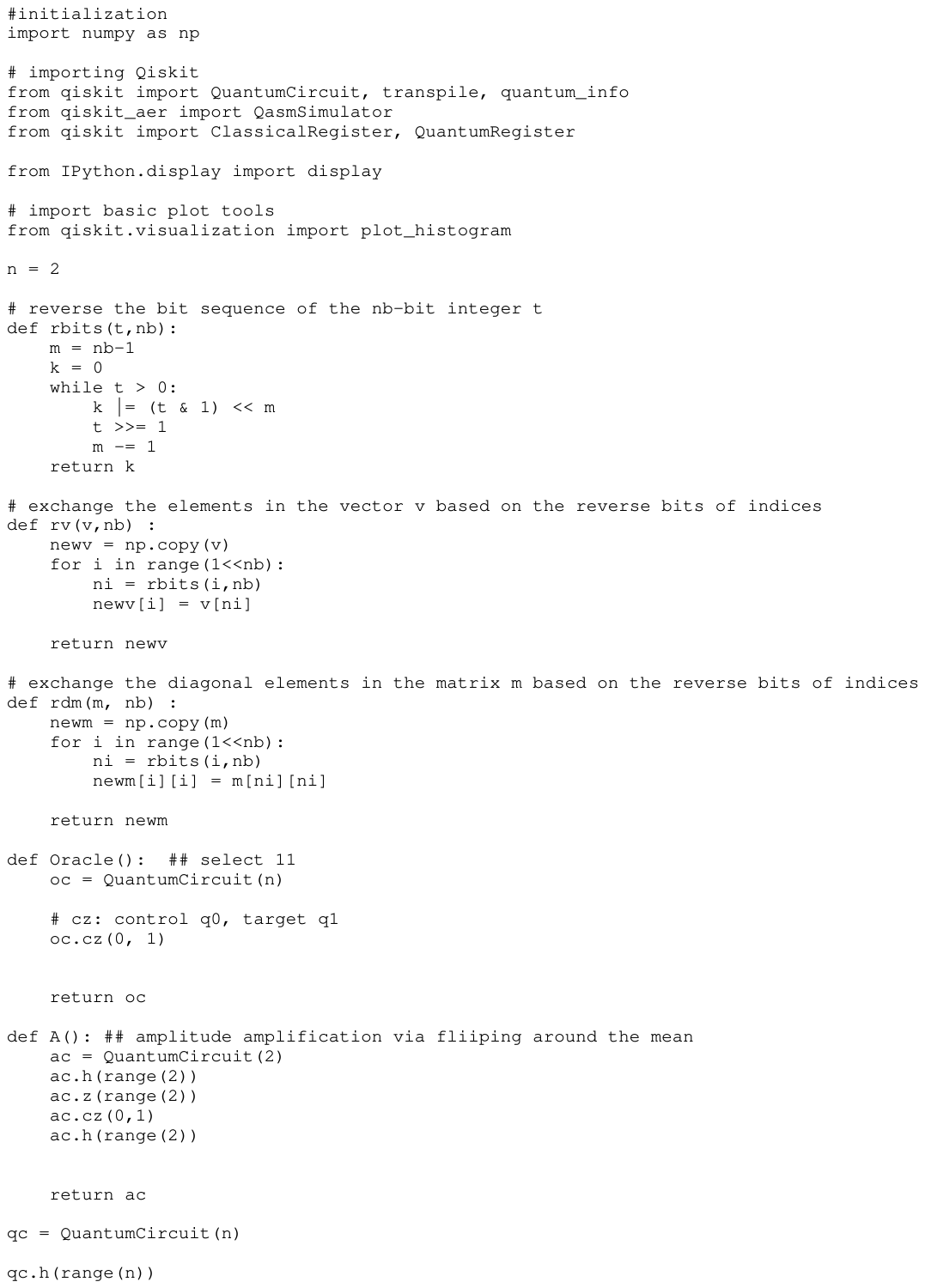}

\fancyhead[L]{\textbf{\textit{Appendix} Template: Grover3.py}}
\includepdf[pages=-,pagecommand={},width=\textwidth]{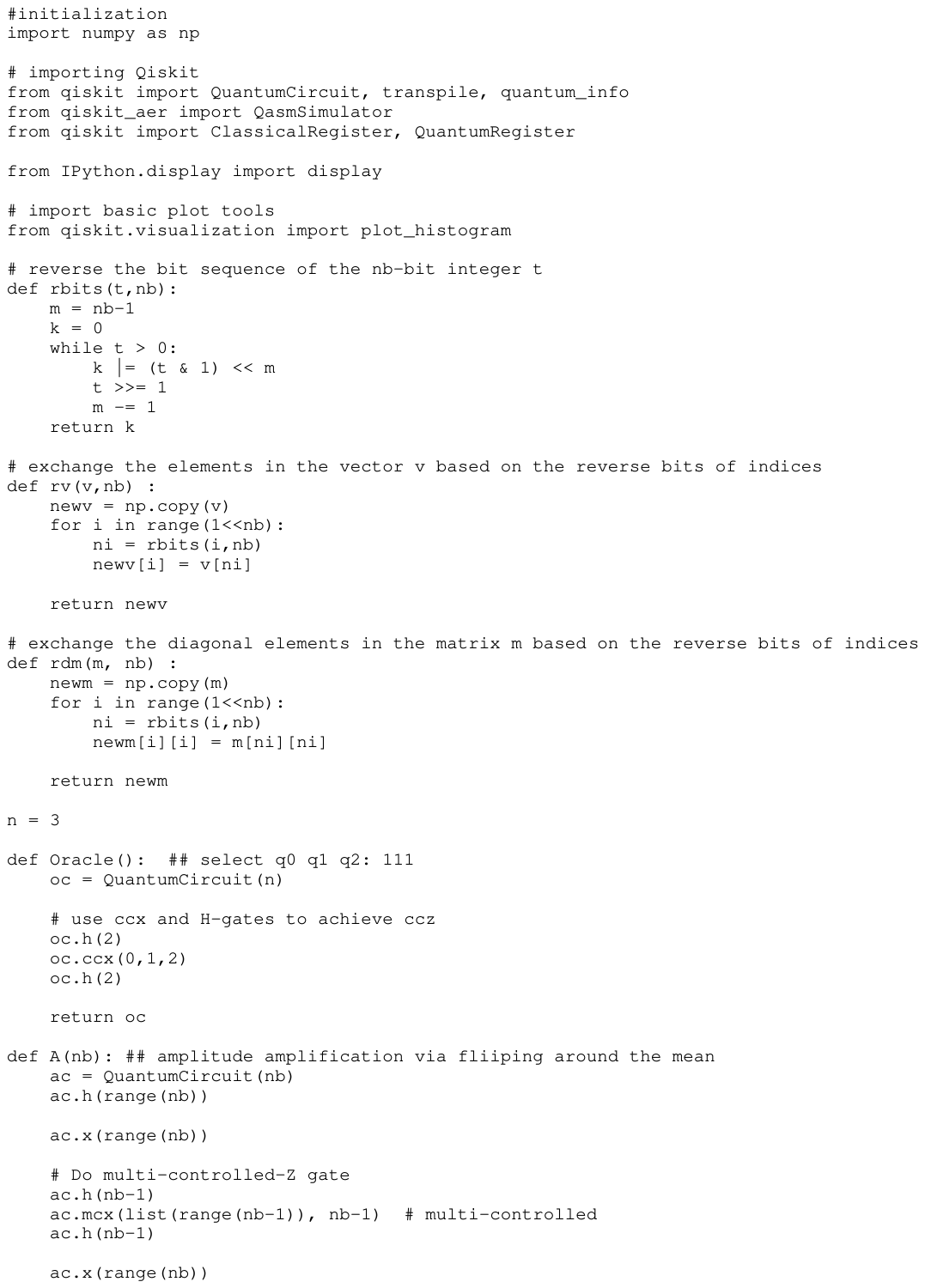}

\fancyhead[L]{\textbf{\textit{Appendix} Qiskit Version}}
\includepdf[pages=-,pagecommand={},width=\textwidth]{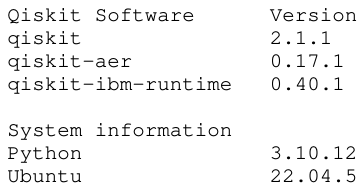}

\setlength{\voffset}{-2.54cm}
\setlength{\hoffset}{-2.54cm}

\end{document}


\begin{center}
{\Large \bf CIS 492/593 $\;\;\;\;\;$ Hands-On Lab 1 $\;\;$ Python and Jupyter $\;\;\;\;\;$ Fall 202x  } \\
\end{center}

\vskip 0.25in

\noindent In the first hands-on lab, you will: 
\begin{itemize}
\parskip 0in
\parsep 0in
\itemsep 0in

\item access your Linux account and change your password 
\item learn the features of Python 
\item practice programming with Jupyter Notebook 
\end{itemize}

\section*{Task I: Linux Environment settings}
Login to a Linux workstation and do the following steps:
\begin{itemize} 
\parskip 0in
\parsep 0in
\itemsep 0in

\item Type CTRL-ALT-T to open a terminal. On the left sidebar, right click
      the Terminal icon and choose "Add to Favorites".  

\item Inside the terminal window, type the command "yppasswd" to change your 
      password. Please remember your new password! Next, type "ssh grail" 
      to login to the server grail using  your new password to make sure it works. 
      Then, type "exit" to logout grail and return back.  
\end{itemize}

\section*{Task II: python3} 
Perform the following steps:
\begin{itemize} 
\parskip 0in
\parsep 0in
\itemsep 0in
\item In the terminal window, use the commands below to create a sub-directory QC 
      (i.e. a folder) and change your current working directory to it :\\
\mbox{\hskip 0.5in \tt mkdir QC}\\
\mbox{\hskip 0.5in \tt cd QC}

\item Use the command below to enter our pre-installed qiskit virtual environment:\\
\mbox{\hskip 0.5in \tt source /opt/qiskit\url{_}env/bin/activate}

\item Use the following command to copy a sample python/qiskit program to the
      current directory:\\
\mbox{\hskip 0.5in \tt cp \url{~}cis492s/pub/Hgate.py  . } \\
(Note: Don't forget the dot (.) at the end. In Linux, the dot means the current directory.)

\item To run the sample qiskit program, type:\\
\mbox{\hskip 0.5in \tt python3 Hgate.py}
       
\end{itemize}

\section*{Task III: jupyter notebook} 
Do the following experiments:

\subsection*{Experiment 1}
\begin{itemize}
\parskip 0in
\parsep 0in
\itemsep 0in

\item Use CTRL-ALT-T (or middle click the terminal icon on the left side bar) to
      open a second terminal. In the second terminal window, type\\
\mbox{\hskip 0.5in \tt cd QC} \\
      to change the current directory. Then, use\\
\mbox{\hskip 0.5in \tt source /opt/qiskit\url{_}env/bin/activate}\\
\mbox{\hskip 0.5in \tt jupyter notebook} \\
      to launch the Jupyter Notebook App, which is installed in our virtual environment. 
      You will find that the jupyter notebook is inside a web browser.  
\item Inside the browser, click "New" and choose "Python 3 (ipykernel)".
      You will see an entry box beginning with "{\tt In [ ]:}" which we call it
      a cell.

\item In the cell, type the following lines of python code:
\begin{verbatim}
  a = 5
  b = 7
  print(a,b)
\end{verbatim}
      (Note: Do NOT type any spaces or tabs at the beginning
      of each line above). Next, press SHIFT-Enter (or click "Run") to
      execute the code in the cell. Record the output result.

\end{itemize}

\subsection*{Experiment 2}
 
\begin{itemize}
\parskip 0in
\parsep 0in
\itemsep 0in

\item After you run the cell, you will have a new empty cell. Type
\begin{verbatim}
  a,b = b,a
  print(a,b)
\end{verbatim}
      to exchange two variable values without using a third temporary variable.
      Run the code and record the result.

\end{itemize}

\subsection*{Experiment 3}
\begin{itemize}
\parskip 0in
\parsep 0in
\itemsep 0in

\item Python uses indentation (i.e. the spaces at the beginning of a line)
      to indicate a block of code. In the third cell, type
\begin{verbatim}
  if a > b : 
      c = a + b
      d = a - b
      print(c,d)
  else:
      x = a * b
      y = a / b  
      print(x,y)
\end{verbatim}
    In C/C++ or Java, a block of statements should be included in a pair of 
    curly braces. Run the code and record the result.
\end{itemize}

\subsection*{Experiment 4}
\begin{itemize}
\parskip 0in 
\parsep 0in
\itemsep 0in

\item Click the first cell and change variable a's value to 9.
      Click the menu "Cell" and choose "Run All".
      Record the result changed in each cell.  
 
\end{itemize}

\subsection*{Experiment 5}
\begin{itemize}
\parskip 0in 
\parsep 0in
\itemsep 0in

\item Python "for" loop can iterate over a range of numbers, lists, or strings. In a new empty cell, type
\begin{verbatim}
  def foo(n):
     print('i   j   k')
     for i in range(n):
         for j in ['Hello','World']:
             for k in "XYZ":
                 print(i,j,k)

  foo(2) ## call the function 
\end{verbatim}
Run the code and record the result of the nested loops.
\end{itemize}

\subsection*{Experiment 6}
\begin{itemize}
\parskip 0in 
\parsep 0in
\itemsep 0in
\item In Python, we can perform the bitwise Exclusive OR operation (i.e. XOR) using the $\; \hat{ } \;$ symbol.
      In a new empty cell, type
\begin{verbatim}
   for i in [0,1]:
       for j in [0,1]:
           p, q = i, j
           p = p ^ q
           q = p ^ q
           p = p ^ q
           print("in p, q:",i,j, "  out p, q:", p,q)
\end{verbatim}
Run the code and record the result. Explain what the three sequential statements containing the 
  XOR operator do.
\end{itemize}

\subsection*{Experiment 7}
\begin{itemize}
\parskip 0in 
\parsep 0in
\itemsep 0in

\item Python's dictionary (also know as an associative array) 
      is a data structure which consists of a collection
      key-value pairs. The feature is that the array index (i.e. the key) can
      be a string. Type 
\begin{verbatim}
      counts = {}  # initialize counts as a dictionary
      counts['00'] = 477
      counts['11'] = 522
      counts['00'] += 1
      print(counts)
\end{verbatim}
      Run the code and record the result.
\end{itemize}

\subsection*{Experiment 8}
\begin{itemize}
\parskip 0in
\parsep 0in
\itemsep 0in

\item In a new empty cell, type 
\begin{verbatim}
     %load Hgate.py
\end{verbatim}
     and click "Run" to load the sample program into the cell.
     Click to select the cell and then click "Run" to execute the sample program.
     Record the result of total counts.

\end{itemize} 

\section*{Turnin}

\noindent Click "File" and choose "Save as". Type "lab1" in the entry box and click "Save"
          to save your work today into the file {\tt lab1.ipynb}. 
          To turn in your file, use CTRL-ALT-T to open a terminal and type\\
\mbox{\hskip 0.5in \tt ssh grail} \\
and type your password to login to the server grail. Then, type\\
\mbox{\hskip 0.5in \tt cd QC} \\
\mbox{\hskip 0.5in \tt turnin -c cis492s -p lab1 lab1.ipynb} \\
          to electronically submit your file {\tt lab1.ipynb}. \\

\noindent To shutdown the jupyter notebook, hit CTRL-C in the terminal window
      which you launch the jupyter notebook. Type {\tt y} and press Enter.
 
\noindent To log out from Ubuntu desktop session, go to the top right corner, click to select "Power Off / Log Out", and
choose "Log Out", or simply use the keyboard shortcut CTRL + ALT + DEL and click "Log Out".   \\

\noindent {\bf Hand in your hard-copy lab report before your leave.}


\begin{center}
{\Large \bf CIS 492/593 $\;$ Hands-On Lab 2 $\;$ Qiskit and Entanglement $\;$ Fall 202x  } \\
\end{center}

\vskip 0.25in

\noindent In this hands-on lab, you will learn:
\begin{itemize}
\parskip 0in
\parsep 0in
\itemsep 0in
\item Qiskit 
\item Hadamard gate 
\item two-qubit entanglement
\end{itemize}

\noindent Login to a workstation and open a terminal (either middle click the 
      terminal icon on the left side bar or press CTRL-ALT-T).
      In the terminal window, type  
\begin{verbatim}
     cd QC 
     source /opt/qiskit_env/bin/activate
     jupyter notebook 
\end{verbatim}

\noindent Inside the browser, click "New" and choose "Python 3 (ipykernel)".
For each experiment, you need to put the experiment number
as a comment (e.g. {\tt \# Experiment 1} ) in the top of the code.

\subsection*{Experiment 1: Qiskit Programming}
\begin{itemize} 
\parskip 0in
\parsep 0in
\itemsep 0in

\item In the cell, type\\
\mbox{\hskip 0.5in \tt \%load Hgate.py }\\
     and click "Run" to load the sample Qiskit program into the cell
	and then put the experiment number as a comment.\\
   
     {\it Below is a brief description of the program.
     The code imports several modules at beginning. Then, it invokes the
     constructor {\tt QuantumCircuit(1,1)} to create a quantum circuit {\tt qc}
     which contains 1 qubit and 1 classical bit. Next, we apply an H gate (i.e.
     {\tt qc.h(0)} to the qubit 0. Note that the qubit is allocated using the 
     array format and the first qubit is indexed with 0. Same for the classical bit.
     Finally, we use {\tt qc.measure(0,0)} to measure the qubit 0 and put the result to the
     classical bit 0.   

     After building the circuit, the program calls the function {\tt transpile()} to compile the abstract 
     circuit {\tt qc} to low-level instructions and then uses the Aer simulator
     to execute the instructions 1000 times. } \\
     
\item When calling {\tt simulator.run()}, change {\tt "shots=1000"} to {\tt "shots=10000"} and also add \\
	\mbox{\hskip 0.5in \tt plot\url{_}histogram(counts) }\\
     at the end of the program.
     Click "Run" to execute the code in the cell. Record the numbers for 
     {\tt '0'} and {\tt '1'}.
\item Click the cell and click "Run" again.
     Do you get EXACTLY the same numbers for {\tt '0'} and {\tt '1'} as in 
     the previous run? Explain your result. Moreover, what is the data structure
     of the variable {\tt counts}? 
\end{itemize}

\subsection*{Experiment 2: Apply two H gates in succession} 
     If we consider the H gate to be analogous to a coin toss, two H gates 
in succession should be tossing the coin twice. We would still expect close to a 50/50 
distribution of zeros and ones for the second flipping. However, quantum randomness 
is not simply like a classical random coin toss. Let's do the following experiment.

\begin{itemize} 
\parskip 0in
\parsep 0in
\itemsep 0in

\item Click the cell which contains the code in Experiment 1. Click the "Edit" button
      and select "Copy Cells". Click "Edit" again and choose "Paste Cells Below".
     
\item Apply the H gates twice in a row, like the diagram below:

\scalebox{1.0}{
\Qcircuit @C=1.0em @R=0.2em @!R { \\
	 	\nghost{{q} :  } & \lstick{{q} :  } & \gate{\mathrm{H}} & \gate{\mathrm{H}} & \meter & \qw & \qw\\
	 	\nghost{\mathrm{{c} :  }} & \lstick{\mathrm{{c} :  }} & \lstick{/_{_{1}}} \cw & \cw & \dstick{_{_{\hspace{0.0em}0}}} \cw \ar @{<=} [-1,0] & \cw & \cw\\
\\ }}

      That is, you can add another {\tt qc.h(0)} immediately after {\tt qc.h(0)} in the code.

\item Click "Run" to see and record the result. To verify, calculate by hand the state vector
	after applying the first H gate in your hard-copy lab report. 
	Note that the qubit is initialized to be 0.
{\it Hint:} 
	$$
     \begin{bmatrix}[1.75]
         \frac{1}{\sqrt{2}} & \frac{1}{\sqrt{2}} \\
         \frac{1}{\sqrt{2}} & \frac{-1}{\sqrt{2}}
     \end{bmatrix}
     \times
     \begin{bmatrix}[1.75]
       \mbox{\hskip 0.2in }  \\
       \mbox{\hskip 0.2in }
     \end{bmatrix}
      =
$$

	Then, 
      calculate the state vector
      after applying the second H gate.

\end{itemize}

\subsection*{Experiment 3: Apply two H gates in succession with initial value 1} 

\begin{itemize}
\parskip 0in
\parsep 0in
\itemsep 0in

\item Click the cell above (i.e. two H gates in succession). Click the "Edit" button
      and select "Copy Cells". Click "Edit" again and choose "Paste Cells Below".
 
\item Instead of using the default value 0 for the qubit, we would like
      to initialize the qubit to 1 before applying the two qubits. To achieve
      this, you can add an X gate (i.e. the NOT gate) by using {\tt qc.x(0)} before the first calling 
      of {\tt qc.h(0)}, like below:   

\scalebox{1.0}{
\Qcircuit @C=1.0em @R=0.2em @!R { \\
	 	\nghost{{q} :  } & \lstick{{q} :  } & \gate{\mathrm{X}} & \gate{\mathrm{H}} & \gate{\mathrm{H}} & \meter & \qw & \qw\\
	 	\nghost{\mathrm{{c} :  }} & \lstick{\mathrm{{c} :  }} & \lstick{/_{_{1}}} \cw & \cw & \cw & \dstick{_{_{\hspace{0.0em}0}}} \cw \ar @{<=} [-1,0] & \cw & \cw\\
\\ }}

\item Click "Run" to see and record the result. Again, calculate by hand the state vector
      after applying the first H gate in your report. 
	Then, calculate the state vector
      after applying the second H gate.

\end{itemize}
     
\subsection*{Experiment 4: Apply an H gate for each of two parallel qubits} 

\begin{itemize}
\parskip 0in
\parsep 0in
\itemsep 0in

\item Click the cell which contains the code done in Experiment 1. Click the "Edit" button
      and select "Copy Cells". Click "Edit" again and choose "Paste Cells Below".

\item Modify the code in the cell to build the circuit below:\\
\scalebox{1.0}{
\Qcircuit @C=1.0em @R=0.2em @!R { \\
	 	\nghost{{q}_{0} :  } & \lstick{{q}_{0} :  } & \gate{\mathrm{H}} & \meter & \qw & \qw & \qw\\
	 	\nghost{{q}_{1} :  } & \lstick{{q}_{1} :  } & \gate{\mathrm{H}} & \qw & \meter & \qw & \qw\\
	 	\nghost{\mathrm{{c} :  }} & \lstick{\mathrm{{c} :  }} & \lstick{/_{_{2}}} \cw & \dstick{_{_{\hspace{0.0em}0}}} \cw \ar @{<=} [-2,0] & \dstick{_{_{\hspace{0.0em}1}}} \cw \ar @{<=} [-1,0] & \cw & \cw\\
\\ }}

      {\it Hint:} The circuit needs 2 qubits and 2 classical bits when calling 
                 {\tt QuantumCircuit()}. You also need to add a H gate on qubit 1.
                 Lastly, use {\tt qc.measure([0,1],[0,1]) } to map the qubits 0 and 1
                 to the classical bits 0 and 1, respectively. 

\item Click "Run" to get and record the result. Explain.

\item What is the initial state vector? 
$$
    q_0 \otimes q_1 = 
    \begin{bmatrix}[1.75]
       \mbox{\hskip 0.2in }  \\
       \mbox{\hskip 0.2in }  
     \end{bmatrix}
             \otimes
     \begin{bmatrix}[1.75]
       \mbox{\hskip 0.2in }  \\
       \mbox{\hskip 0.2in }  
     \end{bmatrix}
     =
     \begin{bmatrix}[1.5]
       \mbox{\hskip 0.2in }  \\
       \mbox{\hskip 0.2in }  \\
       \mbox{\hskip 0.2in }  \\
       \mbox{\hskip 0.2in }  
     \end{bmatrix}
$$

      Calculate the state vector after applying the parallel H gates. {\it Hint:}
$$
      \frac{1}{2} \times
     \begin{bmatrix}[1.75]
	     1 & \phantom{-}1  & \phantom{-}1 &  \phantom{-}1 \\ 
	     1 & -1            & \phantom{-}1 & -1 \\ 
	     1 & \phantom{-}1  & -1           & -1 \\ 
	     1 & -1             & -1           & \phantom{-}1 \\ 
     \end{bmatrix}
	     \times
     \begin{bmatrix}[1.75]
       \mbox{\hskip 0.2in }  \\
       \mbox{\hskip 0.2in }  \\
       \mbox{\hskip 0.2in }  \\
       \mbox{\hskip 0.2in }  
     \end{bmatrix}
     =
$$

\end{itemize}

\subsection*{Experiment 5: Entangle two qubits}

\begin{itemize}
\parskip 0in
\parsep 0in
\itemsep 0in

\item Click the cell which contains the code done in Experiment 4. Click the "Edit" button
      and select "Copy Cells". Click "Edit" again and choose "Paste Cells Below".
 
\item Modify the code in the cell to build the circuit below:\\
\scalebox{1.0}{
\Qcircuit @C=1.0em @R=0.2em @!R { \\
	 	\nghost{{q}_{0} :  } & \lstick{{q}_{0} :  } & \gate{\mathrm{H}} & \ctrl{1} & \meter & \qw & \qw & \qw\\
	 	\nghost{{q}_{1} :  } & \lstick{{q}_{1} :  } & \qw & \targ & \qw & \meter & \qw & \qw\\
	 	\nghost{\mathrm{{c} :  }} & \lstick{\mathrm{{c} :  }} & \lstick{/_{_{2}}} \cw & \cw & \dstick{_{_{\hspace{0.0em}0}}} \cw \ar @{<=} [-2,0] & \dstick{_{_{\hspace{0.0em}1}}} \cw \ar @{<=} [-1,0] & \cw & \cw\\
\\ }}

     {\it Hint:} Replace the H gate on qubit 1 with a CNOT gate using qubit 0 as the control qubit
                 and qubit 1 as the target qubit (i.e. {\tt qc.cx(0,1)}).

\item Click "Run" to get and record the result. Explain.

\item To save the code in this cell into a file {\tt entangle2.py}, put the following\\
\mbox{\hskip 0.5in \tt \%\%writefile entangle2.py }\\
      as the FIRST line of the code. Click "Run". 

\item To check the file {\tt entangle2.py} has been saved correctly, open a terminal and
     type
\begin{verbatim}
     cd QC
     cat entangle2.py
     python3 entangle2.py
\end{verbatim} 

\end{itemize}

\subsection*{Experiment 6: Are they entangled? }

\begin{itemize}
\parskip 0in
\parsep 0in
\itemsep 0in

\item Click the cell which contains the code done in Experiment 5. Click the "Edit" button
      and select "Copy Cells". Click "Edit" again and choose "Paste Cells Below".
      Remove the first line \\
\mbox{\hskip 0.5in \tt \%\%writefile entangle2.py }\\

\item Add the following two lines of code
\begin{verbatim}
     qc.x(1)
     qc.barrier()
\end{verbatim}
     immediately after calling the function {\tt QuantumCircuit()}, just like the diagram below: \\
\scalebox{1.0}{
\Qcircuit @C=1.0em @R=0.2em @!R { \\
	 	\nghost{{q}_{0} :  } & \lstick{{q}_{0} :  } & \qw \barrier[0em]{1} & \qw & \gate{\mathrm{H}} & \ctrl{1} & \meter & \qw & \qw & \qw\\
	 	\nghost{{q}_{1} :  } & \lstick{{q}_{1} :  } & \gate{\mathrm{X}} & \qw & \qw & \targ & \qw & \meter & \qw & \qw\\
	 	\nghost{\mathrm{{c} :  }} & \lstick{\mathrm{{c} :  }} & \lstick{/_{_{2}}} \cw & \cw & \cw & \cw & \dstick{_{_{\hspace{0.0em}0}}} \cw \ar @{<=} [-2,0] & \dstick{_{_{\hspace{0.0em}1}}} \cw \ar @{<=} [-1,0] & \cw & \cw\\
\\ }}

The purpose of using the X gate on qubit 1 is to invert its initial value from 0 to 1.      
		The {\tt barrier()} (i.e. a dashed line in the circuit diagram) 
		is a directive which tells the circuit transpiler not to 
optimize the code crossing the barrier separation line. Here we just use it for more readability.
     
\item Click "Run" to see and record the result. 
      Are these two qubits entangled? Explain. {\it Hint: See Textbook Page 59}

\item Calculate the state vector (i.e. the tensor product of the two qubits)
	before applying the CNOT gate. 
$$
	    \begin{bmatrix}[1.75]
		    \frac{1}{\sqrt{2}} \\
		    \frac{1}{\sqrt{2}} \\
     \end{bmatrix}
             \otimes
     \begin{bmatrix}[1.75]
       \mbox{\hskip 0.2in }  \\
       \mbox{\hskip 0.2in }  \\
     \end{bmatrix}
     =
     \begin{bmatrix}[1.75]
       \mbox{\hskip 0.2in }  \\
       \mbox{\hskip 0.2in }  \\
       \mbox{\hskip 0.2in }  \\
       \mbox{\hskip 0.2in }  
     \end{bmatrix}
$$
\\
	Calculate the state vector after applying the CNOT gate.  
	{\it Hint: similar calculation can be found on Textbook Page 67.} 
$$
     \begin{bmatrix}[1.75]
	    \;\; 1 & \;\; 0  & \;\; 0 & \;\; 0 \; \\ 
	    \;\; 0 & \;\; 1  & \;\; 0 & \;\; 0 \; \\ 
	    \;\; 0 & \;\; 0  & \;\; 0 & \;\; 1 \; \\ 
	    \;\; 0 & \;\; 0  & \;\; 1 & \;\; 0 \; \\ 
     \end{bmatrix}
	     \times
     \begin{bmatrix}[1.75]
       \mbox{\hskip 0.2in }  \\
       \mbox{\hskip 0.2in }  \\
       \mbox{\hskip 0.2in }  \\
       \mbox{\hskip 0.2in }  
     \end{bmatrix}
     =
$$

\end{itemize}

\section*{Turnin}

\noindent Click "File" and choose "Save as". Type "lab2" in the entry box and click "Save"
          to save your work today into the file {\tt lab2.ipynb}.
          To turn in your file, use CTRL-ALT-T to open a terminal and type\\
\mbox{\hskip 0.5in \tt ssh grail} \\
and type your password to login to the server grail. Then, type\\
\mbox{\hskip 0.5in \tt cd QC} \\
\mbox{\hskip 0.5in \tt turnin -c cis492s -p lab2 lab2.ipynb} \\
          to electronically submit your file {\tt lab2.ipynb}. \\

\noindent If you finish early, you can start doing the 
          "IBM QC Homework" posted on the CSU Blackboard
	  to run Qiskit programs on a real IBM quantum computer. \\

\noindent Shutdown the jupyter notebook and logout the workstation. \\

\noindent {\bf Hand in your lab report before your leave.}


\begin{center}
{\Large \bf CIS 492/593 $\;\;\;\;\;$ Hands-On Lab 3 $\;\;$ Quantum Gates and Circuits $\;\;\;\;\;$ Fall 202x  } \\
\end{center}

\vskip 0.25in
\noindent The aim of this hands-on lab is to familiarize yourself with:
\begin{itemize}
\parskip 0in
\parsep 0in
\itemsep 0in
\item entanglement with more than 2 qubits in Qiskit 
\item constructing circuits with quantum gates such as X, H, CNOT, and Toffoli    
\item use of quantum gates to emulate classical gates (AND, NAND, and OR) 
\end{itemize}

\noindent Login to a workstation and open a terminal (either middle click the
terminal icon on the left side bar or press CTRL-ALT-T).
In the terminal window, type
\begin{verbatim}
     cd QC
     source /opt/qiskit_env/bin/activate
     jupyter notebook
\end{verbatim}
Inside the browser, click "New" and choose "Python 3 (ipykernel)".
For each experiment, you need to put the experiment number
as a comment (e.g. {\tt \# Experiment 1} ) in the top of the code.

\subsection*{Experiment 1: 3-qubit Entanglement }
\begin{itemize} 
\parskip 0in
\parsep 0in
\itemsep 0in
\item In the cell, type \\
\mbox{\hskip 0.5in \tt \%load entangle2.py }\\
and click "Run" to load the Qiskit program which you did in Lab 2 into the cell.
   
\item Modify the code in the cell to build the circuit below which entangles 3 qubits: \\
\scalebox{0.9}{
\Qcircuit @C=1.0em @R=0.2em @!R { \\
	 	\nghost{{q}_{0} :  } & \lstick{{q}_{0} :  } & \gate{\mathrm{H}} & \ctrl{1} & \ctrl{2} \barrier[0em]{2} & \qw & \meter & \qw & \qw & \qw & \qw\\
	 	\nghost{{q}_{1} :  } & \lstick{{q}_{1} :  } & \qw & \targ & \qw & \qw & \qw & \meter & \qw & \qw & \qw\\
	 	\nghost{{q}_{2} :  } & \lstick{{q}_{2} :  } & \qw & \qw & \targ & \qw & \qw & \qw & \meter & \qw & \qw\\
	 	\nghost{\mathrm{{c} :  }} & \lstick{\mathrm{{c} :  }} & \lstick{/_{_{3}}} \cw & \cw & \cw & \cw & \dstick{_{_{\hspace{0.0em}0}}} \cw \ar @{<=} [-3,0] & \dstick{_{_{\hspace{0.0em}1}}} \cw \ar @{<=} [-2,0] & \dstick{_{_{\hspace{0.0em}2}}} \cw \ar @{<=} [-1,0] & \cw & \cw\\
\\ }}

    Note that you may call {\tt qc.barrier()} before the measurements.
    This will generate a synchronization point (i.e. a dashed line in the circuit diagram),
    though it will not affect the result.

\item Click "Run" to get, record, and explain the result in your report. 

\end{itemize}

\subsection*{Experiment 2: 4-qubit Entanglement }
\begin{itemize} 
\parskip 0in
\parsep 0in
\itemsep 0in

\item Click the cell which contains the code done in Experiment 1. Click the "Edit" button
      and select "Copy Cells". Click "Edit" again and choose "Paste Cells Below".
     
\item In this experiment, we use an alternative way to entangle 4 qubits.
      Modify the code in the cell to build the circuit below: \\
\scalebox{0.9}{
\Qcircuit @C=1.0em @R=0.2em @!R { \\
	 	\nghost{{q}_{0} :  } & \lstick{{q}_{0} :  } & \gate{\mathrm{H}} & \ctrl{1} & \qw & \qw \barrier[0em]{3} & \qw & \meter & \qw & \qw & \qw & \qw & \qw\\
	 	\nghost{{q}_{1} :  } & \lstick{{q}_{1} :  } & \qw & \targ & \ctrl{1} & \qw & \qw & \qw & \meter & \qw & \qw & \qw & \qw\\
	 	\nghost{{q}_{2} :  } & \lstick{{q}_{2} :  } & \qw & \qw & \targ & \ctrl{1} & \qw & \qw & \qw & \meter & \qw & \qw & \qw\\
	 	\nghost{{q}_{3} :  } & \lstick{{q}_{3} :  } & \qw & \qw & \qw & \targ & \qw & \qw & \qw & \qw & \meter & \qw & \qw\\
	 	\nghost{\mathrm{{c} :  }} & \lstick{\mathrm{{c} :  }} & \lstick{/_{_{4}}} \cw & \cw & \cw & \cw & \cw & \dstick{_{_{\hspace{0.0em}0}}} \cw \ar @{<=} [-4,0] & \dstick{_{_{\hspace{0.0em}1}}} \cw \ar @{<=} [-3,0] & \dstick{_{_{\hspace{0.0em}2}}} \cw \ar @{<=} [-2,0] & \dstick{_{_{\hspace{0.0em}3}}} \cw \ar @{<=} [-1,0] & \cw & \cw\\
\\ }}

\item Click "Run" to get, record, and explain the result in your report. 

\end{itemize}




     
\subsection*{Experiment 3: Toffoli gate}
      In the class, we study the the Toffoli gate (also called CCNOT gate or 
      "controlled-controlled-not" gate) which is a universal reversible logic gate. 
      That is, if the first two control bits are both set to 1, it inverts the third bit, 
      otherwise all bits remain the same.

\begin{itemize}
\parskip 0in
\parsep 0in
\itemsep 0in

\item Copy and paste the code cell done in Experiment 1.
      Modify the code in the cell to build the circuit below: \\
\scalebox{0.9}{
\Qcircuit @C=1.0em @R=0.2em @!R { \\
	 	\nghost{{q}_{0} :  } & \lstick{{q}_{0} :  } & \gate{\mathrm{H}} & \ctrl{1} \barrier[0em]{2} & \qw & \qw & \qw & \qw\\
	 	\nghost{{q}_{1} :  } & \lstick{{q}_{1} :  } & \gate{\mathrm{H}} & \ctrl{1} & \qw & \qw & \qw & \qw\\
	 	\nghost{{q}_{2} :  } & \lstick{{q}_{2} :  } & \qw & \targ & \qw & \meter & \qw & \qw\\
	 	\nghost{\mathrm{{c} :  }} & \lstick{\mathrm{{c} :  }} & \lstick{/_{_{1}}} \cw & \cw & \cw & \dstick{_{_{\hspace{0.0em}0}}} \cw \ar @{<=} [-1,0] & \cw & \cw\\
\\ }}

       Note that we only need one classical bit and it has the value measured from $q_2$.
       In Qiskit, the interface of the Toffoli gate is\\
 \mbox{\hskip 0.5 in \tt ccx(c0,c1,t)} \\
       where {\tt c0} and {\tt c1} are the control qubits and t is the target qubit. 

\item Click "Run" to get and record the result. Explain the result.

\end{itemize}

\subsection*{Experiment 4: Implementing AND and NAND gates }

\begin{itemize}
\parskip 0in
\parsep 0in
\itemsep 0in

\item Type the following in a new empty cell: \\
\mbox{\hskip 0.5in \tt \%load \url{~}cis492s/pub/ANDgate.py }\\
      Click "Run" to load the code template into the cell.

\item The AND function is incomplete. Add one line of code to complete it.
      Describe your circuit design in the report. 

\item Click "Run" to execute the code and verify your result.  
      Is your AND quantum circuit reversible?

\item Copy and paste the code cell into a new cell.
      To implement the NAND gate function, simply add a NOT gate (i.e. the quantum X gate) 
      to flip the result of qubit 2 at the end.  Also, replace every occurrence of "AND"
      with "NAND" in the code, including comments.   

\item Click "Run" to execute the code and verify your result

\end{itemize}

\subsection*{Experiment 5: Implementation of the OR gate }

\begin{itemize}
\parskip 0in
\parsep 0in
\itemsep 0in

\item Type the following in a new empty cell:\\
\mbox{\hskip 0.5in \tt \%load \url{~}cis492s/pub/ORgate.py }\\
      Click "Run" to load the code template into the cell.

\item The OR function is incomplete. Add three lines of code to complete it.
      Describe your circuit design in the report. \\
      {\it Hint:} Use two CNOT gates and one Toffoli gate to flip the target qubit in succession. 
                  An alternative way is to invert both inputs, use a NAND gate to get the 
                  output, and then invert inputs back.

\item Click "Run" to execute the code and verify your result.  

\end{itemize}

\subsection*{Experiment 6: Application of CNOT gates }

\begin{itemize}
\parskip 0in
\parsep 0in
\itemsep 0in

\item Type the following in a new empty cell: \\
\mbox{\hskip 0.5in \tt \%load \url{~}cis492s/pub/UNKNOWNfunc.py }\\
      Click "Run" to load the code template into the cell.

\item The function UNKNOWN is incomplete.
      Add the code between two barrier calls based on the circuit below:\\
\scalebox{0.9}{
\Qcircuit @C=1.0em @R=0.8em @!R { \\
	 	\nghost{{q}_{0} :  } & \lstick{{q}_{0} :  } \barrier[0em]{1} & \qw & \ctrl{1} & \targ & \ctrl{1} \barrier[0em]{1} & \qw & \meter & \qw & \qw & \qw\\
	 	\nghost{{q}_{1} :  } & \lstick{{q}_{1} :  } & \qw & \targ & \ctrl{-1} & \targ & \qw & \qw & \meter & \qw & \qw\\
	 	\nghost{\mathrm{{c} :  }} & \lstick{\mathrm{{c} :  }} & \lstick{/_{_{2}}} \cw & \cw & \cw & \cw & \cw & \dstick{_{_{\hspace{0.0em}0}}} \cw \ar @{<=} [-2,0] & \dstick{_{_{\hspace{0.0em}1}}} \cw \ar @{<=} [-1,0] & \cw & \cw\\
\\ }}

\item Click "Run" to execute the code. Figure out what the the unknown function is. \\
      {\it Hint:} The CNOT function is like an XOR gate. Hence, the circuit is like executing
      the following instructions in sequence:\\
      $q_1 = q_0 \oplus  q_1 $ \\
      $q_0 = q_0 \oplus  q_1 $ \\
      $q_1 = q_0 \oplus  q_1 $ 

\end{itemize}

\subsection*{Experiment 7: Application of one CNOT and four H gates }

\begin{itemize}
\parskip 0in
\parsep 0in
\itemsep 0in

\item Type the following in a new empty cell: \\
\mbox{\hskip 0.5in \tt \%load \url{~}cis492s/pub/UNKNOWNfunc.py }\\
      Click "Run" to load the code template into the cell.

\item Add the code between two barrier calls in the UNKNOWN function based on the circuit below:\\
\scalebox{0.9}{
\Qcircuit @C=1.0em @R=0.2em @!R { \\
	 	\nghost{{q}_{0} :  } & \lstick{{q}_{0} :  } \barrier[0em]{1} & \qw & \gate{\mathrm{H}} & \ctrl{1} & \gate{\mathrm{H}} \barrier[0em]{1} & \qw & \meter & \qw & \qw & \qw\\
	 	\nghost{{q}_{1} :  } & \lstick{{q}_{1} :  } & \qw & \gate{\mathrm{H}} & \targ & \gate{\mathrm{H}} & \qw & \qw & \meter & \qw & \qw\\
	 	\nghost{\mathrm{{c} :  }} & \lstick{\mathrm{{c} :  }} & \lstick{/_{_{2}}} \cw & \cw & \cw & \cw & \cw & \dstick{_{_{\hspace{0.0em}0}}} \cw \ar @{<=} [-2,0] & \dstick{_{_{\hspace{0.0em}1}}} \cw \ar @{<=} [-1,0] & \cw & \cw\\
\\ }}

\item Click "Run" to execute the code. Figure out what the the unknown function is. \\
     {\it Hint:} How does the qubit 1 affect the qubit 0?

\end{itemize}

\vskip 0.25in

\noindent Click "File" and choose "Save as". Type "lab3" in the entry box and click "Save"
          to save your work today into the file {\tt lab3.ipynb}. 
	  To turn in your file, use CTRL-ALT-T to open a terminal and type\\
\mbox{\hskip 0.5in \tt ssh grail} \\
and type your password to login to the server grail. Then, type\\
\mbox{\hskip 0.5in \tt cd QC} \\
\mbox{\hskip 0.5in \tt turnin -c cis492s -p lab3 lab3.ipynb} \\
          to electronically submit your file {\tt lab3.ipynb}. \\

\noindent Shutdown the jupyter notebook and logout the workstation. \\

\noindent {\bf Hand in your lab report before your leave.}


\begin{center}
{\Large \bf CIS 492/593 $\;\;\;\;\;$ Hands-On Lab 4 $\;\;$ Quantum Key Distribution $\;\;\;\;\;$ Fall 202x  } \\
\end{center}

\vskip 0.25in
\noindent In this hands-on lab, you will study and practice two Quantum Key Distribution methods: 
\begin{itemize}
\parskip 0in
\parsep 0in
\itemsep 0in
\item the BB84 protocol (see textbook page 53) 
\item the Ekert protocol (see textbook page 87)    
\end{itemize}

\noindent Login to a workstation and open a terminal (either middle click the
terminal icon on the left side bar or press CTRL-ALT-T).
In the terminal window, type
\begin{verbatim}
     cd QC
     source /opt/qiskit_env/bin/activate
     jupyter notebook
\end{verbatim}
Inside the browser, click "New" and choose "Python 3 (ipykernel)".
For each experiment, you need to put the experiment number
as a comment (e.g. {\tt \# Experiment 1} ) in the top of the code.

\subsection*{Experiment 1: Alice's Encoding and Bob's Measurement using the same basis}
\begin{itemize}
\parskip 0in
\parsep 0in
\itemsep 0in

\item In the cell, type \\
\mbox{\hskip 0.5in \tt \%load \url{~}cis492s/pub/BB84Test.py }\\
and click "Run" to load a simple program. In this program, Alice prepares 
and sends a qubit with value 0 in the vertical direction to Bob.  Bob measures
the qubit in the same direction. The procedure is like the diagram below.\\ 
\scalebox{0.9}{
\Qcircuit @C=2.5em @R=0.2em @!R { \\
                                 &  \;\;\;\;\;\; Alice                       &     &  Bob\\
	 	\nghost{{q} :  } &   \lstick{{q} :  } \barrier[0em]{0} & \qw & \meter & \qw & \qw\\
	 	\nghost{\mathrm{{c} :  }} & \lstick{\mathrm{{c} :  }} & \lstick{/_{_{1}}} \cw & \dstick{_{_{\hspace{0.0em}0}}} \cw \ar @{<=} [-1,0] & \cw & \cw\\
\\ }}

Note that the barrier used here is only an indication of work separation 
between Alice and Bob. Click "Run" to get and record the result.

\item Click the cell done in the previous step. Click the "Edit" button
      and select "Copy Cells". Click "Edit" again and choose "Paste Cells Below".
    
       Assume that Alice encodes the qubit in the horizontal direction instead of the vertical direction.
      That is, she applies the H-gate on the qubit. Also assume that Bob measures the qubit in the horizontal direction.
      Modify the code just like the figure below:\\
\scalebox{0.9}{
\Qcircuit @C=1.0em @R=0.2em @!R { \\
                                 &                  & Alice                              &     & Bob               &        &
     &    \\
	 	\nghost{{q} :  } & \lstick{{q} :  } & \gate{\mathrm{H}} \barrier[0em]{0} & \qw & \gate{\mathrm{H}} & \meter & \qw & \qw\\
	 	\nghost{\mathrm{{c} :  }} & \lstick{\mathrm{{c} :  }} & \lstick{/_{_{1}}} \cw & \cw & \cw & \dstick{_{_{\hspace{0.0em}0}}} \cw \ar @{<=} [-1,0] & \cw & \cw\\
\\ }}

Click "Run" to get and record the result. What is your conclusion about the result if Alice and Bob use the same basis?
\end{itemize}

\subsection*{Experiment 2: BB84 emulation (without Eve's eavesdropping)}
\begin{itemize} 
\parskip 0in
\parsep 0in
\itemsep 0in
\item In a new and empty cell, type \\
\mbox{\hskip 0.5in \tt \%load \url{~}cis492s/pub/BB84.py }\\
and click "Run" to load the Qiskit program which emulates the BB84 protocol into the cell.
Study the code.
   
\item Click "Run" to get the outputs. Record the outputs (i.e. "Alice bits", "Alice bases". "Bob bases", and
      "Bob results") in your report. Mark the bit positions where Alice and Bob use the same bases
      and then extract the corresponding bits from "Alice bits" and "Bob results" into two bit sequences. 
      
\item Alice and Bob will select half of the same-basis bits and compare them over an unencrypted line. For simplicity,
      we just choose the first half of the sequence as the sample bits. Add the code at the end of the cell:\\
\begin{verbatim}
    sample_size = len(alice_bits_samebasis) // 2
    alice_sample = alice_bits_samebasis[0:sample_size]
    bob_sample = bob_bits_samebasis[0:sample_size]

    print("sample size:", sample_size)
    print("alice_sample = ", alice_sample)
    print("  bob_sample = " , bob_sample)

    print(bob_sample == alice_sample)
\end{verbatim}
      Click "Run" to execute the code. What is the size of the sampling bits? 
      Note that if the result is True, Alice and Bob can use the other half of the bit sequence as the key.           
     
\item Modify the code to change the value of the variable {\tt n} from 4 to 1000.  
      Click "Run" to execute the code to make sure that the program works for a larger size of bit sequence.
\end{itemize}

\subsection*{Experiment 3: Alice's Encoding and Bob's Measurement using the same basis (with Eve's eavesdropping) }
\begin{itemize} 
\parskip 0in
\parsep 0in
\itemsep 0in

\item Click the cell done in Experiment 1 which contains the code based on the horizontal direction. 
      Click the "Edit" button and select "Copy Cells". Click "Edit" again and choose "Paste Cells Below".

\item Assume that Eve intercepts the qubit on the way from Alice to Bob and she luckily uses the same basis
      (i.e. the horizontal direction) to measure the qubit. She then prepares and sends Bob a qubit
      based on the horizontal basis. To emulate Eve's interception, add four lines of code 
      (including a barrier call at the end) before Bob's action, like the diagram below. \\
\scalebox{0.9}{
\Qcircuit @C=1.0em @R=0.2em @!R { \\
                                &                  &      Alice                    &     &              &  Eve      & 
                            &     & Bob               &        &     &   \\
	 	\nghost{{q} :  } & \lstick{{q} :  } & \gate{\mathrm{H}} \barrier[0em]{0} & \qw & \gate{\mathrm{H}} & \meter & \gate{\mathrm{H}} \barrier[0em]{0} & \qw & \gate{\mathrm{H}} & \meter & \qw & \qw\\
	 	\nghost{\mathrm{{c} :  }} & \lstick{\mathrm{{c} :  }} & \lstick{/_{_{1}}} \cw & \cw & \cw & \dstick{_{_{\hspace{0.0em}0}}} \cw \ar @{<=} [-1,0] & \cw & \cw & \cw & \dstick{_{_{\hspace{0.0em}0}}} \cw \ar @{<=} [-1,0] & \cw & \cw\\
\\ }}

      Click "Run" to get the result. Is there any difference as compared with the result in Experiment 1 (the horizontal direction)? Can Eve's eavesdropping be detected if Alice, Eve, and Bob use the same basis?

\item Now assume that Eve guesses the wrong basis to measure the qubit.
      Click the cell done above. 
      click the "Edit" button and select "Copy Cells". Click "Edit" again and choose "Paste Cells Below".
      Modify the code so Eve will use the vertical direction to measure and encode the qubit before
      sending to Bob, like the figure below:\\ 
\scalebox{0.9}{
\Qcircuit @C=1.0em @R=0.2em @!R { \\
                                 &                  &  Alice                             &     & Eve                     &
  & Bob               &        &     & \\ 
	 	\nghost{{q} :  } & \lstick{{q} :  } & \gate{\mathrm{H}} \barrier[0em]{0} & \qw & \meter \barrier[0em]{0} & \qw & \gate{\mathrm{H}} & \meter & \qw & \qw\\
	 	\nghost{\mathrm{{c} :  }} & \lstick{\mathrm{{c} :  }} & \lstick{/_{_{1}}} \cw & \cw & \dstick{_{_{\hspace{0.0em}0}}} \cw \ar @{<=} [-1,0] & \cw & \cw & \dstick{_{_{\hspace{0.0em}0}}} \cw \ar @{<=} [-1,0] & \cw & \cw\\
\\ }}

       Click "Run" to get and explain the result.

\end{itemize}

\subsection*{Experiment 4: BB84 emulation (with Eve's eavesdropping)}
\begin{itemize} 
\parskip 0in
\parsep 0in
\itemsep 0in

\item Click the cell which contains the code done in Experiment 2. Click the "Edit" button
      and select "Copy Cells". Click "Edit" again and choose "Paste Cells Below".
     
\item In this experiment, you are asked to add some code between Step 2 and Step 3 to emulate
      the actions of Eve. That is, she firstly generates a random choice of basis for each bit. Next, 
      she measures each qubit based on the generated random choice of basis. Finally, she encodes
      the message again based on the generated random choice of basis
      and then sends the message to Bob.  Click "Run" to execute your code and get the result.
        
\item In addition to getting the result False at the end of the previous step, we are also interested
      in getting the percentage of the sample bits in which Alice and Bob disagree. Add a few lines of code
      at the end to get the percentage and verify it with the textbook (page 55, lines 23--26).  \\
      {\it Hint:} Use a counter and a loop to compare each Alice's sample bit with Bob's sample bit.
                   Increment the counter if they disagree. Output the ratio of the counter over the size
                   of the sample bits. 
\end{itemize}

\subsection*{Experiment 5: Ekert -- Measuring two entangled bits with different bases  }

Based on the textbook description about the Ekert91 protocol, Alice 
and Bob receive a stream of qubits. For each entangled pair of qubits,    
Alice receives one and randomly chooses a direction to measure her qubit,  
while Bob gets the other and also randomly selects a direction to measure it.
There are three directions Alice and Bob can use: $0^{\circ}$,$120^{\circ}$,
or $240^{\circ}$. If they choose the same direction, they will get the same 
result (i.e. both get 0 or both get 1). However,  if they use different directions,
they may not get the same result. In this experiment, we are interested
in getting the probability that they will AGREE the result,
and the probability that they will DISAGREE the result, under the
condition that thy will use different directions to measure.  

\begin{itemize} 
\parskip 0in
\parsep 0in
\itemsep 0in

\item Note that these three directions $0^{\circ}$,$120^{\circ}$,
and $240^{\circ}$ can be implemented by a Qiskit U-gate via
passing the first parameter with a value 0, $2*pi/3$, $4*pi/3$, respectively. 
The fourth parameter is the index of the qubit where the the U-gate will be applied.
 
In a new cell, type \\
\mbox{\hskip 0.5in \tt \%load \url{~}cis492s/pub/EkertTest.py }\\
and click "Run" to load a simple program. In this program, Alice always
measures her qubit in the direction $0^{\circ}$, while Bob always uses the 
direction $120^{\circ}$ to measure. Below is the circuit diagram: \\

\scalebox{0.9}{
\Qcircuit @C=1.0em @R=0.2em @!R { \\
	 	\nghost{{q}_{0} :  } & \lstick{{q}_{0} :  } & \gate{\mathrm{H}} & \ctrl{1} & \gate{\mathrm{U}\,(\mathrm{0,0,0})} & \meter & \qw & \qw & \qw\\
	 	\nghost{{q}_{1} :  } & \lstick{{q}_{1} :  } & \qw & \targ & \gate{\mathrm{U}\,(\mathrm{\frac{2\pi}{3},0,0})} & \qw & \meter & \qw & \qw\\
	 	\nghost{\mathrm{{c} :  }} & \lstick{\mathrm{{c} :  }} & \lstick{/_{_{2}}} \cw & \cw & \cw & \dstick{_{_{\hspace{0.0em}0}}} \cw \ar @{<=} [-2,0] & \dstick{_{_{\hspace{0.0em}1}}} \cw \ar @{<=} [-1,0] & \cw & \cw\\
                                     &                      &                   &          &                                    & Alice       & Bob    &     &        \\
\\ }}

\item Study the program.  
       Click "Run" to execute the code. 
       Calculate the probability they AGREE (i.e. combining the percentages of
       '00' and '11') and the probability they DISAGREE (i.e. adding the percentages
       of '10' and '01') . 
       
\item Repeat previous step with Alice using $0^{\circ}$ and Bob using $240^{\circ}$.
      Calculate the two probabilities.\\ 
       Repeat previous step with Alice using $120^{\circ}$ and Bob using $240^{\circ}$.
      Calculate the two probabilities.\\ 
       Repeat previous step with Alice using $120^{\circ}$ and Bob using $0^{\circ}$.
      Calculate the two probabilities.\\ 
       Repeat previous step with Alice using $240^{\circ}$ and Bob using $0^{\circ}$.
      Calculate the two probabilities.\\ 
       Repeat previous step with Alice using $240^{\circ}$ and Bob using $120^{\circ}$.
      Calculate the two probabilities.\\ 

\end{itemize}

\subsection*{Experiment 6: Ekert emulation (without Eve's eavesdropping) }
\begin{itemize} 
\parskip 0in
\parsep 0in
\itemsep 0in

\item In a new cell, type \\
\mbox{\hskip 0.5in \tt \%load \url{~}cis492s/pub/Ekert.py }\\
and click "Run" to load the program which emulates the Ekert protocol. 

\item Study the program. Add a line of code at the end of the program to print  
      the probability that Alice and Bob AGREE the result when they use different bases.

\item Click "Run" to execute the code. Verify your result with the equation on Textbook page 81,
      line 5.

\end{itemize}

\subsection*{Experiment 7: Ekert emulation (with Eve's eavesdropping) }
\begin{itemize} 
\parskip 0in
\parsep 0in
\itemsep 0in

\item Click the cell which contains the code done in Experiment 6. Click the "Edit" button
      and select "Copy Cells". Click "Edit" again and choose "Paste Cells Below".

\item To emulate Eve's eavesdropping, add one line of code \\
      \mbox{\hskip 0.5in \tt message[q].measure(1,1)} \\
      before Alice measures her entangled qubit in the function {\tt measure\_message()}.

\item Click "Run" to execute the code. Compare the probability that Alice and Bob AGREE 
      when they use different bases with the corresponding probability  in Experiment 6.
      Verify your result with the description on Textbook page 87, lines 14--16. 

\item Is the probability that Alice and Bob AGREE the result when they use the same basis
      still 1 as in the Experiment 6? 
     
\end{itemize}

\vskip 0.25in

\noindent Click "File" and choose "Save as". Type "lab4" in the entry box and click "Save"
          to save your work today into the file {\tt lab4.ipynb}. 

	   To turn in your file, use CTRL-ALT-T to open a terminal and type\\
\mbox{\hskip 0.5in \tt ssh grail} \\
and type your password to login to the server grail. Then, type\\
\mbox{\hskip 0.5in \tt cd QC} \\
\mbox{\hskip 0.5in \tt turnin -c cis492s -p lab4 lab4.ipynb} \\
          to electronically submit your file {\tt lab4.ipynb}. \\

           Shutdown the jupyter notebook and logout the workstation. \\

\noindent {\bf Hand in your lab report before your leave.}


\begin{center}
{\Large \bf CIS 492/593 $\;\;\;\;\;$ Lab 5 $\;\;$ Deutsch and Deutsch-Jozsa Algorithms $\;\;\;\;\;$ Fall 202x  } \\
\end{center}

\vskip 0.25in
\noindent The purpose of this hands-on lab is to familiarize yourself with:
\begin{itemize}
\parskip 0in
\parsep 0in
\itemsep 0in
\item the Deutsch algorithm 
\item the Deutsch-Jozsa algorithm
\end{itemize}

\noindent Login to a workstation and open a terminal (either middle click the
terminal icon on the left side bar or press CTRL-ALT-T).
In the terminal window, type
\begin{verbatim}
     cd QC
     source /opt/qiskit_env/bin/activate
     jupyter notebook
\end{verbatim}
Inside the browser, click "New" and choose "Python 3 (ipykernel)".
For each experiment, you need to put the experiment number
as a comment (e.g. {\tt \# Experiment 1} ) in the top of the code.

\subsection*{Experiment 1: Deutsch -- the Constant 0 function }
\begin{itemize}
\parskip 0in
\parsep 0in
\itemsep 0in

\item In the cell, type \\
\mbox{\hskip 0.5in \tt \%load \url{~}cis492s/pub/DeutschConst0.py }\\
and click "Run" to load a program in which the black box function $F$ is a 
constant 0 function (i.e. f(0) = 0, and f(1) = 0). Instead of using the
default name 'q' for the qubits, we call the function {\tt QuantumRegister()}
twice to label the top qubit as $'x'$ and the output qubit as $'y'$, respectively.
The corresponding circuit diagram is like below:\\
\scalebox{1.0}{
\Qcircuit @C=1.0em @R=0.2em @!R { 
                                 &                  &                      &     &                                    &  & F \\ 
	 	\nghost{{x} :  } & \lstick{{x} :  } & \qw \barrier[0em]{1} & \qw & \gate{\mathrm{H}} \barrier[0em]{1} & \qw  & \qw \barrier[0em]{1} & \qw & \gate{\mathrm{H}} & \meter & \qw & \qw\\
	 	\nghost{{y} :  } & \lstick{{y} :  } & \gate{\mathrm{X}} & \qw & \gate{\mathrm{H}} & \qw & \qw & \qw & \qw & \qw & \qw & \qw\\
	 	\nghost{\mathrm{{c} :  }} & \lstick{\mathrm{{c} :  }} & \lstick{/_{_{1}}} \cw & \cw & \cw & \cw & \cw & \cw & \cw & \dstick{_{_{\hspace{0.0em}0}}} \cw \ar @{<=} [-2,0] & \cw & \cw\\
\\ }}

Note that because $f(x) = 0$, the output of the bottom qubit becomes $y \oplus f(x) = y \oplus 0 = y$.
Hence, there are only two straight wires in the black box $F$ and there is no connection between 
the qubit $x$ and the qubit $y$. 

\item Click "Run" to execute the code. What is the output of the top qubit $x$?
      Refer the textbook page 149 about how to determine the the black box
      whether it is a constant or a balanced function. Explain the result.  

\end{itemize}
     
\subsection*{Experiment 2: Deutsch -- the Constant 1 function }
\begin{itemize}
\parskip 0in
\parsep 0in
\itemsep 0in

\item Click the cell which contains the code done in Experiment 1. Click the "Edit" button
      and select "Copy Cells". Click "Edit" again and choose "Paste Cells Below".

\item Assume that the black box $F$ is a constant 1 function (i.e. f(0) = 1, and f(1) = 1).
      Hence, $y \oplus f(x) = y \oplus 1 = \overline{y}$. That is, you need to modify the code 
      (i.e. putting a NOT gate on the qubit $y$ and changing the corresponding comment), like below:\\
\scalebox{1.0}{
\Qcircuit @C=1.0em @R=0.2em @!R { 
                                 &                  &                      &     &                                    &  & F \\
	 	\nghost{{x} :  } & \lstick{{x} :  } & \qw \barrier[0em]{1} & \qw & \gate{\mathrm{H}} \barrier[0em]{1} & \qw & \qw \barrier[0em]{1} & \qw & \gate{\mathrm{H}} & \meter & \qw & \qw\\
	 	\nghost{{y} :  } & \lstick{{y} :  } & \gate{\mathrm{X}} & \qw & \gate{\mathrm{H}} &  \qw & \gate{\mathrm{X}} & \qw & \qw & \qw & \qw & \qw\\
	 	\nghost{\mathrm{{c} :  }} & \lstick{\mathrm{{c} :  }} & \lstick{/_{_{1}}} \cw & \cw & \cw & \cw & \cw & \cw & \cw & \dstick{_{_{\hspace{0.0em}0}}} \cw \ar @{<=} [-2,0] & \cw & \cw\\
\\ }}
 
\item Click "Run" to execute the code. What is the output of the top qubit $x$?
	Explain the result.

\end{itemize}

\subsection*{Experiment 3: Deutsch -- the Identity function (balanced) }
\begin{itemize}
\parskip 0in
\parsep 0in
\itemsep 0in

\item Click the cell which contains the code done in Experiment 1. Click the "Edit" button
      and select "Copy Cells". Click "Edit" again and choose "Paste Cells Below".

\item Assume that the black box $F$ is the identity  function (i.e. f(0) = 0, and f(1) = 1).
      To realize $y \oplus f(x)$, you need to modify the code (i.e. adding a CNOT gate with the qubit 
      $x$ as the control while the qubit $y$ as the target, and changing the corresponding comment), 
      just like below:\\
\scalebox{1.0}{
\Qcircuit @C=1.0em @R=0.2em @!R { 
                                 &                  &                      &     &                                    &  & F \\
	 	\nghost{{x} :  } & \lstick{{x} :  } & \qw \barrier[0em]{1} & \qw & \gate{\mathrm{H}} \barrier[0em]{1} & \qw & \ctrl{1} \barrier[0em]{1} & \qw & \gate{\mathrm{H}} & \meter & \qw & \qw\\
	 	\nghost{{y} :  } & \lstick{{y} :  } & \gate{\mathrm{X}} & \qw & \gate{\mathrm{H}} & \qw & \targ & \qw & \qw & \qw & \qw & \qw\\
	 	\nghost{\mathrm{{c} :  }} & \lstick{\mathrm{{c} :  }} & \lstick{/_{_{1}}} \cw & \cw & \cw & \cw & \cw & \cw & \cw & \dstick{_{_{\hspace{0.0em}0}}} \cw \ar @{<=} [-2,0] & \cw & \cw\\
\\ }}

\item Click "Run" to execute the code. What is the output of the top qubit $x$?
	  Explain the result.

\end{itemize}

\subsection*{Experiment 4: Deutsch -- the Inversion function (balanced) }
\begin{itemize}
\parskip 0in
\parsep 0in
\itemsep 0in

\item Click the cell which contains the code done in Experiment 3. Click the "Edit" button
      and select "Copy Cells". Click "Edit" again and choose "Paste Cells Below".

\item Assume that the black box $F$ is the inversion function (i.e. f(0) = 1, and f(1) = 0).
      You need to modify the code (i.e. adding a NOT gate to the qubit $x$, and changing the corresponding comment),
      like the figure below:\\
\scalebox{1.0}{
\Qcircuit @C=1.0em @R=0.2em @!R { 
                                 &                  &                      &     &                                    &  & F \\
	 	\nghost{{x} :  } & \lstick{{x} :  } & \qw \barrier[0em]{1} & \qw & \gate{\mathrm{H}} \barrier[0em]{1} & \qw & \gate{\mathrm{X}} & \ctrl{1} \barrier[0em]{1} & \qw & \gate{\mathrm{H}} & \meter & \qw & \qw\\
	 	\nghost{{y} :  } & \lstick{{y} :  } & \gate{\mathrm{X}} & \qw & \gate{\mathrm{H}} & \qw & \qw & \targ & \qw & \qw & \qw & \qw & \qw\\
	 	\nghost{\mathrm{{c} :  }} & \lstick{\mathrm{{c} :  }} & \lstick{/_{_{1}}} \cw & \cw & \cw & \cw & \cw & \cw & \cw & \cw & \dstick{_{_{\hspace{0.0em}0}}} \cw \ar @{<=} [-2,0] & \cw & \cw\\
\\ }}

\item Click "Run" to execute the code. What is the output of the top qubit $x$?
	Explain the result.

\end{itemize}

\subsection*{Experiment 5: Deutsch-Jozsa -- the Constant function}
\begin{itemize}
\parskip 0in
\parsep 0in
\itemsep 0in

\item Using a new and empty cell, type \\
\mbox{\hskip 0.5in \tt \%load \url{~}cis492s/pub/DJ-Const0.py }\\
and click "Run" to load a program in which the black box function $F$ is a 
constant 0 function with three parameters $x_0$, $x_1$, and $x_2$. The corresponding 
circuit diagram is like below:\\
\scalebox{1.0}{
\Qcircuit @C=1.0em @R=0.2em @!R { 
                                     &                  &                      &     &                                    &  & F \\
	 	\nghost{{x}_{0} :  } & \lstick{{x}_{0} :  } & \qw \barrier[0em]{3} & \qw & \gate{\mathrm{H}} \barrier[0em]{3} & \qw & \qw \barrier[0em]{3} & \qw & \gate{\mathrm{H}} & \meter & \qw & \qw & \qw & \qw\\
	 	\nghost{{x}_{1} :  } & \lstick{{x}_{1} :  } & \qw & \qw & \gate{\mathrm{H}} & \qw & \qw & \qw & \gate{\mathrm{H}} & \qw & \meter & \qw & \qw & \qw\\
	 	\nghost{{x}_{2} :  } & \lstick{{x}_{2} :  } & \qw & \qw & \gate{\mathrm{H}} & \qw & \qw & \qw & \gate{\mathrm{H}} & \qw & \qw & \meter & \qw & \qw\\
	 	\nghost{{y} :  } & \lstick{{y} :  } & \gate{\mathrm{X}} & \qw & \gate{\mathrm{H}} & \qw & \qw & \qw & \qw & \qw & \qw & \qw & \qw & \qw\\
	 	\nghost{\mathrm{{c} :  }} & \lstick{\mathrm{{c} :  }} & \lstick{/_{_{3}}} \cw & \cw & \cw & \cw & \cw & \cw & \cw & \dstick{_{_{\hspace{0.0em}0}}} \cw \ar @{<=} [-4,0] & \dstick{_{_{\hspace{0.0em}1}}} \cw \ar @{<=} [-3,0] & \dstick{_{_{\hspace{0.0em}2}}} \cw \ar @{<=} [-2,0] & \cw & \cw\\
\\ }}

\item Click "Run" to execute the code. What is the output of the top 3-qubit $x$?
      Do you see any output string other than '000' during the 1000 times of executions?
      (Note that in fact we only need one execution to tell the black box is a constant or a balanced function).

\item Click the cell which contains the code done above. Click the "Edit" button
      and select "Copy Cells". Click "Edit" again and choose "Paste Cells Below".

\item Modify the code to make the black box $F$ a constant 1 function.
	{\it Hint: see Experiment 2}.

\item Click "Run" to execute the code. What is the output of the top 3-qubit $x$?
      Do you see any output string other than '000' during the 1000 times of executions?

\end{itemize}

\subsection*{Experiment 6: Deutsch-Jozsa -- the balanced function $f(x_0,x_1,x_2) = x_0 \oplus x_1 \oplus x_2$ }
\begin{itemize}
\parskip 0in
\parsep 0in
\itemsep 0in

\item Construct the truth table for $ f(x_0,x_1,x_2) = x_0 \oplus x_1 \oplus x_2$ to verify it is a balanced function.  

\item Using a new and empty cell, type \\
\mbox{\hskip 0.5in \tt \%load \url{~}cis492s/pub/DJ-Const0.py }\\
and click "Run" to load a template program. 

\item Modify the code to implement $x_0 \oplus x_1 \oplus x_2$, like the circuit below:\\   
\scalebox{1.0}{
\Qcircuit @C=1.0em @R=0.2em @!R { 
                                     &                  &                      &     &                                    &  &    &  F \\
	 	\nghost{{x}_{0} :  } & \lstick{{x}_{0} :  } & \qw \barrier[0em]{3} & \qw & \gate{\mathrm{H}} \barrier[0em]{3} & \qw & \ctrl{3} & \qw & \qw \barrier[0em]{3} & \qw & \gate{\mathrm{H}} & \meter & \qw & \qw & \qw & \qw\\
	 	\nghost{{x}_{1} :  } & \lstick{{x}_{1} :  } & \qw & \qw & \gate{\mathrm{H}} & \qw & \qw & \ctrl{2} & \qw & \qw & \gate{\mathrm{H}} & \qw & \meter & \qw & \qw & \qw\\
	 	\nghost{{x}_{2} :  } & \lstick{{x}_{2} :  } & \qw & \qw & \gate{\mathrm{H}} & \qw & \qw & \qw & \ctrl{1} & \qw & \gate{\mathrm{H}} & \qw & \qw & \meter & \qw & \qw\\
	 	\nghost{{y} :  } & \lstick{{y} :  } & \gate{\mathrm{X}} & \qw & \gate{\mathrm{H}} & \qw & \targ & \targ & \targ & \qw & \qw & \qw & \qw & \qw & \qw & \qw\\
	 	\nghost{\mathrm{{c} :  }} & \lstick{\mathrm{{c} :  }} & \lstick{/_{_{3}}} \cw & \cw & \cw & \cw & \cw & \cw & \cw & \cw & \cw & \dstick{_{_{\hspace{0.0em}0}}} \cw \ar @{<=} [-4,0] & \dstick{_{_{\hspace{0.0em}1}}} \cw \ar @{<=} [-3,0] & \dstick{_{_{\hspace{0.0em}2}}} \cw \ar @{<=} [-2,0] & \cw & \cw\\
\\ }}

\item Click "Run" to execute the code. What is the output of the top 3-qubit $x$?
      Do you see '000' appeared during the 1000 times of executions?
      (Note that in fact we only need one execution to tell the black box is a constant or a balanced function).

\end{itemize}

\subsection*{Experiment 7: Deutsch-Jozsa -- the balanced function $f(x_0,x_1,x_2) = x_0  x_1 \oplus x_2$ }
\begin{itemize}
\parskip 0in
\parsep 0in
\itemsep 0in

\item Construct the truth table for $ f(x_0,x_1,x_2) = x_0 x_1 \oplus x_2$ to verify it is a balanced function.  

\item Modify the code to implement the balanced function $ f(x_0,x_1,x_2) = x_0 x_1 \oplus x_2$.   

\item Click "Run" to execute the code. What is the output of the top 3-qubit $x$?
      Do you see '000' appeared during the 1000 times of executions?
      (Note that in fact we only need one execution to tell the black box is a constant or a balanced function).

\end{itemize}

\noindent Click "File" and choose "Save as". Type "lab5" in the entry box and click "Save"
          to save your work today into the file {\tt lab5.ipynb}. \\
\vskip 0.02in
\noindent Open a terminal and type 
\begin{verbatim}
     ssh grail 
\end{verbatim}
          to login to the server grail. Then, type
\begin{verbatim}
     cd QC
     turnin -c cis492s -p lab5 lab5.ipynb
\end{verbatim}
          to electronically turn in your file {\tt lab5.ipynb}. \\ 
\vskip 0.02in
\noindent  Shutdown the jupyter notebook and logout the workstation. \\
\noindent {\bf Hand in your lab report before your leave.}

\newpage 

\subsection*{Questions:}

Below are some questions which will be included in the
Homework 3 (Due: Dec. 5). You may start working on these questions now.

\begin{itemize}

\item If the black box is a constant function as in the Experiments 1, 2, and 5, 
\begin{enumerate}[a)] 
\parskip 0in
\parsep 0in
\itemsep 0in

\item Is there any wire connection between the qubit(s) $x$ and qubit $y$ ?   

\item What is the initial value for the $x$ qubit(s)?   

\item Explain why the measured value for each $x$ qubit(s) is 0.\\
      ($Hint:$ How do the gates applied on each $x$ qubit(s) work? Review the Experiment 2 in Lab2 )
         
\end{enumerate}

\vskip 0.1in

\item If the black box is a balanced function as in the Experiments 3, 4, 6 and 7, 
\begin{enumerate}[a),start=4]
\parskip 0in
\parsep 0in
\itemsep 0in

\item Is there any connection via the CNOT gate between the qubit(s) $x$ and qubit $y$ ?   

\item If the answer above is Yes, what happen if we put an H-gate for the corresponding $x$ qubit
      and the qubit $y$? ($Hint:$ Review the Experiment 7 in Lab3 )

\item What is the initial value for the $y$ qubit before applying the H-gate?

\item Explain why the measured value for each $x$ qubit(s) is not 0 (as in the Experiments 3 and 4) 
      or not '000' (as in the Experiments 6 and 7).

\end{enumerate}

\end{itemize}


\begin{center}
{\Large \bf CIS 492/593 $\;\;\;\;\;$ Lab 6 $\;\;$ Simon's Algorithm $\;\;\;\;\;$ Fall 202x  } \\
\end{center}

\vskip 0.25in
\noindent In this hands-on lab, you will study and practice: 
\begin{itemize}
\parskip 0in
\parsep 0in
\itemsep 0in
\item the Simon's algorithm 
\item the Oracle construction method for the Simon's algorithm
\end{itemize}

\noindent Login to a workstation and open a terminal (either middle click the
terminal icon on the left side bar or press CTRL-ALT-T).
In the terminal window, type
\begin{verbatim}
     cd QC
     source /opt/qiskit_env/bin/activate
     jupyter notebook
\end{verbatim}
Inside the browser, click "New" and choose "Python 3 (ipykernel)".
For each experiment, you need to put the experiment number
as a comment (e.g. {\tt \# Experiment 1} ) in the top of the code.

\subsection*{Experiment 1: Simon's Algorithm with a 2-bit secret string }
\begin{itemize}
\parskip 0in
\parsep 0in
\itemsep 0in

\item In the empty cell, type \\
\mbox{\hskip 0.5in \tt \%load \url{~}cis492s/pub/Simon.py }\\
and click "Run" to load a program which will give us a 2-bit string
whose dot product with the secret string $s$ is 0. 
 
\item Study the code. Note that the {\tt Oracle()} function builds and returns
the circuit for the oracle function $F$. We also use the method {\tt compose()}
in Qiskit to combine two quantum circuits together. That is, we add the 
oracle circuit {\tt Fc} to the circuit {\tt qc}.

\item Click the cell containing the code and then click "Run" to execute it. 
      Record the output string of the top 2-qubit $x$.

\item Repeat above step to query the oracle until you get two different strings.
      How many times have you queried the oracle?

\item Ignore the output string '00' which wouldn't give us any information.
      Use the other output string to calculate its dot product with the secret 
		string $s$ (i.e. $s_0 s_1$) and the result should be 0. 
		Hence, you can get one equation regarding $s_0$
      and/or $s_1$.  Based on the equation, figure out the bit pattern of the 
      secret string $s$.

\end{itemize}
     
\subsection*{Experiment 2: Simon's Algorithm with another 2-bit secret string }
\begin{itemize}
\parskip 0in
\parsep 0in
\itemsep 0in

\item Click the cell which contains the code done in Experiment 1. 
      Click the "Edit" button
      and select "Copy Cells". Click "Edit" again and choose "Paste Cells Below".

\item Modify the code in the {\tt Oracle()} function to construct the circuit of $F$
       (i.e. between two barrier lines) , like below:\\
\scalebox{1.0}{
\Qcircuit @C=1.0em @R=0.2em @!R { 
                                     &                      &                                    &     &          &  F \\    
	 	\nghost{{x}_{0} :  } & \lstick{{x}_{0} :  } & \gate{\mathrm{H}} \barrier[0em]{3} & \qw & \ctrl{2} & \qw & \ctrl{2} & \ctrl{3} \barrier[0em]{3} & \qw & \gate{\mathrm{H}} & \meter & \qw & \qw & \qw\\
	 	\nghost{{x}_{1} :  } & \lstick{{x}_{1} :  } & \gate{\mathrm{H}} & \qw & \qw & \ctrl{2} & \qw & \qw & \qw & \gate{\mathrm{H}} & \qw & \meter & \qw & \qw\\
	 	\nghost{{y}_{0} :  } & \lstick{{y}_{0} :  } & \qw & \qw & \targ & \qw & \targ & \qw & \qw & \qw & \qw & \qw & \qw & \qw\\
	 	\nghost{{y}_{1} :  } & \lstick{{y}_{1} :  } & \qw & \qw & \qw & \targ & \qw & \targ & \qw & \qw & \qw & \qw & \qw & \qw\\
	 	\nghost{\mathrm{{c} :  }} & \lstick{\mathrm{{c} :  }} & \lstick{/_{_{2}}} \cw & \cw & \cw & \cw & \cw & \cw & \cw & \cw & \dstick{_{_{\hspace{0.0em}0}}} \cw \ar @{<=} [-4,0] & \dstick{_{_{\hspace{0.0em}1}}} \cw \ar @{<=} [-3,0] & \cw & \cw\\
\\ }}

\item Click "Run" to execute the code. 
      Record the output string of the top 2-qubit $x$.

\item Repeat above step to query the oracle until you get two different strings.
      How many times have you queried the oracle?

\item Use the output string which is not '00. We know that this string's dot product with the secret 
      string $s$ (i.e. $s_0 s_1$ should be 0. Hence, you can obtain one equation related
      to $s_0$ and/or $s_1$. 
      Based on the equation, derive the bit pattern of the 
      secret string $s$. Note that the secret string cannot be '00'.

\end{itemize}
     
\subsection*{Experiment 3: Simon's Algorithm with a 3-bit secret string }
\begin{itemize}
\parskip 0in
\parsep 0in
\itemsep 0in

\item Click the cell which contains the code done in Experiment 1. Click the "Edit" button
      and select "Copy Cells". Click "Edit" again and choose "Paste Cells Below".

\item Change the line of the code 'n = 2' to be 'n = 3'.

\item Modify the code in the {\tt Oracle()} function to construct the circuit of $F$
       (i.e. between two barrier lines) , like below:\\
\scalebox{1.0}{
\Qcircuit @C=1.0em @R=0.2em @!R { 
                                     &                      &                                    &     &          &     &   F \\
	 	\nghost{{x}_{0} :  } & \lstick{{x}_{0} :  } & \gate{\mathrm{H}} \barrier[0em]{5} & \qw & \ctrl{3} & \qw & \qw & \qw & \qw \barrier[0em]{5} & \qw & \gate{\mathrm{H}} & \meter & \qw & \qw & \qw & \qw\\
	 	\nghost{{x}_{1} :  } & \lstick{{x}_{1} :  } & \gate{\mathrm{H}} & \qw & \qw & \ctrl{3} & \qw & \ctrl{3} & \ctrl{4} & \qw & \gate{\mathrm{H}} & \qw & \meter & \qw & \qw & \qw\\
	 	\nghost{{x}_{2} :  } & \lstick{{x}_{2} :  } & \gate{\mathrm{H}} & \qw & \qw & \qw & \ctrl{3} & \qw & \qw & \qw & \gate{\mathrm{H}} & \qw & \qw & \meter & \qw & \qw\\
	 	\nghost{{y}_{0} :  } & \lstick{{y}_{0} :  } & \qw & \qw & \targ & \qw & \qw & \qw & \qw & \qw & \qw & \qw & \qw & \qw & \qw & \qw\\
	 	\nghost{{y}_{1} :  } & \lstick{{y}_{1} :  } & \qw & \qw & \qw & \targ & \qw & \targ & \qw & \qw & \qw & \qw & \qw & \qw & \qw & \qw\\
	 	\nghost{{y}_{2} :  } & \lstick{{y}_{2} :  } & \qw & \qw & \qw & \qw & \targ & \qw & \targ & \qw & \qw & \qw & \qw & \qw & \qw & \qw\\
	 	\nghost{\mathrm{{c} :  }} & \lstick{\mathrm{{c} :  }} & \lstick{/_{_{3}}} \cw & \cw & \cw & \cw & \cw & \cw & \cw & \cw & \cw & \dstick{_{_{\hspace{0.0em}0}}} \cw \ar @{<=} [-6,0] & \dstick{_{_{\hspace{0.0em}1}}} \cw \ar @{<=} [-5,0] & \dstick{_{_{\hspace{0.0em}2}}} \cw \ar @{<=} [-4,0] & \cw & \cw\\
\\ }}

\item Click "Run" to execute the code. 
      Record the output string of the top 3-qubit $x$.

\item Repeat above step to query the oracle until you get three different strings.
      How many times have you queried the oracle?

\item Use the output strings except the string '000'. We know that each of these strings has a dot product 
      of 0 with the secret string $s$ (i.e. $s_0 s_1 s_2$). Hence, you can obtain two (or three) 
      equations related to $s_0$, $s_1$, and/or $s_2$. 
      Based on the equations, figure out the bit pattern of the 
      secret string $s$. Note that the secret string cannot be '000'.

\end{itemize}

\subsection*{Experiment 4: The Oracle Construction method for Simon's Algorithm }

In this experiment, you will learn how to construct the Oracle circuit for a 
given string $s$. There are two steps in the construction method:

\begin{description}
\item[Step I:] Apply the CNOT gates from qubits of the first register (i.e. $x$)
           to qubits of the second register (i.e. $y$).  
           So the content of each qubit in the first register, which is the classical information encoded as either a $\ket{0}$ or a $\ket{0}$, 
           will be copied to the corresponding qubit in the second register.
           In fact, this can be simply implemented by using one line of code:\\
           \mbox{\hskip 1in fc.cx(x,y)} \\
           instead of using a loop : \\
           \mbox{\hskip 1in for i in range n:}\\
           \mbox{\hskip 1.25in  fc.cx(x[i],y[i])}
          
\item[Step II:] Because the bits in the string $s$ cannot be all 0, find the least
            index $k$ such that $s_k = 1$. Next, apply the CNOT gates from the qubit $x_k$ 
            to any qubit $y_i$ if $s_i = 1$. \\
            For example, if $s = 110$, the least index $k$
            is 0 because $s_0 = 1$. Therefore, we apply two CNOT gates from $x_0$ to $y_0$ and
            from $x_0$ to $y_1$. 

            In fact, the Oracle circuits in Experiment 2 and Experiment 3 are also based
            on the two-step procedure. 

\end{description}

\vskip 0.25in

\begin{itemize}
\parskip 0in
\parsep 0in
\itemsep 0in

\item Click the cell which contains the code done in Experiment 3. Click the "Edit" button
      and select "Copy Cells". Click "Edit" again and choose "Paste Cells Below".

\item Assume $s$ is '110'. Replace the line 
\begin{verbatim}
     n = 3  
\end{verbatim}
            with
\begin{verbatim}
     s = '110'  
     n = len(s) 
\end{verbatim}

\item Modify the code in the {\tt Oracle()} function to construct the circuit of $F$ for $s = 110$
      using the two-step procedure described above.     

\item Increase the parameter shots from 1 to 1000 when calling {\tt aer\_sim.run()} because
      we want to collect all possible strings whose dot product with $s$ is 0.

\item To verify that those output strings which have a dot product of 0 with $s$, add the code below
      at the end of the program.  
\begin{verbatim}
   # Calculate the dot product of the results
   def bdotz(s, a):
       accum = 0
       for i in range(len(s)):
           accum += int(s[i]) * int(a[i])
       return (accum % 2)

   for a in answer:
       print( '{}.{} = {} (mod 2)'.format(s, a, bdotz(s,a)) )
\end{verbatim}

\item Click "Run" to execute the code and record the result. Use the first two output strings which are not '000'
      to figure out the string $s$. 

\end{itemize}

\subsection*{Experiment 5: The Oracle Construction method for Simon's Algorithm -- II }
\begin{itemize}
\parskip 0in
\parsep 0in
\itemsep 0in

\item Click the cell which contains the code done in Experiment 4. Click the "Edit" button
      and select "Copy Cells". Click "Edit" again and choose "Paste Cells Below".

\item Change the string $s$ from '110' to '01011'.

\item Modify the code in the {\tt Oracle()} function to construct the circuit of $F$ for $s = 01011$
      using the two-step procedure.     

\item Click "Run" to execute the code and record the result. Use the four linear independent output strings which are not '000'
      to figure out the string $s$. 


\end{itemize}

\vskip 0.1in

\noindent Click "File" and choose "Save as". Type "lab6" in the entry box and click "Save"
          to save your work today into the file {\tt lab6.ipynb}.  

\noindent To turn in your file, use CTRL-ALT-T to open a terminal and type\\
\mbox{\hskip 0.5in \tt ssh grail} \\
and type your password to login to the server grail. Then, type\\
\mbox{\hskip 0.5in \tt cd QC} \\
\mbox{\hskip 0.5in \tt turnin -c cis492s -p lab6 lab6.ipynb} \\
          to electronically turn in your file {\tt lab6.ipynb}. \\

\noindent  Shutdown the jupyter notebook and logout the workstation. \\
\noindent {\bf Hand in your lab report before your leave.}


\begin{center}
{\Large \bf CIS 492/593 $\;\;\;\;\;$ Lab 7 $\;\;$ Grover's Search Algorithm $\;\;\;\;\;$ Fall 202x  } \\
\end{center}

\vskip 0.25in
\noindent In this hands-on lab, you will learn and practice: 
\begin{itemize}
\parskip 0in
\parsep 0in
\itemsep 0in
\item the Grover's algorithm, including the Oracle construction and the Amplitude magnification method 
\item the Qiskit program for solving the 3-SAT problem via Grover's Search
\end{itemize}

\noindent Login to a workstation and open a terminal (either middle click the
terminal icon on the left side bar or press CTRL-ALT-T).
In the terminal window, type
\begin{verbatim}
     cd QC
     source /opt/qiskit_env/bin/activate
     jupyter notebook
\end{verbatim}
Inside the browser, click "New" and choose "Python 3 (ipykernel)".
For each experiment, you need to put the experiment number
as a comment (e.g. {\tt \# Experiment 1} ) in the top of the code.

\subsection*{Experiment 1: Grover's Algorithm with a 2-bit oracle }
\begin{itemize}
\parskip 0in
\parsep 0in
\itemsep 0in

\item In the empty cell, type \\
\mbox{\hskip 0.5in \tt \%load \url{~}cis492s/pub/Grover2.py }\\
and click "Run" to load the program which will find the desired element location 11 
among four binary strings: 00, 01, 10, and 11.  
In fact, this is a very simple SAT problem that
determines the satisfiability of the expression $q_0 \wedge q_1$.
 
\item Study the code. Note that the {\tt Oracle()} function builds and returns
the circuit which flips the sign of the probability amplitude of
the location we are trying to find. The function {\tt A()} performs
the work of amplitude amplification by flipping a sequence of numbers about their mean. 
Below is the circuit generated from the program:\\
\scalebox{1.0}{
\Qcircuit @C=2em @R=0.2em @!R { 
                                     &                      &                                    &        & Oracle &   &      & \;\;\;\;\;\;\;\;\;\; A \\
	 	\nghost{{q}_{0} :  } & \lstick{{q}_{0} :  } & \gate{\mathrm{H}} \barrier[0em]{1} & \qw & \ctrl{1} \barrier[0em]{1} & \qw & \gate{\mathrm{H}} & \gate{\mathrm{Z}} & \ctrl{1} & \gate{\mathrm
{H}} \barrier[0em]{1} & \qw & \qw & \qw\\
	 	\nghost{{q}_{1} :  } & \lstick{{q}_{1} :  } & \gate{\mathrm{H}} & \qw & \control\qw & \qw & \gate{\mathrm{H}} & \gate{\mathrm{Z}} & \control\qw & \gate{\mathrm{H}} & \qw & \qw & \qw\\
\\ }} \\

Recall that
\begin{equation*}
Z =
\begin{bmatrix}
1 & 0 \\
0 & -1 
\end{bmatrix}
, \;\;\;\;\;\;\;\;\;\;\;\;\;\;
Controlled{\text -}Z =
\begin{bmatrix}
1 & 0 & 0 & 0 \\
0 & 1 & 0 & 0 \\
0 & 0 & 1 & 0 \\
0 & 0 & 0 & -1 
\end{bmatrix}
\end{equation*}

\item We are interested in the operation matrices for the oracle and for the amplitude amplification
      circuits. Instead of calculating the matrices by hand, you can use the 
      {\tt Operator} class defined in the Qiskit {\tt quantum\_info} library.  
      Add the following line of the code \\
      \mbox{\hskip 0.75in \tt  display(rdm(quantum\_info.Operator(oc).data,n)) }  \\
      to the {\tt Oracle()} function before it returns. Also, Add the following line of the code \\ 
      \mbox{\hskip 0.75in \tt  display(rdm(quantum\_info.Operator(ac).data,n)) }  \\
      to the {\tt A()} function before it returns.
      Because the Qiskit Operator uses a reverse
      bit sequence (i.e. $q_1 q_0$), we implement and call the function {\tt rdm()}
      to reverse the bit sequence to be consistent with the order 
      (i.e. $q_0 q_1$) used in the Bernhardt's textbook.  

\item Click the cell containing the code and then click "Run" to execute it. 
      Record the matrix for the oracle. You may ignore the imaginary part
      (i.e. the term associated with the letter 'j'). Show which element
      location has the value {\tt -1}. Note that {\tt -1} will flip the
      probability amplitude of the element we are trying to find.

      Record the matrix for the amplitude amplification. Also ignore the imaginary part.
      Is the matrix the same as the the matrix A in the textbook (page 179, line 18)?

      Record and explain the result of the state vector (i.e. probability amplitude )
      (with the transpose operator).

      What is the probability for each location?

\end{itemize}
     
\subsection*{Experiment 2: Grover's Algorithm with another 2-bit oracle }
\begin{itemize}
\parskip 0in
\parsep 0in
\itemsep 0in

\item Click the cell which contains the code done in Experiment 1. 
      Click the "Edit" button
      and select "Copy Cells". Click "Edit" again and choose "Paste Cells Below".

\item Instead of finding '11' as in the Experiment 1,  you are asked to search the
       element location '01'. Note that this can be seen as a SAT problem that determines 
       the satisfiability of the expression $\neg q_0 \wedge q_1$.
       To achieve this, you only need to add two X-gates,
       one is before the Controlled-Z and the other is after the the Controlled-Z,
       on the qubits that should be zero. In this experiment, it is the qubit $q_0$
       which has to be enclosed by a pair of X-gates, just like the circuit below: \\ 
\scalebox{1.0}{
\Qcircuit @C=2.0em @R=0.2em @!R { 
                                     &                      &                                    &     &                   & Oracle   &                                    &     &     & \;\;\;\;\;\;\;\;\;\;\; A (same \; as \; Exp. \; 1) \\
                \nghost{{q}_{0} :  } & \lstick{{q}_{0} :  } & \gate{\mathrm{H}} \barrier[0em]{1} & \qw & \gate{\mathrm{X}} & \ctrl{1} & \gate{\mathrm{X}} \barrier[0em]{1} & \qw & \qw & \qw\\
                \nghost{{q}_{1} :  } & \lstick{{q}_{1} :  } & \gate{\mathrm{H}} & \qw & \qw & \control\qw & \qw & \qw & \qw & \qw\\
\\ }}

      Note that the first X-gate in the above diagram will swap the columns '00' and '10',
      and also swap the columns '01' and '11', in the diagonal matrix defined by the Controlled-Z,
      while the second X-gate will exchange the rows '00' and '10', and also exchange the rows
      '01' and '11'. These swappings will move the value {\tt -1} to the desired element location.\\    
      Modify the code in the {\tt Oracle()} function based on the diagram above.

\item Click "Run" to execute the code. 
      Record the matrix for the oracle. Again, ignore the imaginary part.
      Indicate the location which contains the value {\tt -1}.

      Record and explain the result of the state vector (i.e. probability amplitude )
      (with the transpose operator).

      What is the probability for each location?

\end{itemize}
     
\subsection*{Experiment 3: Grover's Algorithm with a 3-bit Oracle }
\begin{itemize}
\parskip 0in
\parsep 0in
\itemsep 0in

\item In the empty cell, type \\
      \mbox{\hskip 0.5in \tt \%load \url{~}cis492s/pub/Grover3.py }\\
      and click "Run" to load a program which will find the desired element location 111 
      among eight binary strings: 000, 001, 010, 011. 100, 101, 110, and 111.  
      That is, this is a SAT problem that determines the satisfiability of the expression 
      $q_0 \wedge q_1 \wedge q_2 $.

      Qiskit does not support the ccz gate. As shown in the {\tt Oracle()} function,
      we use a ccx gate and put a pair of H-gates before and after the X gate in ccx 
      to emulate the ccz operation. Note that $HXH = Z$.   

\item Click "Run" to execute the code. 
      Record the state vector result (with the transpose operator). Ignore the imaginary part.  
      You should get two different probability amplitudes.
      Compare these two probability amplitudes with the two probability amplitudes
      mentioned in the textbook (page 180, line 4). Before you compare, the two probability 
      amplitudes in the textbook should be divided by $\sqrt{8}$ due to $H^{\otimes 3}$.    

\item To see whether we can get a higher probability of the correct answer, we may    
      flip the sign of the probability amplitude associated with the desired location
      and then perform the amplitude amplification again. That is,
      we need to execute the function {\tt Oracle()} and {\tt A()} again. 
      This can be implemented by invoking the {\tt repeat(iter)} method in Qiskit where {\tt iter}
      is the number of repetition times.  Modify the code
\begin{verbatim}
# put the OA after the H-gates
qc.compose(OA, inplace=True)
\end{verbatim}
      to be
\begin{verbatim}
# put the repeated OA after the H-gates
qc.compose( OA.repeat(2), inplace=True)
\end{verbatim}

\item Click "Run" to execute the code. 

      Record the result of the state vector (i.e. probability amplitude )
      (with the transpose operator). 

      What is the probability for each location? 
      Does the probability of the desired location become higher?
       
\item Modify the code again to repeatedly execute the function {\tt Oracle()}
      and {\tt A()} three times. Record the result of the state vector (i.e. probability amplitude )
      (with the transpose operator). Does the probability of the desired location become higher? 
      If not, that means we 'overcook' the numbers. Change the repetition times back to 2.

\end{itemize}

\subsection*{Experiment 4: Grover's Algorithm with another 3-bit Oracle }

\begin{itemize}
\parskip 0in
\parsep 0in
\itemsep 0in

\item Click the cell which contains the code done in Experiment 3. Click the "Edit" button
      and select "Copy Cells". Click "Edit" again and choose "Paste Cells Below".

\item Using the method you learned in Experiment 2, modify the code in the 
      {\tt Oracle()} function to make the desired location be '001'.  
      Namely, it can be seen as a SAT problem that determines the satisfiability of the 
      expression $\neg q_0 \wedge \neg q_1 \wedge q_2$

\item Click "Run" to execute the code. 
      Record and explain the result of the state vector (i.e. probability amplitude )
      (with the transpose operator). 

\end{itemize}



\vskip 0.5in

\noindent Click "File" and choose "Save as". Type "lab7" in the entry box and click "Save"
          to save your work today into the file {\tt lab7.ipynb}. 

	   To turn in your file, use CTRL-ALT-T to open a terminal and type\\
\mbox{\hskip 0.5in \tt ssh grail} \\
and type your password to login to the server grail. Then, type\\
\mbox{\hskip 0.5in \tt cd QC} \\
\mbox{\hskip 0.5in \tt turnin -c cis492s -p lab7 lab7.ipynb} \\
          to electronically submit your file {\tt lab7.ipynb}. \\

           Shutdown the jupyter notebook and logout the workstation. \\

\noindent {\bf Hand in your lab report before your leave.}